%% file: main.tex
\documentclass[journal,twocolumn]{IEEEtran}

\usepackage{amsmath}
\usepackage{epsfig}
\usepackage{amssymb}
\usepackage{amsthm}
\usepackage{mathrsfs}
\usepackage{amsfonts}
\usepackage{amssymb}
\usepackage{subfigure}
\usepackage{epsfig}
\usepackage{epsf}
\usepackage{xcolor}
\usepackage{algorithm}
\usepackage{algpseudocode}
\usepackage[utf8]{inputenc}
\usepackage[english]{babel}
\usepackage{cite}

\usepackage{textcomp}
\usepackage{tabularx}
\usepackage{tikz}
\usepackage{enumitem}
\usepackage{url}
\usepackage{booktabs}
\usepackage{array}
\usepackage{multirow}
\usepackage{booktabs} 
\usepackage{siunitx}  
\usepackage{xfrac}
\usepackage{nicefrac}
\usepackage{psfrag}
\usepackage{times}
\usepackage{balance} 
\usepackage{amssymb}
\usepackage{amsfonts}
\usepackage{bm}
\usepackage[bottom]{footmisc}
\usepackage{mwe}
\usepackage{color}
\usepackage{booktabs}
\usetikzlibrary{patterns,arrows}

\definecolor{color1}{RGB}{237, 191, 193}
\definecolor{color2}{RGB}{229, 153, 157}
\definecolor{color3}{RGB}{225, 123, 116}
\definecolor{color4}{RGB}{217,87,77}
\definecolor{color5}{RGB}{203,52,38}

\definecolor{color6}{rgb}{0.56, 0.74, 0.56}
\definecolor{color7}{rgb}{0.31, 0.78, 0.47}
\definecolor{color8}{RGB}{60,192,158}
\definecolor{color10}{RGB}{30,192,128}
\definecolor{color9}{rgb}{0.0, 0.66, 0.47}
\definecolor{color11}{rgb}{0.0, 1.0, 0.5}

\def\l{\left}
\def\r{\right}
\def\({\l(}
\def\){\r)}
\def\[{\l[}
\def\]{\r]}

\def\BibTeX{{\rm B\kern-.05em{\sc i\kern-.025em b}\kern-.08em
    T\kern-.1667em\lower.7ex\hbox{E}\kern-.125emX}}

 {
    \theoremstyle{plain}
    
 }

\begin{document}

\title{ Predictive Closed-Loop Remote Control over Wireless Two-Way Split Koopman Autoencoder }

\author{Abanoub M.~Girgis,~\IEEEmembership{Member,~IEEE,} $^\dagger$Hyowoon~Seo,~\IEEEmembership{Member,~IEEE,} 
        $^\ddagger$Jihong~Park,~\IEEEmembership{Senior Member,~IEEE,}
        Mehdi~Bennis,~\IEEEmembership{Fellow,~IEEE,}
        and~$^\ddagger$Jinho~Choi,~\IEEEmembership{Senior Member,~IEEE} 
\thanks{This work is supported by Academy of Finland $6$G Flagship (grant no. $318927$) and project SMARTER, projects EU-ICT IntellIoT and EU-CHISTERA LeadingEdge , and CONNECT, Infotech-NOOR, and NEGEIN. This work was supported by the National Research Foundation of Korea(NRF) grant funded by the Korea government(MSIT) (No. 2022R1F1A1075078). A preliminary conference version of this work appeared in the proceedings of IEEE PIMRC-2021~\cite{girgis2021split}. \emph{(Corresponding author: Hyowoon Seo)}}        
\thanks{A. Girgis and M. Bennis are with the Centre for Wireless Communications, University of Oulu, 90014 Oulu, Finland (e-mail: abanoub.pipaoy@oulu.fi, mehdi.bennis@oulu.fi).}
\thanks{$^\dagger$H. Seo is with the Department of Electronics and Communications Engineering, Kwangwoon University, Seoul 01897, Korea (e-mail: hyowoonseo@kw.ac.kr).}
\thanks{$^\ddagger$J. Park and J. Choi are with the School of Information Technology, Deakin University, Geelong, VIC 3220, Australia (e-mail: jihong.park@deakin.edu.au, jinho.choi@deakin.edu.au).}}

\maketitle

\begin{abstract}
Real-time remote control over wireless is an important-yet-challenging application in 5G and beyond due to its mission-critical nature under limited communication resources. Current solutions hinge on not only utilizing ultra-reliable and low-latency communication (URLLC) links but also predicting future states, which may consume enormous communication resources and struggle with a short prediction time horizon. To fill this void, in this article we propose a novel two-way Koopman autoencoder (AE) approach wherein: 1) a sensing Koopman AE learns to understand the temporal state dynamics and predicts missing packets from a sensor to its remote controller; and 2) a controlling Koopman AE learns to understand the temporal action dynamics and predicts missing packets from the controller to an actuator co-located with the sensor. Specifically, each Koopman AE aims to learn the Koopman operator in the hidden layers while the encoder of the AE aims to project the non-linear dynamics onto a lifted subspace, which is reverted into the original non-linear dynamics by the decoder of the AE. The Koopman operator describes the linearized temporal dynamics, enabling long-term future prediction and coping with missing packets and closed-form optimal control in the lifted subspace. Simulation results corroborate that the proposed approach achieves a 38x lower mean squared control error at 0 dBm signal-to-noise ratio (SNR) than the non-predictive baseline.  
\end{abstract}


\IEEEpeerreviewmaketitle 

\section{Introduction}

One of the visions of beyond fifth generation (5G) and 6G communication systems \cite{yang20196g,saad2019vision,latva2020key} is to leverage sensing, communication, and connectivity in a closed-loop integrated manner. The basic principle of closed-loop remote control have been studied in the literature of wireless networked control systems~\cite{park2017wireless}. Nonetheless, the feasibility of these methods becomes questionable in the presence of mission-critical remote control applications including remote surgery~\cite{meng2004remote}, industrial internet of things (IIoT) control in a smart factory~\cite{liu2019taming}, autonomous vehicular platooning~\cite{zeng2019joint}, low-earth-orbit (LEO) satellite maneuvering to avoid space debris \cite{razzaghi2021real}, and so forth. These remote control applications not only impose extremely stringent communication latency requirements but also are unable to afford missing any single control command.
 
To address these concerns, one could exploit wireless communication that guarantee extremely high reliability with low latency. Indeed, supporting mission-critical applications over wireless links has been investigated in the context of ultra-reliable and low-latency communication (URLLC) \cite{Docomo:18,angjelichinoski2019statistical}. However, URLLC is originally intended to support short packets~\cite{PetarURLLC:17}, in which the guaranteed reliability and latency are challenged by the long packets transmission requirements in the aforementioned mission-critical applications. Furthermore, even one-way URLLC requires an enormous amount of bandwidth to reserve dedicated channels  while consuming a significant amount of energy to increase transmit power~\cite{bennis2018ultrareliable}. This challenge is aggravated when supporting closed-loop control applications where errors may propagate during the two-way closed-loop interaction.

 On the other hand, by leveraging recent advances in data-driven machine learning (ML), one could learn to predict future states (e.g., position, velocity, and temperature) based on previous data samples. This enables to carry out proactive decision-making for reduced control latency or equivalently for improved reliability by communicating the same messages multiple times \cite{park2017coverage}. Time-series prediction falls in this category, in which recurrent neural networks (RNNs) such as long short-term memory (LSTM) \cite{bahdanau2014neural} and gated recurrent unit (GRU) \cite{cho2014learning} are widely used for this purpose. While effective, their prediction accuracy often decreases sharply with the forecast time horizon, limiting their applicability only to short-term prediction such as forecasting the next video frame within a few milliseconds~\cite{Virtual_Reality}. Such a time horizon may be too short to evaluate an optimal control action, particularly under non-linear and complex state dynamics. 

 To fill this void, in this article we propose a novel ML-based closed-loop remote control framework inspired by Koopman theory \cite{koopman1931hamiltonian, birkhoff1932recent} and split learning \cite{Vepakomma:2018:Splita,park2019wireless} using an autoencoder (AE) architecture, i.e., Koopman AE. In particular, we consider an AE split into its encoder and decoder parts, whereby the smashed data or latent representations propagate from the encoder to the decoder while the prediction errors or gradients propagate back in the reverse direction \cite{Vepakomma:2018:Splita,park2019wireless}. The encoder part stored at the transmitter learns to map the input data into its lower-dimensional latent representation which is then reconstructed by the decoder stored at the receiver. This mapping reduces the communication payload size in the case of discovering the Koopman invariant subspace that ensures long-term prediction accuracy and control system stability\footnote{Ideally, according to Koopman theory, the lower-dimensional mapping in the Koopman operator amounts to lossless compression. However, the Koopman AE based approach might lose system reliability depending on how low dimensional space the input data are mapped into. \cite{theis2017lossy}.}

 Subsequently, rather than predicting individual future states, we learn the temporal state dynamics as a whole. However, discovering the dynamics in the raw state space is difficult notably when the dynamics are non-linear. Instead, inspired by Koopman operator theory \cite{koopman1931hamiltonian, birkhoff1932recent}, we aim to identify the dynamics in a lifted subspace in which the latent dynamics are linearized such that multiplying the Koopman operator $M$ times yields the state prediction in $M$ time slots. This provides an additional benefit of applying a closed-form optimal control derived from the linearized subspace, thereby reducing the control overhead in terms of computation. Additionally, the control errors are reduced compared to the linear approximation method around an equilibrium point that becomes vacuous when the initial condition goes far from the equilibrium point. Recent works have demonstrated that the matrix form of the Koopman operator can be found in the hidden layers of an AE \cite{lusch2018deep}, which coincides with the aforementioned split AE architecture. 

 Lastly, in closed-loop remote control, we avoid error propagation by additionally learning the temporal action dynamics as illustrated in Fig.~1. Consequently, in the forward link, any missing state reception due to poor channel conditions or excessive latency can be predicted by the remote controller that understands the state dynamics. Likewise, in the reverse link, any missing control action command receptions can be replaced by the actions predicted by the actuator that understands the action dynamics.

 \subsection{Backgrounds and Related Works}
 
 In the literature of remote control over wireless links, one central question is how to cope with communication imperfections on the control system operations. These works are often termed communication and control co-design (CoCoCo) that deal with the trade-off between wireless resource consumption and control stability~\cite{CoCoCo,Gigis21_twoGPR}. In this direction, dynamic sensor scheduling approaches were proposed to cope with time-varying control and channel states, thereby improving control stability and communication efficiency~\cite{yu2019event,WSN}. In~\cite{gatsis2015opportunistic,BoChang}, a channel state information (CSI) aware scheduling and power allocation method was introduced to minimize the total power consumption while guaranteeing a target control performance under limited communication resources. In order to meet stringent latency and reliability requirements in the context of URLLC and time-sensitive control systems, control-dynamics and CSI aware resource allocation and scheduling solutions were developed in~\cite{eisen2019control,eisen2019control2}, which can partly relax URLLC requirements without compromising the control performance. In our prior work~\cite{Gigis21_twoGPR}, a predictive and control-aware scheduler was proposed, in which the future states and actions are locally predicted using a Gaussian process regression approach (GPR) at the controller and actuator, respectively. Nevertheless, these works only consider linear control systems, questioning their feasibility when the system dynamics are non-linear.

  Non-linear control system dynamics have been extensively studied in the field of control theory. A periodic event-triggered control is proposed in~\cite{wang2016stabilization,aranda2017design,6761063} to stabilize non-linear control systems under communication constraints. According to this scheme, the event-triggered condition is only updated at some sampling instances to preserve control stability and reduce the number of transmissions. However, the previously mentioned scheme in~\cite{wang2016stabilization,aranda2017design,6761063} suffers from high computational complexity due to the system's non-linearity. Furthermore, it suffers from long-term savings of the wireless communication resources as it fails to identify the non-linear system dynamics at the controller. In~\cite{li2013network,de2008lyapunov}, a model predictive control approach is proposed for non-linear control systems to compensate for two-channel packet loss. However,~\cite{li2013network,de2008lyapunov} assume that the accurate non-linear system dynamics are given at the remote controller to compute the control action based on a non-linear optimization problem. Hence, this approach suffers from impracticality due to the previously specified non-linear dynamics and inaccuracy due to the uncertainty inherited in practical systems. In addition, it has a high computational burden due to the non-convex optimization problem. Finally, this approach requires transmitting a finite number of predicted control actions in one packet to the actuator, affecting the transmission delay.

 To make the non-linearity amenable to stability analysis, 
 the Jacobian linearization is a well-known method that provides a linear approximation around the equilibrium point using the Taylor series expansion~\cite{fadali2013digital,tailor2011linearization}. However, such an approximation becomes vacuous when the states of interest are far from the equilibrium point~\cite{fadali2013digital,tailor2011linearization}. As an alternative, Koopman operator methods have been suggested in~\cite{koopman1931hamiltonian,mezic2017koopman}, which represent finite-dimensional and non-linear system states using infinite-dimensional and linear forms via linearly evolving functions of states, termed observables. Such Koopman representations can be obtained using the dynamic mode decomposition (DMD) algorithm as shown in open-loop~\cite{schmid2010dynamic} and closed-loop~\cite{proctor2016dynamic} scenarios. The main challenge in this direction is to find the minimum number of observables that sufficiently describe given non-linear system dynamics~\cite{schmid2010dynamic,proctor2016dynamic}. In~\cite{lusch2018deep}, a deep learning-based method was proposed for providing finite-dimensional Koopman representations using an AE, i.e., Koopman AE. Leveraging and extending this method that originally ignores wireless connectivity, in this work we propose a two-way Koopman AE framework using a pair of two AE architectures to cope with wireless communication outages that may propagate control errors in closed-loop systems, hindering control stability.
 
 \begin{figure*}[t]
 \centering
 \subfigure[Wireless Split Sensing Koopman Autoencoder.]{\includegraphics[width=0.85\textwidth]{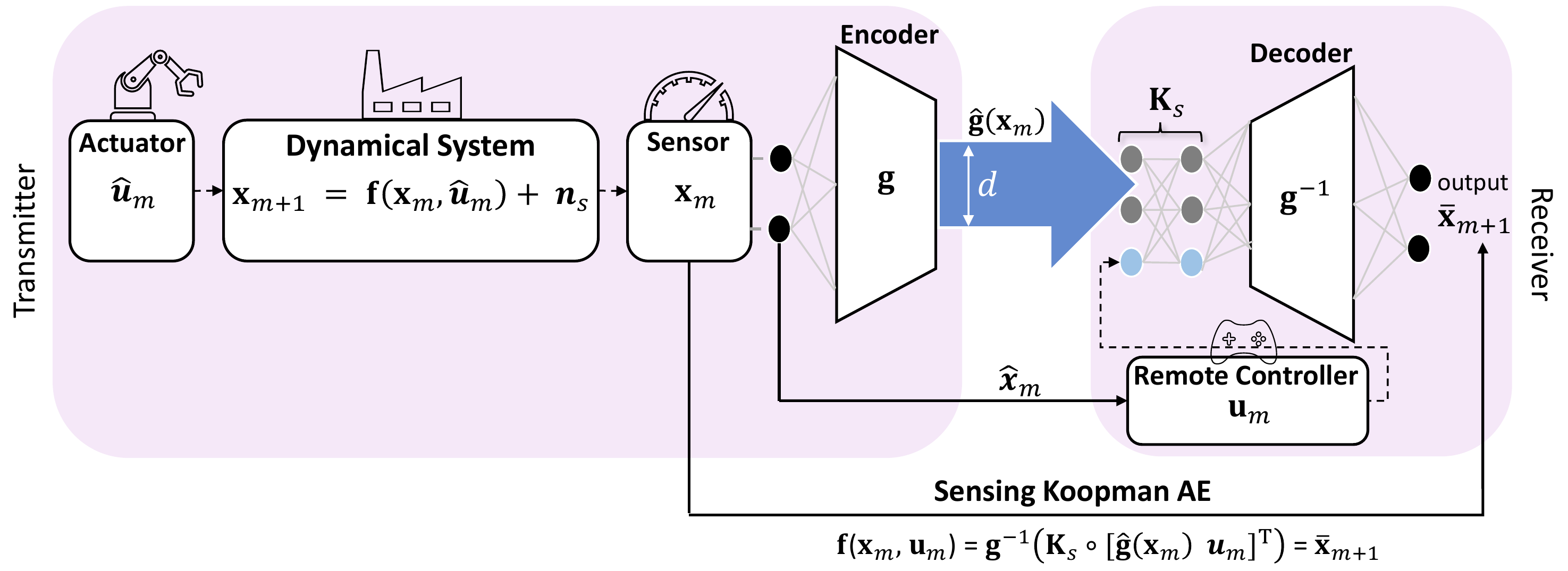}} 
    \subfigure[Wireless Split Controlling Koopman Autoencoder.]{\includegraphics[width=0.85\textwidth]{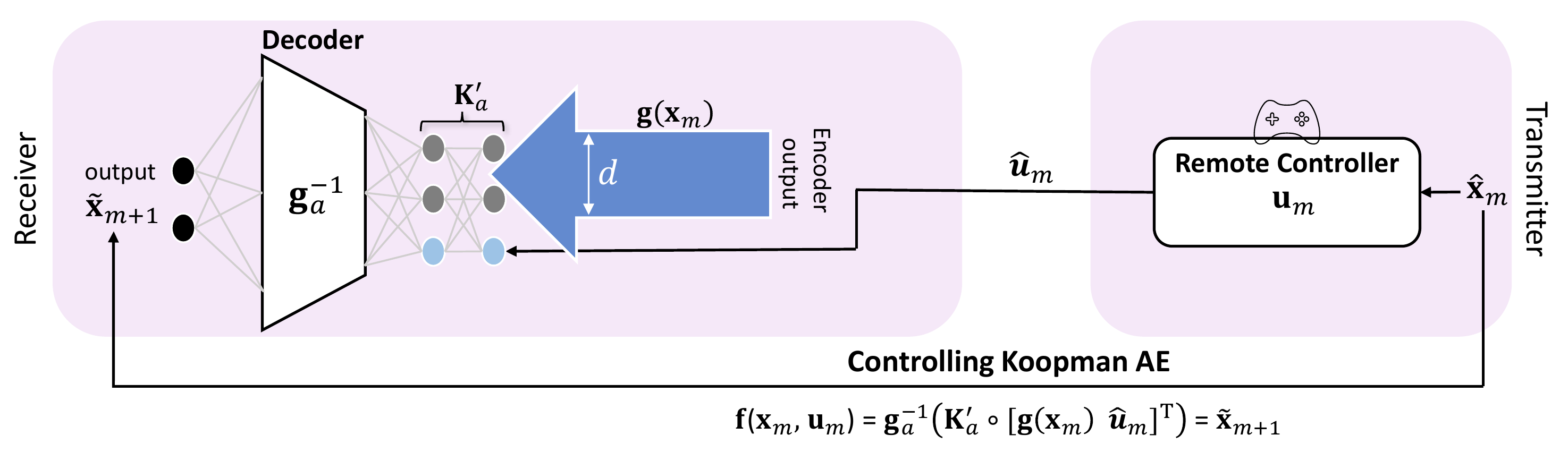}} 
\caption{Schematic representation of WNCS via two-way split Koopman autoencoder architecture.}
    \label{fig_system}
 \end{figure*}

 \subsection{Contributions and Organization}

  We propose a split learning of Koopman operator for real-time remote control systems. The major contributions of this paper can be summarized as follows.
 \begin{itemize}
\item For predictive closed-loop remote control, we propose a two-way split Koopman AE architecture comprising: (i) \emph{a sensing Koopman AE} from a sensor to its remote controller in the forward link, and (ii) \emph{a controlling Koopman AE} from the controller to an actuator co-located with the sensor in the reverse link (see. Fig.~1).

\item To train the proposed architecture, for the sensing Koopman AE, we first derive the temporal state evolution with the Koopman operator (see (16) in Sec. III-B), and provide the training loss function (see (26) in Sec. III-C). For the controlling Koopman AE, we derive the temporal action evolution with the Koopman operator (see (27) in Sec. III-E), and provide the training loss function (see (29) in Sec. III-E).

\item By simulation in an inverted cart-pole system we corroborate that the proposed two-way split Koopman AE is robust to consecutive packet losses. At $0$ dBm signal-to-noise ratio (SNR) in both forward and reverse links, we observe that the mean squared control error (MSCE) of the proposed architecture is $38$x lower than that of the remote control without prediction (see Fig. 8 in Sec. IV).

\item Given the linearized temporal state evolution via the Koopman operator, we derive a closed-form optimal control action (see (19) in Sec. III-C) and observe that the proposed approach has better control performance than the standard Jacobian linearization technique for non-linear control (see Fig. 9 in Sec. IV).

\item Finally, with an extensive set of simulation results, we show the impact of SNR, latent representation dimension, number of control trajectories, and training dataset size on the state/action prediction error (see Tables I and II), Koopman AE convergence (Fig. 4), and training completion time (Figs. 5-7 in Sec.~IV). 

 \end{itemize}
 
 Note that our preliminary work has shown the effectiveness of the one-way sensing Koopman AE in a remote monitoring scenario~\cite{girgis2021split}. Extending this to the closed-loop remote control scenario under study is a daunting task mainly due to the error propagation within the loop. We address this challenge by developing a two-way Koopman AE framework that additionally introduces another Koopman AE for control action prediction. Furthermore, compared to the preliminary version, in this article we provide a more extensive set of simulation results, advocating the feasibility of the Koopman AE framework under a wide range of future prediction intervals, different control system parameters, and different communication costs in terms of SNR and payload sizes.

 The remainder of this paper is organized as follows. In Section~\ref{System_Model}, we specify the wireless networked control systems (WNCS) architecture including the models for control and communication systems. In Section~\ref{Split_Closed_Loop_Koopman}, we present the proposed two-way split Koopman AE architecture for predictive and linear control. In Section~\ref{Simulation_Results} and Section~\ref{conclusion}, we present the simulation results, and conclude the paper.

\input{Sec2_System_Model.tex}
\input{Sec3_Two_Way_Predictive_Control.tex}
\input{Sec4_Simulation_Results.tex}

\section{Conclusion}
\label{conclusion}
 In this article, we proposed a two-way Koopman AE split learning framework for closed-loop real-time remote control with a sensing Koopman AE and a controlling Koopman AE that predict missing packets of state and action information, respectively. Numerical results demonstrate that the proposed approach predicts the future control system state and action with high accuracy for a practical range of target SNR, latent representation dimensions, and training periods. Leveraging the proposed method, developing multiple access schemes, and extending its applicability to multiple closed-loop control systems could be interesting topics for future research. To further improve the communication efficiency, transmitting partial state observations emanating from utilizing multiple distributed sensors exploiting a scheduling scheme could be another interesting topic for future work. 

\ifCLASSOPTIONcaptionsoff
  \newpage
\fi

\bibliographystyle{IEEEtran}
\bibliography{IEEEabrv,bibliography}

\newpage

\begin{IEEEbiography}[{\includegraphics[width=1in,height=1.25in,clip,keepaspectratio]{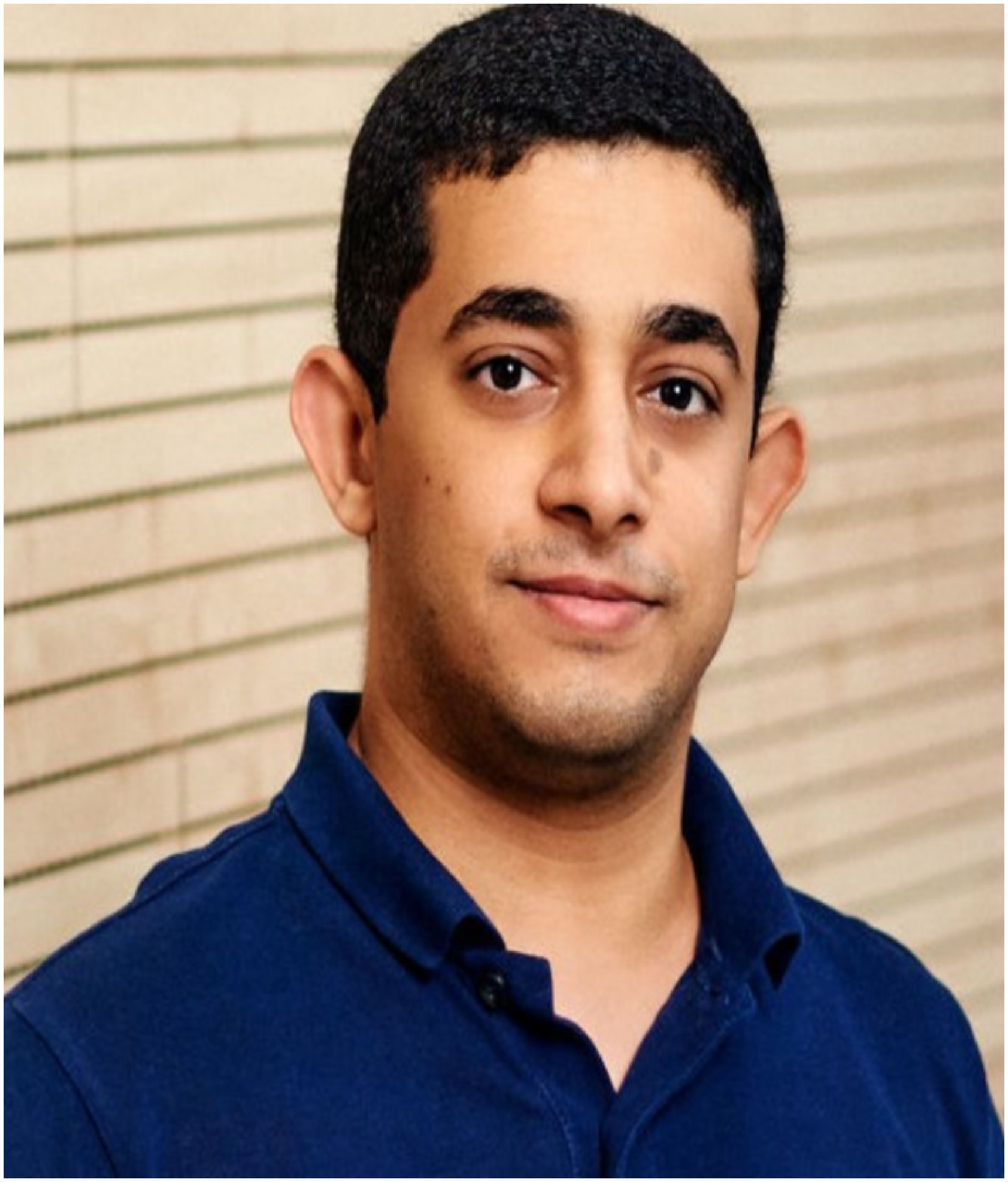}}]{Abanoub~M.~Girgis} (Member, IEEE) received his B.Sc. degree (Hons) in Electronics and Communications Engineering from Thebes Higher Institute of Engineering, Egypt, in 2013 and the M.Sc. degree in Electronics and Communications Engineering from Ain Shams University, Egypt, in 2018. He was a teaching assistant at Thebes Higher Institute of Engineering, Egypt, from 2013 to 2018. He is currently pursuing his Ph.D. degree at the Centre for Wireless Communication, University of Oulu, Finland. His main research interests include Cyber-physical systems, communication and control co-design, semantic communication, massive MIMO, and signal processing.
\end{IEEEbiography}

\begin{IEEEbiography}[{\includegraphics[width=1in,height=1.25in,clip,keepaspectratio]{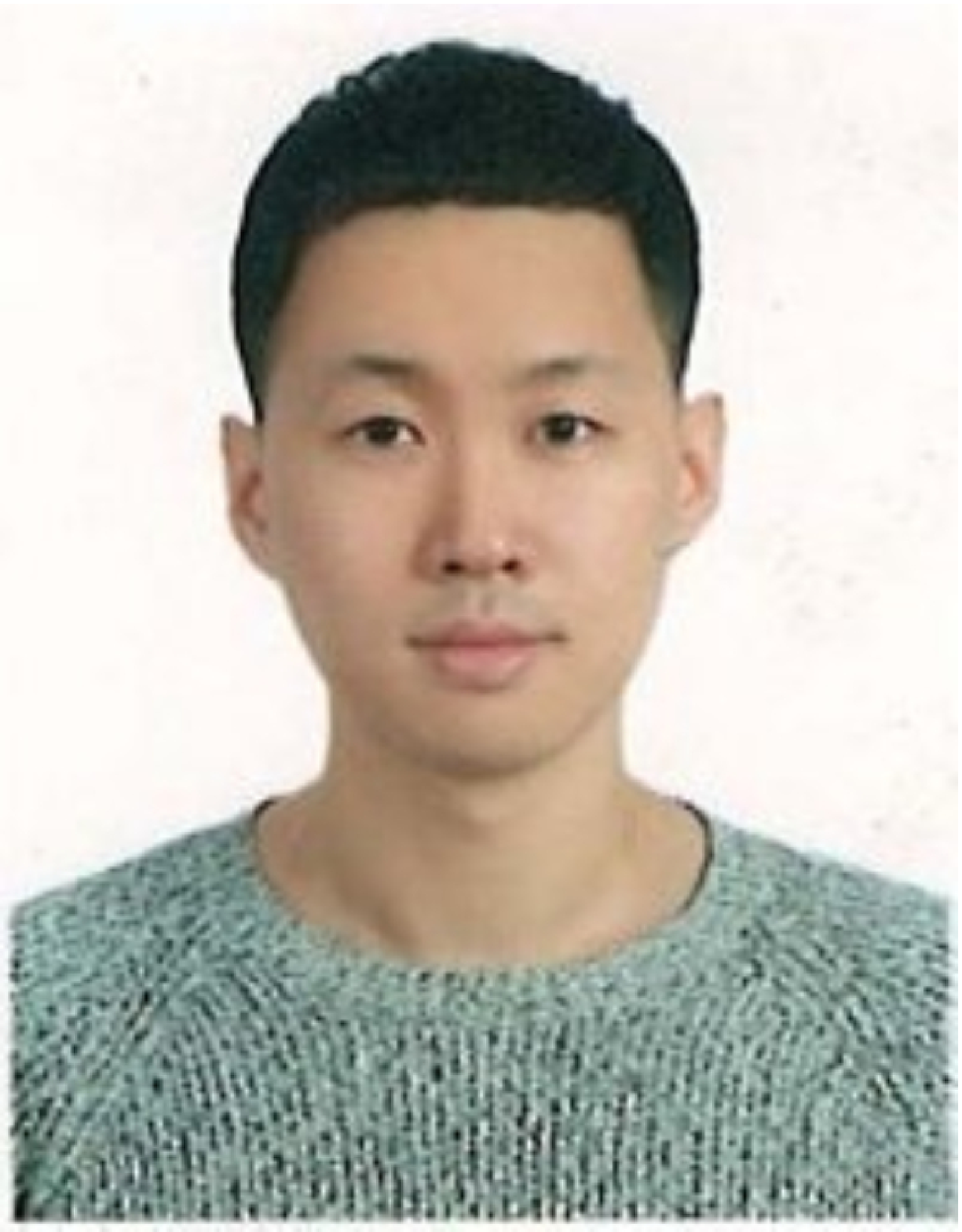}}]{Hyowoon~Seo} (Member, IEEE) is an assistant professor of the Department of Electronics and Communications Engineering of Kwangwoon University, Seoul, Korea. He received the B.S, M.S, and Ph.D. degree from the School of Electrical Engineering (EE), Korea Advanced Institute of Science and Technology (KAIST), Daejeon, Korea, in 2012, 2014, and 2020, respectively. He was a post-doctoral research fellow at Seoul National University, Seoul, Korea, and University of Oulu, Oulu, Finland. His research interests include wireless communication, Internet-of-Things, physical layer security and privacy, vehicle-to-everything (V2X) communications, and deep and distributed learning. He received Gold Prize in the 24th Samsung Humantech Paper Award in 2018.
\end{IEEEbiography}

\begin{IEEEbiography}[{\includegraphics[width=1in,height=1.25in,clip,keepaspectratio]{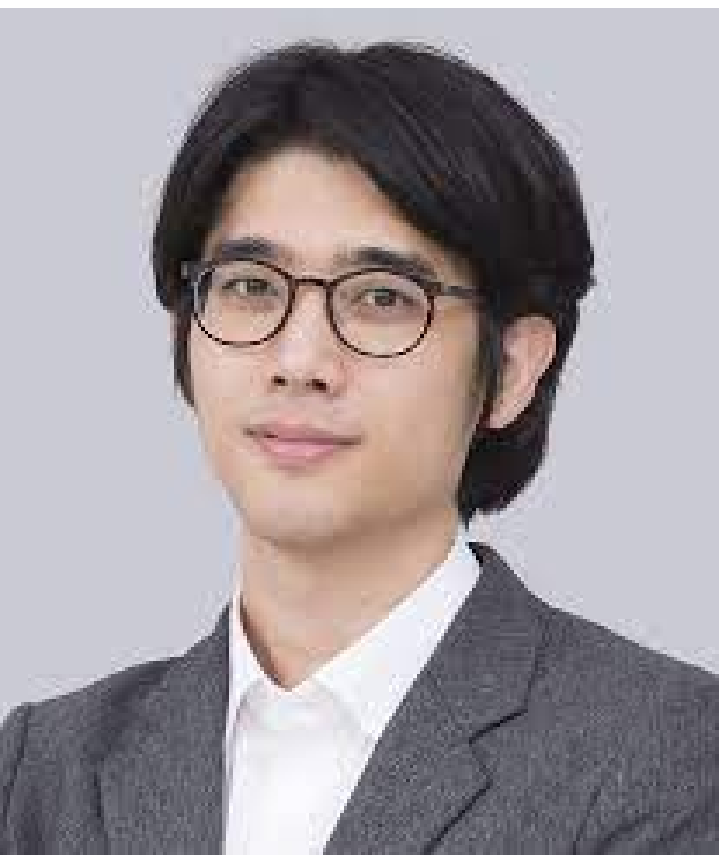}}]{Jihong~Park} (Senior Member, IEEE) received the B.S. and Ph.D. degrees from Yonsei University, South Korea. He is currently a Lecturer (Assistant Professor) with the School of Information Technology, Deakin University, Australia. His research interests include His recent research focus includes distributed machine learning, control, and resource management, as well as their applications to 6G semantic, AI-native, and non-terrestrial communications. He served as a Conference/Workshop Program Committee Member for IEEE GLOBECOM, ICC, and INFOCOM, and for NeurIPS, ICML, and IJCAI. He received the IEEE GLOBECOM Student Travel Grant and the IEEE Seoul Section Student Paper Contest Bronze Prize in 2014, the 6th IDIS-ETNEWS Paper Award, and FL-IJCAI Best Student Paper Award in 2022. Currently, he is an Associate Editor of Frontiers in Data Science for Communications and in Signal Processing for Communications. He is a Senior Member of IEEE and a Member of ACM.
\end{IEEEbiography}

\begin{IEEEbiography}[{\includegraphics[width=1in,height=1.25in,clip,keepaspectratio]{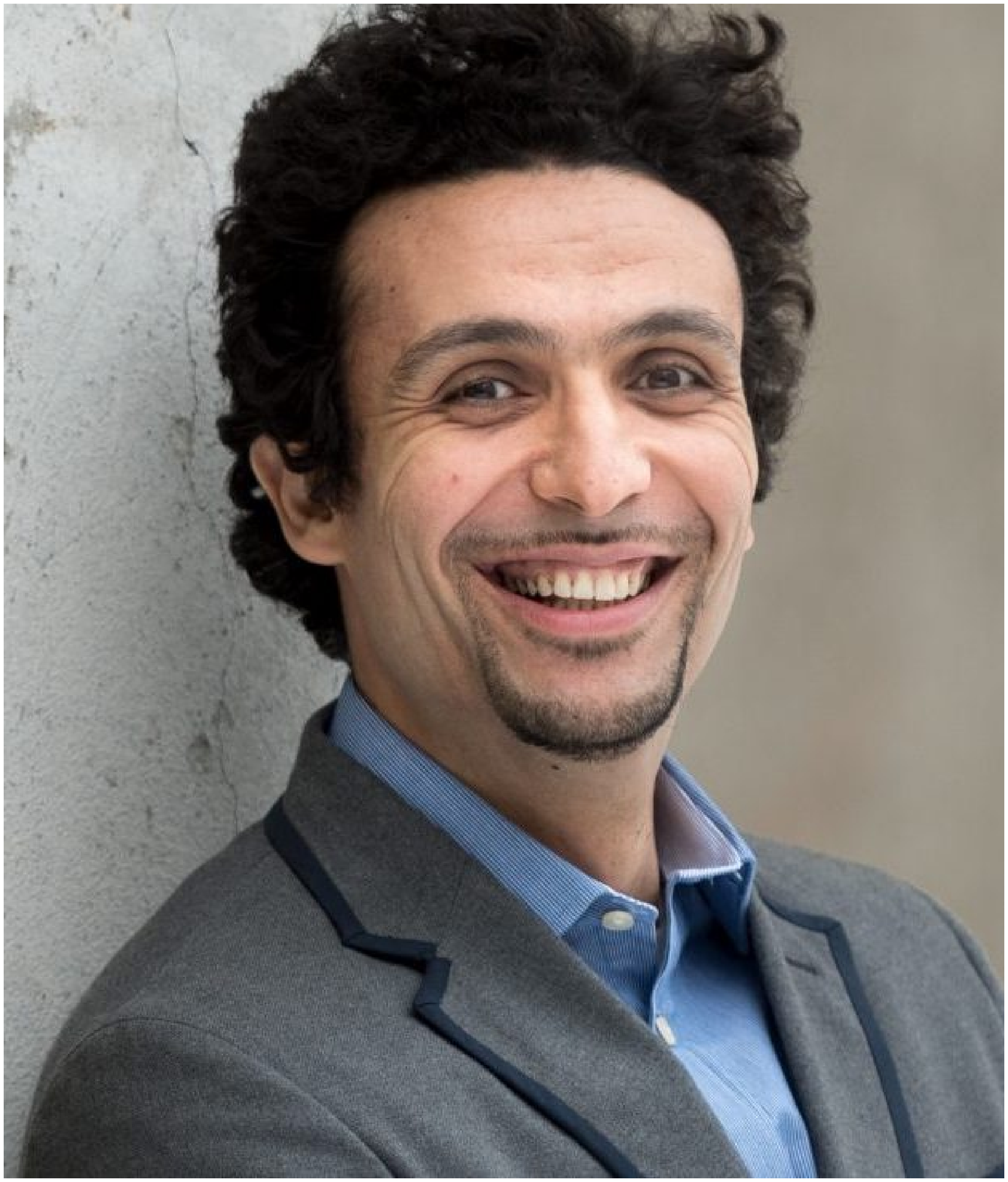}}]{Mehdi~Bennis} (Fellow, IEEE) is a full (tenured) Professor at the Centre for Wireless Communications, University of Oulu, Finland and head of the intelligent connectivity and networks/systems group (ICON). His main research interests are in radio resource management, game theory and distributed AI in 5G/6G networks. He has published more than 200 research papers in international conferences, journals and book chapters. He has been the recipient of several prestigious awards including the 2015 Fred W. Ellersick Prize from the IEEE Communications Society, the 2016 Best Tutorial Prize from the IEEE Communications Society, the 2017 EURASIP Best paper Award for the Journal of Wireless Communications and Networks, the all-University of Oulu award for research, the 2019 IEEE ComSoc Radio Communications Committee Early Achievement Award and the 2020 Clarviate Highly Cited Researcher by the Web of Science. Dr. Bennis is an editor of IEEE TCOM and Specialty Chief Editor for
Data Science for Communications in the Frontiers in Communications and
Networks journal. Dr. Bennis is an IEEE Fellow.
\end{IEEEbiography}

\begin{IEEEbiography}[{\includegraphics[width=1in,height=1.25in,clip,keepaspectratio]{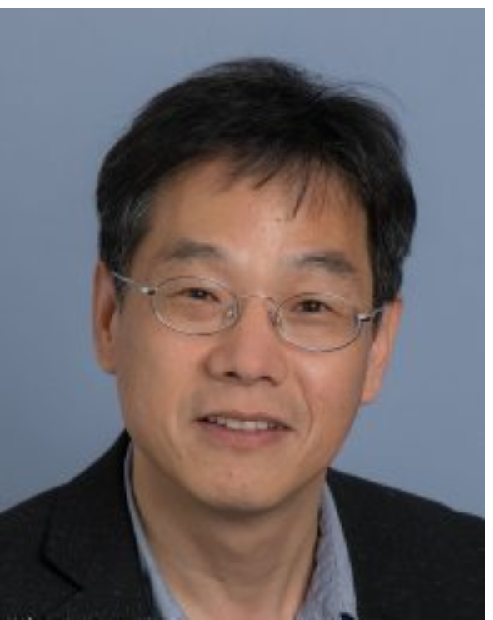}}]{Jinho~Choi} (Senior Member, IEEE) is with the School of Information Technology, Burwood, Deakin University, Australia, as
a Professor. Prior to joining Deakin in 2018, he was with Swansea University, United Kingdom, as a Professor/Chair in Wireless, and Gwangju Institute of Science and Technology (GIST), Korea, as a Professor. His research interests include the Internet of Things (IoT), wireless communications, and statistical signal processing. He authored two books published by Cambridge University Press in 2006 and 2010. Prof. Choi received a number of best paper awards including the 1999 Best Paper Award for Signal Processing from EURASIP. He is on the list of Worlds Top 2\% Scientists by Stanford University in 2020 and 2021. Currently, he is an Editor of IEEE Wireless Communications Letters and a Division Editor of Journal of Communications and Networks (JCN). He has also served as an Associate Editor or Editor of other journals including IEEE Trans. Communications, IEEE Communications
Letters, IEEE Trans. Vehicular Technology, JCN, and ETRI journal.
\end{IEEEbiography}

\end{document}

%% file: Sec2_System_Model.tex

\section{System Model}
\label{System_Model}

 \label{Wireless_Networked_Control_System_Architecture}
 Consider a closed-loop \emph{WNCS} which comprises a \emph{plant} that exhibits non-linear dynamics, a \emph{sensor} that samples the plant's state, a \emph{controller} that computes optimal control actions based on the sampled states, and an \emph{actuator} that applies the control action command on the plant received from the controller. Specifically, as illustrated in Fig.~\ref{fig_system} we assume that the sensor and actuator are co-located and share the same transceiver. On the other hand, the controller is located away from them with a computational capability to produce control action commands to ensure system stability. On this account, the controller operates remotely over wireless channels, by receiving the sensors' sampled states via the forward link and feeding back the computed optimal control action commands to the actuators via the reverse link.
 
 \subsection{Control System Architecture}
 Consider that the control system operates at a fixed \emph{control period} $\tau_o>0$ seconds to keep the plant stable over time. Hence the sensor samples the state of the plant at a fixed sampling rate $1/\tau_o$ and transmits the samples to the controller over wireless links. The $m$-th sampled state of the plant at time $t = m\tau_o$ is denoted by a $p$-dimensional vector $\mathbf{x}_{m} = [x_{m,1}\ x_{m,2}\ \cdots\ x_{m,p}] \in \mathbb{R}^{p}$. On the other hand, an optimal control action command computed at the controller based on the $m$-th sample $\mathbf{x}_{m}$ is denoted by a $q$-dimensional vector $\mathbf{u}_{m} = [u_{m,1}\ u_{m,2}\ \cdots\ u_{m,q}] \in \mathbb{R}^{q}$. Moreover, the state evolution of the plant's non-linear dynamics at time $t = (m+1)\tau_o$ is given as
 \begin{align}
 \label{eq_nonlinear_fn}
 \mathbf{x}_{m+1} &=  \mathbf{f} ( \mathbf{x}_{m},  \mathbf{u}_{m} ) + \mathbf{n}_{s}\\
 &= \mathbf{f}_s( \mathbf{x}_{m} ) + \sum^{q}_{i=1} \mathbf{f}_{u,i} ( \mathbf{x}_{m} ) u_{m,i}  + \mathbf{n}_{s},\label{eq_affine_Control}
 \end{align}
 where $\mathbf{n}_{s} \in \mathbb{R}^{p}$ is a $p$-dimensional random system noise vector at $t = (m+1)\tau_o$. The entries of the noise vector are assumed to be independently and identically distributed (i.i.d.) Gaussian random variables with zero mean and variance $N_{s}$ for all $m\in\mathbb{Z}_{+}$. The function $\mathbf{f}:\mathbb{R}^{p}\times\mathbb{R}^{q} \to \mathbb{R}^{p}$ is a non-linear state transition function of the plant's current state and control action command that steers the plant's state forward in time. Specifically, we introduce a model that disambiguates the state dynamics from the effect of the control actuation by reformulating the non-linear dynamics into a control-affine non-linear system \cite{7799268,korda2018linear}  as \eqref{eq_affine_Control}, where $\mathbf{f}_s(\cdot): \mathbb{R}^{p} \rightarrow \mathbb{R}^{p} $ is the unforced system dynamics and $\mathbf{f}_{u,i}(\cdot): \mathbb{R}^{p} \rightarrow \mathbb{R}^{p}$ is a state dependant control coupling term for $u_{m,i}$ for all $i \in \{1,\dots,q\}$.

 The time duration between $t = m\tau_o$ and the time point at which an actuator receives the estimated control action command $\hat{\mathbf{u}}_m$ corresponding to the plant's state $\mathbf{x}_m$ is referred to as \emph{control loop duration} and denoted by
 \begin{align}
 \tau_m = \tau_{m,comm} + \tau_{m,comp}
 \end{align}
 for all $m\in\mathbb{Z}_+$. This control loop duration incorporates the communication delay $\tau_{m,comm}$ through forward and reverse links and computing delay $\tau_{m,comp}$ for obtaining the optimal control action command. On the one hand, in the Jacobian baseline, this calls for the forward link to deliver the sampled states to the controller, while the reverse link are also needed for the controller to deliver the obtained control action command to the actuator. On the other hand, in the proposed approach explained in the following section, the system state can be predicted using the trained sensing Koopman AE, thereby only requiring the reverse link as long as the state prediction is accurate.
 
 \subsection{Wireless Communication Model}
 \label{Sensing_and_Actuating_Communications}
Suppose the forward and reverse links follow a time division multiple access (TDMA) approach, and assume that channel reciprocity holds between forward and reverse channels under the assumption that the sensor and actuator are co-located. Moreover, we consider a Rayleigh block fading channel model, where the coherence time is strictly longer than the control loop time. The path-loss of both forward and reverse links is given as
 \begin{align}
     \label{eq_Path_loss}
     \mathrm{PL}_{\mathrm{dB}} \left( D \right)  = \mathrm{PL}_{\mathrm{dB}} \left( D_0 \right)  + 10 \eta\log_{10} \left( \frac{D}{D_0} \right),
 \end{align}
  where $D$ is the distance from sensor and actuator to the controller, $\mathrm{PL}_{\mathrm{dB}} \left( D_o \right)$ denotes the path-loss at the reference distance $D_o$, and $\eta \geq 2$ is the path-loss exponent. Hence the received SNR at the remote controller is given as
 \begin{align}
     \label{eq_SNR_UL}
     \text{SNR}_{m} = 10^{-\frac{\mathrm{PL}_{\mathrm{dB}}\left( D_o \right)}{10}}\frac{P_t | h_{m}|^{2}}{ N_c }\left(\frac{D_o}{D}\right)^{\eta},
 \end{align}
 where $h_{m}$ is the Rayleigh fading channel gain from the sensor to the controller (or the controller to the actuator) at the $m$-th control loop duration, where $|h_{m}|^2$ is exponentially distributed with parameter $\lambda = 1$ for all $m \in \mathbb{Z}_+$, $P_t$ is the transmission power used at both communication ends, and $N_{c}$ is the channel noise power. In addition, we suppose that the channel gains are i.i.d. over time. Having \eqref{eq_SNR_UL}, the transmission rate of both forward and reverse link communications in the control system during the $m$-th control loop can be expressed as
 \begin{align}
     \label{eq_Shannon_data_rate}
     R_{m} = W \log_{2} \left( 1 + \text{SNR}_{m} \right),
 \end{align}
 where $W$ is the communication bandwidth.
 
 To stabilize the control system, the control loop duration should be at least shorter than the control period, that is $\tau_{m} < \tau_{o}$ for all $m \in \mathbb{Z}_{+}$. Denote by $L_{m}$ the overall communicating information bit length in the $m$-th control loop duration, such that $\frac{L_{m}}{R_{m}}$ indicates the communication time. Then, the outage probability of the $m$-th control loop can be expressed as
 \begin{align}
 \epsilon_{m} &= \mathbb{P}\left[ \tau_{m,comm} > \tau_{o} - \tau_{m,comp} \right]\\
 &= 1 - \text{exp} \left[ -10^{\frac{\mathrm{PL}_{\mathrm{dB}}(D_0)}{10}} \frac{ N_c D^{\eta} }{ P_t} \left(2^{ \frac{L_{m}}{W (\tau_{o} - \tau_{m,comp}) } } -1\right)  \right]\label{eq:exp_cumulative}
 \end{align}
 where \eqref{eq:exp_cumulative} holds from the cumulative distribution function (c.d.f.) of the exponential random variable. Note that the control outage probability for each control loop is affected by the communication payload size $L_m$ and computation time $\tau_{m,comp}$ while the other communication parameters in~\eqref{eq:exp_cumulative} are given. Therefore, to reduce the outage probability, the control system should be designed to operate at small communication and computation costs.

%% file: Sec3_Two_Way_Predictive_Control.tex
\section{Two-Way Split Koopman Autoencoder Architecture for Predictive Linear Control}
 \label{Split_Closed_Loop_Koopman}
 
 Toward reducing communication and computation costs for closed-loop remote control, we propose a two-way Koopman AE architecture comprised of \emph{sensing and controlling Koopman AEs}. To this end, this section first revisits fundamentals of Koopman operator theory. Then, a split learning method for obtaining a sensing Koopman AE that linearizes the non-linear state dynamics enabling future state prediction at the controller is proposed. Moreover, armed with a well-trained sensing Koopman AE, a linear quadratic regulator (LQR) is applied over the linearized Koopman subspace to obtain the optimal control action command at a small computational cost. In addition, we propose a split learning-based controlling Koopman AE that predicts future control action commands at the actuator.
 
\subsection{Preliminary: Koopman Operator for Closed-Loop Control}
\label{Basics_Theory_of_Closed-loop_Koopman_operator}

 Typically, a data sample observed from a closed-loop controlled dynamical system can be seen as a function of a system state and control action. Define functions $g:\mathbb{R}^{p}\times\mathbb{R}^{q}\rightarrow \mathbb{R}$, dubbed \emph{observables} which span an infinite-dimensional Hilbert space $\mathcal{H}$. The Koopman operator $\mathcal{K}:\mathcal{H}\to\mathcal{H}$ is a linear operator that acts on $\mathcal{H}$ \cite{proctor2018generalizing,brunton2021modern} such that
\begin{align}
     \label{eq_Koopman_Composition}
     \mathcal{K} g ( \mathbf{x}_{m}, \mathbf{u}_{m} ) = g ( \mathbf{f}( \mathbf{x}_{m}, \mathbf{u}_{m} ),  \mathbf{u}_{m+1} ),
 \end{align}
 for some observable $g$. Note that by ignoring the additive system noise in \eqref{eq_nonlinear_fn}, the expression \eqref{eq_Koopman_Composition} can be rewritten as 
 \begin{align}\label{eq:noiselessKoopman}
 \mathcal{K}  g ( \mathbf{x}_{m}, \mathbf{u}_{m} ) = g ( \mathbf{x}_{m+1}, \mathbf{u}_{m+1}),
 \end{align}
 and in other words, the Koopman operator in \eqref{eq:noiselessKoopman} enables to express the linear evolution of the non-linear system dynamics in the function space $\mathcal{G}$. However, to obtain the Koopman operator, representation and computation issues may raise, since the operator acts on the infinite dimensional space $\mathcal{H}$.
 
 One approach to resolve such problems is to find an invariant subspace spanned by a finite set of functions. That is, finding a span of $d$ different functions $\{g_1,g_2, \dots g_d\}$, such that a function $g$ in this subspace
 \begin{align}
 \label{eq_Eigenfunctions_Span}
 g(\mathbf{x}_{m}, \mathbf{u}_{m}) = a_1g_1(\mathbf{x}_{m}, \mathbf{u}_{m}) +  \cdots + a_d g_d(\mathbf{x}_{m}, \mathbf{u}_{m}),
 \end{align}
 for some real-valued coefficients $a_1,\dots,a_d \in \mathbb{R}$, is still in the same subspace after acted on by the Koopman operation
 \begin{align}
 \label{eq_Koopman_Eigs}
 \mathcal{K}g(\mathbf{x}_{m}, \mathbf{u}_{m}) = b_1 g_1(\mathbf{x}_{m},\mathbf{u}_{m}) + \cdots+ b_d g_d(\mathbf{x}_{m}, \mathbf{u}_{m}),
 \end{align}
 for some real-valued coefficients $b_1,\dots,b_d \in \mathbb{R}$. Note that the Koopman operator is linear and thus allows eigendecomposition, and any finite set of eigenfunctions of the Koopman operator will span an Koopman invariant subspace \cite{kaiser2017data}. Therefore, by introducing a finite-dimensional Koopman matrix representation $\mathbf{K} \in \mathbb{R}^{d \times d}$ for a given Koopman invariant subspace, we can obtain a global linearization expression of a non-linear system dynamics in~\eqref{eq_nonlinear_fn} as 
 \begin{align}
     \label{eq_Koopman_Linear}
     \mathbf{g}(\mathbf{x}_{m+1}, \mathbf{u}_{m+1}) = \mathbf{K} \mathbf{g}(\mathbf{x}_{m}, \mathbf{u}_{m}),
 \end{align}
 where $\mathbf{g}$ is the $d$-dimensional vector of which the elements are the Koopman eigenfunctions. Note that for a given current system state $\mathbf{x}_{m}$ and its computed control action command $\mathbf{u}_{m}$, if the concatenated eigenfunctions $\mathbf{g}$, its element-wise inverse $\mathbf{g}^{-1}$, and the Koopman matrix $\mathbf{K}$ are known, the future system state $\mathbf{x}_{m+1}$ and control action command $\mathbf{u}_{m+1}$ can be readily obtained. However, since discovering the Koopman eigenfunctions from finite samples of a dynamical system is still challenging, it calls for an autoencoder-based deep learning based approach as will be explained in the following subsections.
 
 In the meantime, since the control system exhibits control-affine non-linear dynamics in \eqref{eq_affine_Control} in this article, the global linearization expression \eqref{eq_Koopman_Linear} can be recast as~\cite{kaiser2017data}
 \begin{align}
     \label{eq_Koopman_non_Affine}
     \begin{bmatrix}
     \mathbf{g}_s( \mathbf{x}_{m+1} ) \\ 
     \mathbf{g}_u( \mathbf{x}_{m+1},\mathbf{u}_{m+1} )
     \end{bmatrix} =  \mathbf{K}  \begin{bmatrix}
     \mathbf{g}_s ( \mathbf{x}_{m} ) \\ 
     \mathbf{g}_u ( \mathbf{x}_{m},\mathbf{u}_{m})
     \end{bmatrix},
 \end{align}
 where $\mathbf{g}_s$ is the $d$-dimensional vector of Koopman eigenfunctions that depend only on the system states, $\mathbf{g}_u$ is the $f$-dimensional vector of Koopman eigenfunctions that depend on both system states and control action commands. For the purpose of utilizing linear control theory, it is required to ensure that the linear evolution of the Koopman eigenfunctions is explicitly related to the control action commands; hence we simplify $\mathbf{g}_u$ in~\eqref{eq_Koopman_non_Affine} into $\mathbf{g}_u ( \mathbf{x}_{m},\mathbf{u}_{m}) \approx \mathbf{u}_{m}$, and thus \eqref{eq_Koopman_non_Affine} becomes \cite{mamakoukas2019local,bonnert2020estimating, kaiser2017data}
 \begin{align}
     \label{eq_Koopman_Simplification}
     \begin{bmatrix}
     \mathbf{g}_s(\mathbf{x}_{m+1})\\ \mathbf{u}_{m+1}
     \end{bmatrix}
     = \begin{bmatrix}
      \mathbf{K}_{11}  \; \;  \mathbf{K}_{12} \\
      \mathbf{K}_{21}  \; \;  \mathbf{K}_{22}
     \end{bmatrix} \begin{bmatrix}
     \mathbf{g}_s(\mathbf{x}_{m})\\ \mathbf{u}_{m}
     \end{bmatrix},
 \end{align} where $\mathbf{g}_s$ and $\mathbf{g}$ are used interchangeably thereafter, $\mathbf{K}_{11} \in \mathbb{R}^{d \times d}$ is the state transition matrix in the Koopman matrix $\mathbf{K} \in \mathbb{R}^{d+q}\times\mathbb{R}^{d+q}$, $\mathbf{K}_{12} \in \mathbb{R}^{d \times q}$ is the control action matrix of the state dynamics, $\mathbf{K}_{21} \in \mathbb{R}^{q \times d}$ is the state-dependent control action matrix, and $\mathbf{K}_{22} \in \mathbb{R}^{q \times q}$ is the control action matrix of the control dynamics. Moreover, we define $\mathbf{K}_{s} = \left[ \mathbf{K}_{11} \; \mathbf{K}_{12} \right]$ as the state Koopman matrix representing the state dynamics, while $\mathbf{K}_{a} = \left[ \mathbf{K}_{21} \;  \mathbf{K}_{22} \right]$ as the action Koopman matrix representing the control action dynamics. Note that the simplification of~\eqref{eq_Koopman_non_Affine} in~\eqref{eq_Koopman_Simplification} comes at the cost of less accurate approximation of the system dynamics. For instance, if the term $\sin(x_{m,1}) \mathbf{u}_{m}$ appears in the system dynamics, it will be approximated in the Koopman model as $c_{1}\mathbf{u}_{m}$, where $c_{1} \in \mathbb{R}$ is a constant.

 \subsection{Sensing Koopman Autoencoder for State Prediction}
 \label{Wireless_Split_Closed-loop_Koopman_Auto-encoder}
 The first component of the two-way Koopman AE architecture is the \emph{sensing Koopman AE}, which plays two roles: (i) linearizing the non-linear system dynamics and (ii) predicting the future system states based on the linearized system dynamics.
 Note that linearization of non-linear system dynamics helps reduce the complexity in terms of minimizing the computational time required to analyze the system and calculate its optimal control action command, while state prediction helps reduce the communication cost~\cite{guastello2009introduction}. The proposed sensing Koopman AE is a tripartite neural network that consists of an \textit{encoder} that is related to the concatenated Koopman eigenfunctions
 $\mathbf{g}$, two fully-connected hidden layers that constitute a finite-dimensional matrix representation of the Koopman operator $\mathbf{K}$, and the \textit{decoder} which is an inverse function vector of the encoder $\mathbf{g}^{-1}$. Since we consider a remote control scenario, the Koopman AE is split into two parts, where the encoder is located at the sensor and actuator side, while the Koopman hidden layers and the decoder are situated at the controller side. Thus, the sensing Koopman AE is trained through split learning~\cite{Vepakomma:2018:Splita,koda2020communication} which will be detailed in Section \ref{Split_Closed-loop_Koopman_Training}.
 
 Once the sensing Koopman AE is trained, the sensor sends an encoded representation $\mathbf{g}(\mathbf{x}_{m})$ of the sampled system state $\mathbf{x}_{m}$ and the controller obtains an estimated latent state representation $\hat{\mathbf{g}}(\mathbf{x}_{m})$ and concatenates it with the computed control action command $\mathbf{u}_{m}$ as $\mathbf{y}_{m} = [ \hat{\mathbf{g}}(\mathbf{x}_{m})\  \mathbf{u}_{m}]^{\mathsf{T}} $. Then, by multiplying $\mathbf{y}_{m}$ with the sensing Koopman operator matrix $\mathbf{K}_{s}$, the remote controller obtains a linear state representation evolution as \begin{align}
   \label{eq_linear_evol_Representation}
      \hat{\mathbf{g}}(\mathbf{x}_{m+1})  = \mathbf{K}_{11} \hat{\mathbf{g}}(\mathbf{x}_{m}) + \mathbf{K}_{12} \mathbf{u}_{m}, 
   \end{align}
   where $\mathbf{K}_{11}$ and $\mathbf{K}_{12}$ are the Koopman submatrices mentioned in \eqref{eq_Koopman_Simplification}. Moreover, by passing $\mathbf{y}_{m}$ through the trained sensing Koopman matrix and the decoder, the remote controller obtains the predicted future system state as $\bar{\mathbf{x}}_{m+1} = \mathbf{g}^{-1}(\mathbf{K}_{s} \mathbf{y}_{m})$. Moreover, by multiplying the sensing Koopman matrix $m'$ times, the controller can predict the future system states at $t = (m+m')\tau_o$ as
   \begin{align}
       \label{eq_predicted_states}
       \bar{\mathbf{x}}_{m + m'} = \mathbf{g}^{-1} (\mathbf{K}_{s}^{m'} \mathbf{y}_{m}),
   \end{align} for all target prediction depth $m' \in \mathbb{Z}_+$ which will be detailed in Section~\ref{Koopman-based_Controller}. 
   
   To this end, the proposed real-time remote control of a closed-loop control system, aided by the Koopman AE, consists of two-phases. In the first phase, the remote controller receives the measured system states in real-time from the sensor, computes its control action command, and sends it back to the actuator. Meanwhile, the sensor and controller train the sensing Koopman AE and the first phase is maintained until it is well-trained. In the second phase, the remote controller sends back a control signal informing the sensor to stop sending the system states once it is well-trained, and since the sensing Koopman AE is trained, the remote controller can predict the future system states.

 \subsection{Linear Control over Koopman Subspace}
 \label{Koopman-based_Controller}
 
  Having a closed-loop Koopman AE, an optimal control action can be computed over the linearized Koopman subspace by readily applying linear control theory. That is, formulating an infinite-horizon discrete-time linear quadratic regulator (LQR) over the linearized Koopman subspace rather than over the non-linear system dynamics. Following~\cite{brunton2016koopman}, the control objective is a quadratic cost function given as \begin{align}
       \label{eq_lqr_objective}
        \mathcal{J}(\hat{\mathbf{g}} (\mathbf{x}_{m}),\mathbf{u}_{m}) = \frac{1}{2} \sum^{\infty}_{m=0} \hat{\mathbf{g}}(\mathbf{x}_{m})^{\mathsf{T}} \mathbf{Q}_{g} \hat{\mathbf{g}}(\mathbf{x}_{m})  + \mathbf{u}^{ \mathsf{T}}_{m} \mathbf{R} \mathbf{u}_{m},
  \end{align}
  where $\mathbf{Q}_{g} \in \mathbb{S}_{+}^{d \times d}$ is a positive semi-definite weight matrix of the latent state representation deviation cost, and $\mathbf{R} \in \mathbb{S}_{++}^{q \times q}$ is a positive definite weight matrix of the control action cost. Then, the optimization problem of an infinite-horizon discrete-time LQR to calculate the optimal control action is formulated as \vspace{-8pt} \begin{subequations} \label{eq_LQR_Optimization_1} \begin{gather} \label{eq_LQR_Optimization}
  \underset{\hspace{-95pt}\mathbf{u}_{m}}{ \hspace{-100pt} \text{Minimize}} \hspace{-10pt} \mathcal{J}(\hat{\mathbf{g}}(\mathbf{x}_{m}),\mathbf{u}_{m}) \\
  \hspace{15pt} \text{subject to:}\; \;   \hat{\mathbf{g}}(\mathbf{x}_{m+1}) = \mathbf{K}_{11} \hat{\mathbf{g}}(\mathbf{x}_{m}) + \mathbf{K}_{12} \mathbf{u}_{m}. \label{eq_LQR_Optimization_a}  
  \end{gather}\end{subequations} 
  The problem in~\eqref{eq_LQR_Optimization_1} is a convex optimization problem with a quadratic cost function and linear constraint, hence the optimal control action can be readily obtained as
  \begin{align}
     \label{eq_optimal_control_law}
     \mathbf{u}_{m} = - \mathbf{K}_{\mathrm{LQR}} \; \hat{\mathbf{g}}(\mathbf{x}_{m}), 
  \end{align} 
  where $\mathbf{K}_{\mathrm{LQR}} \hspace{-20pt} = \hspace{-20pt} \left( \mathbf{R} + \mathbf{K}_{12}^{\mathsf{T}} \mathbf{P} \mathbf{K}_{12} \right)^{-1}  \mathbf{K}_{12}^{\mathsf{T}}  \mathbf{P} \mathbf{K}_{11}^{\mathsf{T}} $ is the feedback gain matrix, and $\mathbf{P} \hspace{-15pt} = \hspace{-15pt} \mathbf{K}_{11}^{\mathsf{T}}  \mathbf{P} \mathbf{K}_{11}  - \mathbf{K}_{11}^{\mathsf{T}}  \mathbf{P} \mathbf{K}_{12}  \left( \mathbf{K}_{12}^{\mathsf{T}}  \mathbf{P} \mathbf{K}_{12}  + \mathbf{R} \right)^{-1} \mathbf{K}_{12}^{\mathsf{T}}  \mathbf{P} \mathbf{K}_{11}  + \mathbf{Q}_{g}$ is a unique positive definite matrix which satisfies the discrete-time algebraic Riccati equation (DARE)~\cite{bemporad2002explicit}. The Koopman-based controller outperforms the locally linearized-based controller in terms of control performance. This is because the locally linearized-based controller performs optimally near the equilibrium point, but the performance becomes poor for the points that are located far from the equilibrium point. In contrast, since the Koopman operator globally linearizes the system dynamics, the controller over the linearized Koopman subspace performs optimally for all points.

  \subsection{Split Learning-based Koopman Autoencoder}
  \label{Split_Closed-loop_Koopman_Training}
  
  An optimization problem for obtaining the closed-loop Koopman AE in the considered WNCS can be formulated as
  \begin{subequations}  \label{eq_training_Optimization_ab} \begin{gather} \label{eq_training_Optimization}
  \underset{\mathbf{K}, \mathbf{g},\mathbf{g}^{-1} }{ \textrm{minimize}} \;  \frac{1}{M_{d}} \sum_{m' = 1}^{M_d} \left\lVert \hat{\mathbf{x}}_{m+m'} - \sum_{l = 0}^{M_d - 1} \mathbf{g}^{-1} \left( \mathbf{K}_{s}^{ M_{d} - l } \mathbf{y}_{m + l} \right) w_{l} \right\rVert_{2}^{2} \\
  \hspace{-130pt} \text{subject to:}\; \;   \bar{\mathbf{x}}_{m} = \mathbf{g}^{-1} \left( \mathbf{y}_{m} \right) , \label{eq_training_Optimization_a}  
  \end{gather}\end{subequations} where $\hat{\mathbf{x}}_{m+m'}$ is the estimated system state at time $m+m'$, $M_d$ is the target prediction depth considered in the training that represents the Koopman operator prediction time horizon checked in the training, and $w_{l}$ is a weight parameter of the $l$-th predicted system state at the controller side. On the one hand, when $l = 0 $ and $w_{l} = 1$, we call it a \textit{special case} of the optimization problem in~\eqref{eq_training_Optimization_ab} suggested in~\cite{lusch2018deep,xiao2020deep}. On the other hand, when $l > 0$ and $w_{l} = 1 / M_{d}$, we call it a \textit{general case} of the optimization problem in~\eqref{eq_training_Optimization_ab} proposed in this paper. The \textit{general case} of the optimization problem in~\eqref{eq_training_Optimization_ab} utilizes all the received latent state representations to predict the future system states compared to the \textit{special case}. Since it is analytically hard to solve the formulated optimization problem in~\eqref{eq_training_Optimization_ab} for long-term target prediction depth, we solve it by exploiting deep learning over three separated unconstrained optimization sub-problems formulated as follow. 

  \begin{enumerate}
        \item A \textit{reconstruction loss} is formulated to accurately reconstruct the system states. This loss measures the mean squared error (MSE) between the estimated system states $\hat{\mathbf{x}}_{m}$ and the decoded estimated state representation adjoined with the control action command $\mathbf{g}^{-1}(\mathbf{y}_{m})$, given as 
        \begin{align}
            \label{eq_Reconst_loss}
            \mathcal{L}_{1} = \frac{1}{M_s} \sum_{m = 1}^{M_{s}} || \hat{\mathbf{x}}_{m} - \mathbf{g}^{-1} \left(\mathbf{y}_{m} \right) ||^{2}_2, 
        \end{align} where $M_{s}$ denotes the length of the received time-series signals in the first phase of the remote control of a closed-loop control system.
        
        \item A \textit{linear Dynamics loss} is formulated to obtain the linear finite-dimensional sensing Koopman matrix $\mathbf{K}_{s}$ that ensures the linearity in the Koopman invariant subspace. This loss measures the MSE between the estimated state representation of the future system state adjoined with its future control action commands as $\mathbf{y}_{m + m'}$ and the weighted sum of the multiplication of the sensing Koopman matrix with different target prediction depth during training and the estimated state representation of the received system states adjoined with the calculated control action commands $ \sum_{l = 0}^{M_d - 1} \left( \mathbf{K}_{s}^{ M_{d} - l } \mathbf{y}_{m + l} \right) w_{l} $, i.e., \begin{align}
            \label{eq_Linear_Loss}
            \mathcal{L}_{2} = \frac{1}{M_{d}} \sum_{m' = 1}^{M_d} ||  \mathbf{y}_{m+m'} - \sum_{ l = 0}^{M_d - 1} \left( \mathbf{K}_{s}^{ M_{d} - l} \mathbf{y}_{m + l} \right) w_{l} ||_{2}^{2}.
        \end{align} Note that the linear latent state representation evolution in~\eqref{eq_Linear_Loss} for $l = 0$ is represented as  $\mathbf{K}_{s}^{M_{d}} \mathbf{y}_{m} = \mathbf{K}_{11}^{M_{d}} \mathbf{g}(\mathbf{x}_{m}) + \sum^{M_{d}}_{m'=1} \mathbf{K}_{11}^{m'-1} \mathbf{K}_{12} \mathbf{u}_{m+m'-1}$.

        \item A \emph{prediction loss} is formulated to accurately predict the future system states. The prediction loss measures the MSE between the estimated future system states $\hat{\mathbf{x}}_{m+m'}$ and decoded weighted sum of the multiplication of the sensing Koopman matrix with different target prediction depth considered in training and the estimated state representation of the received system states adjoined with the calculated control action commands $ \mathbf{g}^{-1}(\sum_{ l = 0}^{M_d - 1} \left( \mathbf{K}_{s}^{ M_{d} - l } \mathbf{y}_{m + l} \right) w_{l})$, i.e., 
         \begin{align}
            \label{eq_Pred_loss}
            \mathcal{L}_{3} = \frac{1}{M_{d}} \sum_{m' = 1}^{M_d} || \hat{ \mathbf{x}}_{m+m'} - \bar{\mathbf{x}}_{m + m'}  ||_{2}^{2},
         \end{align}
        where the future predicted system state $ \bar{\mathbf{x}}_{m + m'}$ in~\eqref{eq_Pred_loss} is given as  $ \bar{\mathbf{x}}_{m + m'} = \sum_{ l = 0}^{M_d - 1}  \mathbf{g}^{-1} \left( \mathbf{K}_{s}^{ M_{d} - l } \mathbf{y}_{m + l} \right)  w_{l}$.

        \item A \textit{representation cost loss} is formulated to guarantee that the state representation deviation cost is equivalent to the system state deviation cost in the Jacobian linearization method. This loss measures the MSE between the state representation deviation cost $\tilde{\mathbf{Q}}_{g}$ and the system state deviation cost $\mathbf{Q}_{x}$, given as
  \begin{align}
            \label{eq_representation_cost}
            \mathcal{L}_{4} = \frac{1}{M_{s}} \sum_{m = 1}^{M_{s}} || \hat{ \mathbf{x}}_{m}  \mathbf{Q}_{x}  \hat{\mathbf{x}}_{m} -  \hat{ \mathbf{g}}(\mathbf{x}_{m}) \tilde{\mathbf{Q}}_{g} \hat{\mathbf{g}}(\mathbf{x}_{m}) ||_{2}^{2}.
    \end{align} Here, we set $\mathbf{Q}_{g} = [\tilde{\mathbf{Q}}_{g}]_{+}$ as the final estimator, where $[.]_{+}$ truncates the non-positive eigenvalues to zero to ensure the state representation deviation cost is a positive semi-definite weight matrix.
    \end{enumerate} 
    
 Overall, the total weighted-sum loss function to train the closed-loop Koopman AE for predicting the future system states and calculating the optimal control action is given as 
    \begin{align}
       \label{eq_Updated_Overall_loss}
       \mathcal{L} = c_{1} \mathcal{L}_{1} + c_{2} \mathcal{L}_{2} + c_{3} \mathcal{L}_{3} +  c_{4} \mathcal{L}_{4},
    \end{align}
    for some positive coefficients $c_{1}, c_{2}, c_{3}, c_4 \in \mathbb{Z}_+$. The weights of the sensing Koopman AE can be trained based on the stochastic gradient descent (SGD) and backpropagation via feedback channels from the controller to the sensor. An early stopping strategy is utilized to avoid model overfitting, which enhances both the prediction accuracy and communication efficiency. Note that the sensor must consistently sample the system states and send them along with the latent state representations to the controller over wireless fading channels during the first phase of remote control to compute the total weighted-sum loss at the remote controller. Otherwise, if the system states are missed due to adverse channel conditions in the forward link, the remote controller predicts the future system states based on the last observed system states.
    
 \subsection{Controlling Koopman Autoencoder for Control Prediction}
 \label{Wireless_Split_Input_Closed-loop Koopman_Auto-encoder}
 
  To compensate for the missing control action command in case of adverse channel conditions in the reverse link, we propose controlling Koopman AE to linearize the action dynamics and predict the future control action commands at the actuator side. The proposed controlling Koopman AE is composed of three parts of the encoder that is located at the sensor side is the same as the encoder utilized in the sensing Koopman AE $\mathbf{g}$, the Koopman hidden layers at the actuator side $\mathbf{K}'$, and the decoder are located at the actuator side $\mathbf{g}_{a}^{-1}$. After the controlling Koopman AE is well-trained, if the remote controller sends a control action command $\mathbf{u}_{m}$, the actuator obtains a noisy control action command $\hat{\mathbf{u}}_{m}$, applies it on the plant, and concatenates it with the latent state representations $\mathbf{g}(\mathbf{x}_{m})$ as follows $\mathbf{z}_{m} = [ \mathbf{g}(\mathbf{x}_{m}) \; \;  \hat{\mathbf{u}}_{m}]^{\mathsf{T}} $. Then, by passing $\mathbf{z}_{m}$ through the finite-dimensional controlling Koopman matrix $\mathbf{K}'_{a}$, the linear control action command evolution at the actuator is given as
  \begin{align}
   \label{eq_linear_evol_input}
    \hat{\mathbf{u}}_{m+1} =
    \mathbf{K}'_{21} \mathbf{g}(\mathbf{x}_{m}) + \mathbf{K}'_{22} \hat{\mathbf{u}}_{m},  
   \end{align} where $\mathbf{K}'_{21} \in \mathbb{R}^{q \times d}$ and $\mathbf{K}'_{22} \in \mathbb{R}^{q \times q}$ are the trained Koopman submatrices at the actuator side representing the non-linear control action dynamics. 
   
   Since the objective of the controlling Koopman AE is to predict the future control action commands for a long time, it is required to have the measurement functions, its inverse functions, and the finite-dimensional Koopman matrix. Therefore, we train the controlling Koopman AE through an optimization problem formulated as 
   \begin{subequations}  \label{eq_actuating_Optimization_ab} \begin{gather} \label{eq_actuating_Optimization}
  \underset{\mathbf{K}', \mathbf{g},\mathbf{g}_{a}^{-1} }{ \text{minimize}} \hspace{8pt}  \frac{1}{M_{d}}  \sum_{m' = 1}^{M_d} ||\mathbf{x}_{m+m'} - \tilde{\mathbf{x}}_{m + m'} ||_{2}^{2} \\
  \hspace{-80pt} \text{subject to:}\; \;   \tilde{\mathbf{x}}_{m} = \mathbf{g}_{a}^{-1} \left( \mathbf{z}_{m} \right) , \label{eq_actuating_Optimization_a}  
  \end{gather}\end{subequations}  where  $\tilde{\mathbf{x}}_{m + m'} = \sum_{l = 0}^{M_d - 1} \mathbf{g}_{a}^{-1} \left( \mathbf{K}_{a}'^{{M_{d} - l} } \mathbf{z}_{m + l} \right) w'_{l}$ is the future predicted system states at the actuator side, and $w'_{l}$ is the $l$-th weight parameter of the predicted system state at the actuator. The problem in~\eqref{eq_actuating_Optimization_ab} is similar to the problem of the sensing Koopman AE in~\eqref{eq_training_Optimization_ab} except that the controlling Koopman AE receive noisy control action commands and the original system states along with their state representations. As a result, we minimize a total weighted-sum of three separated loss functions similar to the sensing Koopman AE given as  \begin{align}
       \label{eq_Overall_input_loss}
       \mathcal{L}' = c_{1}' \mathcal{L}'_{1} + c_{2}' \mathcal{L}'_{2} + c_{3}' \mathcal{L}'_{3},
   \end{align} where $c_{1}'$, $c_{2}'$, and $c_{3}'$ are positive weighting hyperparameters of the controlling Koopman AE. The loss functions $\mathcal{L}_{1}'$, $\mathcal{L}_{2}'$, and $\mathcal{L}_{3}'$ of the controlling Koopman AE in~\eqref{eq_Overall_input_loss} are similar to the loss functions $\mathcal{L}_{1}$, $\mathcal{L}_{2}$, and $\mathcal{L}_{3}$ of the sensing Koopman AE,  respectively except that $\hat{\mathbf{x}}_{m}$ is substituted by $\mathbf{x}_{m}$ and $\mathbf{y}_{m}$ by $\mathbf{z}_{m}$.
   

%% file: Sec4_Simulation_Results.tex
 \section{Simulation Results and Discussion}
 \label{Simulation_Results}

  In this section, we investigate the performance of both sensing Koopman AE and controlling Koopman AE over wireless fading channels in an inverted cart-pole system. This control system is non-linear and multi-dimensional. For a given time, the inverted cart-pole system is described by a four-dimensional state vector $\mathbf{x} = [ x \; v \;  \theta \; \omega ]^{T}$, where $x$ and $v$ represent the horizontal position and velocity of the cart, respectively. The terms $\theta$ and $\omega$ are the vertical angle and angular velocity of the pendulum, respectively. The control action is described as a horizontal force applied on the cart $u$. Accordingly, the full non-linear system dynamics is described as follows.    
  \begin{align}\label{CIP_Dynamics}
  \begin{split}
  \frac{dx}{dt} &= v,\\
  \frac{d \theta}{dt}  &= \omega, \\
  \frac{dv}{dt} &= \frac{-m_{p}^{2} L^{2} \nu \; c\theta \; s\theta + m_{p} L^{2} (m_{p} L\omega^{2} s\theta - \delta v )  + m_{p} L^{2} u }{m_{p} L^{2} \left(m_{c} + m_{p} (1 - c\theta^{2} ) \right)},\\
  \frac{d \omega}{dt} & = \frac{ m_{pm} m_{p} \nu L s\theta - m_{p} L c\theta ( m_{p} L\omega^{2} s\theta - \delta v ) + m_{p} L c\theta u}{ m_{p} L^{2} \left( m_{c} + m_{p} ( 1 - c\theta^{2} )\right)},
 \end{split}
 \end{align} with $m_{pm} = m_{p} + m_{m}$, $c\theta = \cos(\theta)$,  and $s\theta = \sin(\theta)$. Here, we consider the following simulation parameters unless stated otherwise: pendulum mass $m_{p} = 1$ $\mathrm{Kg}$, cart mass $m_{c} = 5$ $\mathrm{Kg}$, pendulum length $L = 0.2$ $\mathrm{m}$, gravitational acceleration $\nu = -10$  $\mathrm{m/} \mathrm{s}^{2}$, and cart damping $\delta = -2$ $\mathrm{N s} /\mathrm{m}$. 
 
 \subsection{Data Generation and Training}
 \label{Dataset}
 
 The training dataset of the sensing Koopman AE is generated at the controller based on the calculated control action commands and the received system states from the inverted-cart pole system with a sampling rate of $\tau_{0} =  10$ $\mathrm{ms}$ in the time interval $m \in [0, |\mathcal{M}_{\mathrm{Train}}|]$ and with a random initial condition in the range of $[-0.5,0.5]$ for each state dimension. For each initial condition, we solve the system of differential equations in~\eqref{CIP_Dynamics} using the fourth-order Runge Kutta numerical method with $0.01$ equally-spaced step size and trajectory length of $\mathcal{M}_{\mathrm{Train}} = 250$ $\mathrm{s}$. The training dataset consists of $T_{t} = 70$, the validation dataset of $T_{v} = 20$, and the test dataset of $T_{e} = 10$ trajectories. The AE weights are initialized with a normal distribution of the form $w \sim \mathcal{N} ( 0, \frac{2}{d_{in}} )$ where $d_{in}$ is the number of neurons of the input layer. The autoencoder weights are trained to minimize the total weighted sum loss in~\eqref{eq_Updated_Overall_loss} for the sensing Koopman AE, via Adam optimizer~\cite{sutskever2013importance} with batch size of $64$ and learning rate of $10^{-4}$. The encoder of the sensing Koopman AE consists of three fully-connected layers that contain $128, 64, 32$ neurons with rectified linear unit (ReLu) activation, respectively, and the final layer contains $d$ neurons with linear activation. The decoder begins with $2$ fully-connected layers with the width $(d+q) \in \{2, 3, 4, 5\}$ (i.e., latent state representation dimension plus control action command dimension) and the rest follows the same structure of the encoder. The weighting hyperparameters of the sensing Koopman AE are set as $c_{1} = c_{3} = 0.5,$ $c_{2} = c_{4} = 1$. The wireless communication channels are assumed to follow Rayleigh fading channels. Here, the results are obtained based on the model with the lowest validation error of several training runs. We consider a maximum transmission power value of $P_{t} = 20$ $\mathrm{dBm}$, different SNR of $\mathrm{SNR} \in \{-10, 0, 10, 20\}$ $\mathrm{dBm}$, distance between transmitter-receiver pair of $D = 100 \, \mathrm{m}$, and the path-loss exponent of $\eta = 3$. The dataset of the controlling Koopman AE is generated at the actuator based on the system states and the received control action commands from the controller. The rest follows the same structure of the sensing Koopman AE with hyperparameters set as 

  \begin{figure*}[t]
    \centering
    \subfigure[Cart Position.]{\includegraphics[width=0.4\textwidth]{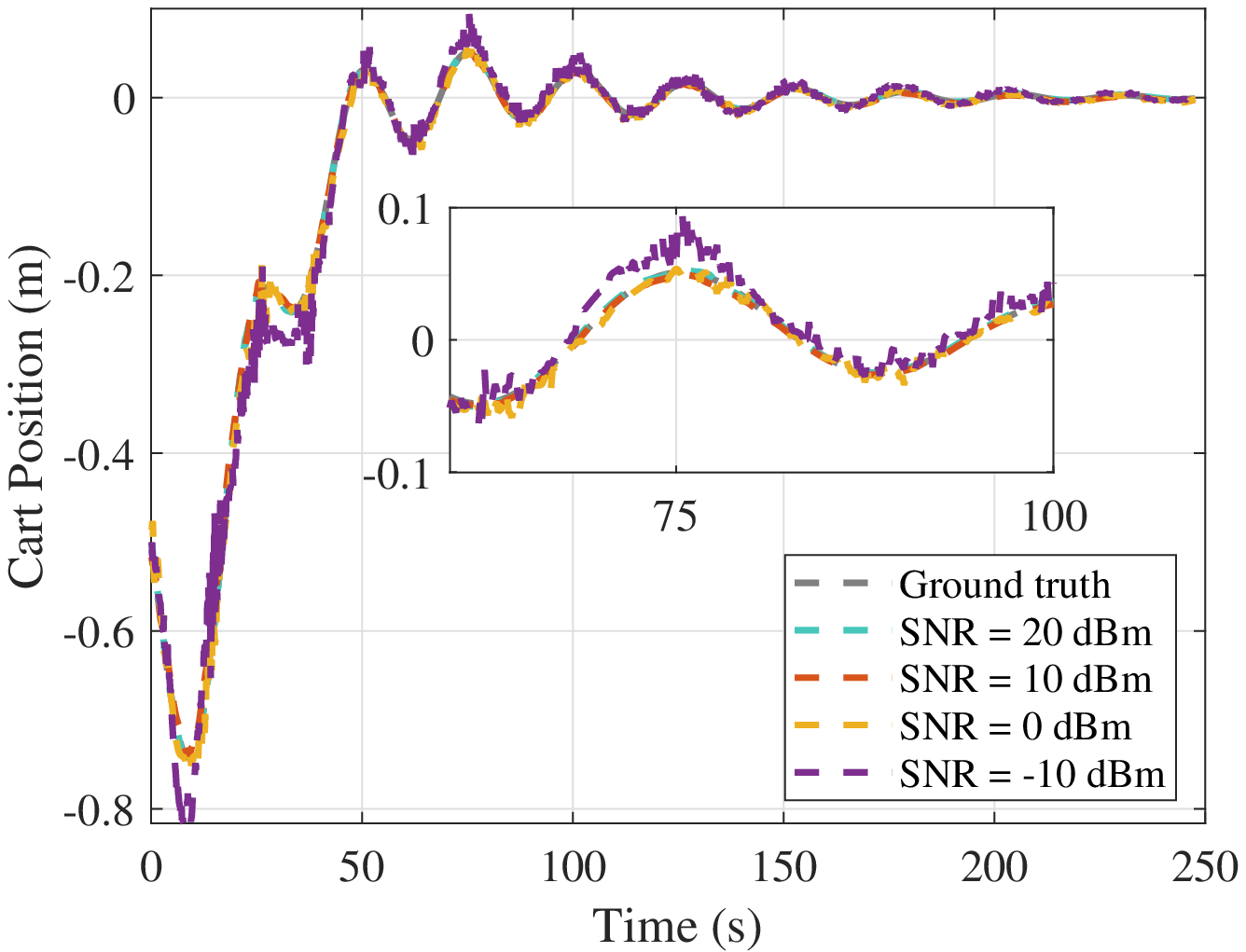}} 
    \subfigure[Cart Velocity.]{\includegraphics[width=0.4\textwidth]{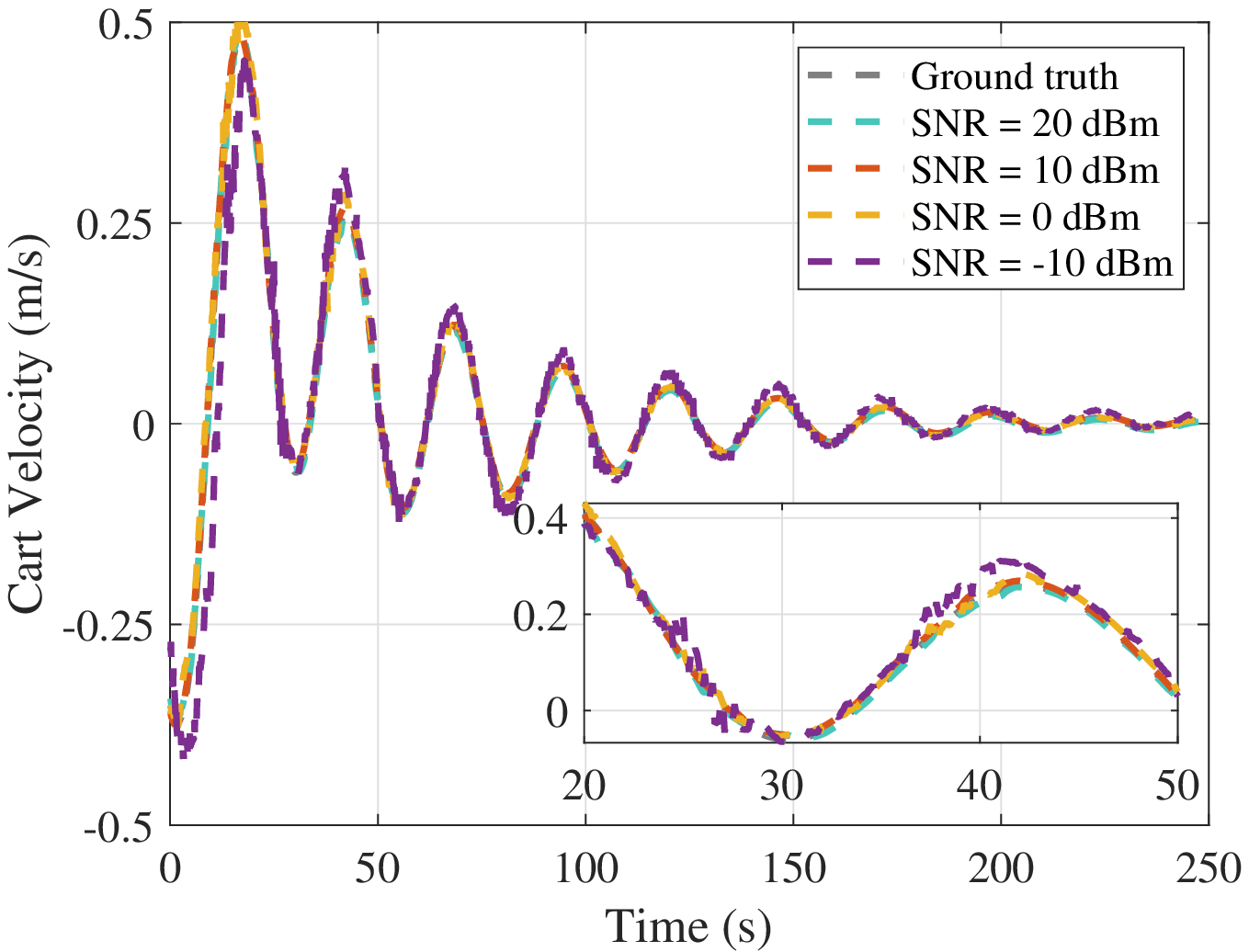}} 
   \subfigure[Pendulum Angular Position.]{\includegraphics[width=0.4\textwidth]{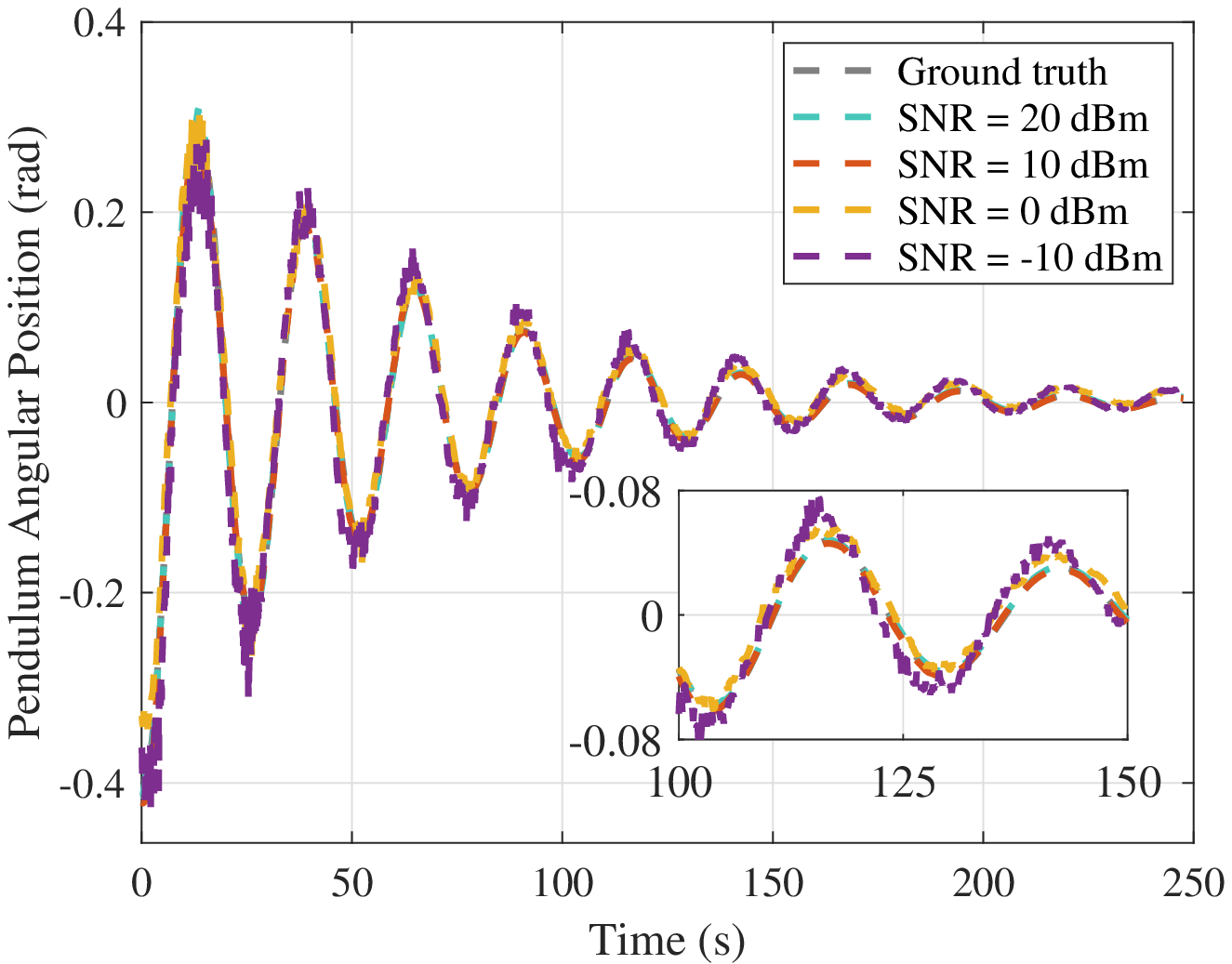}}
    \subfigure[Pendulum Angular Velocity.]{\includegraphics[width=0.4\textwidth]{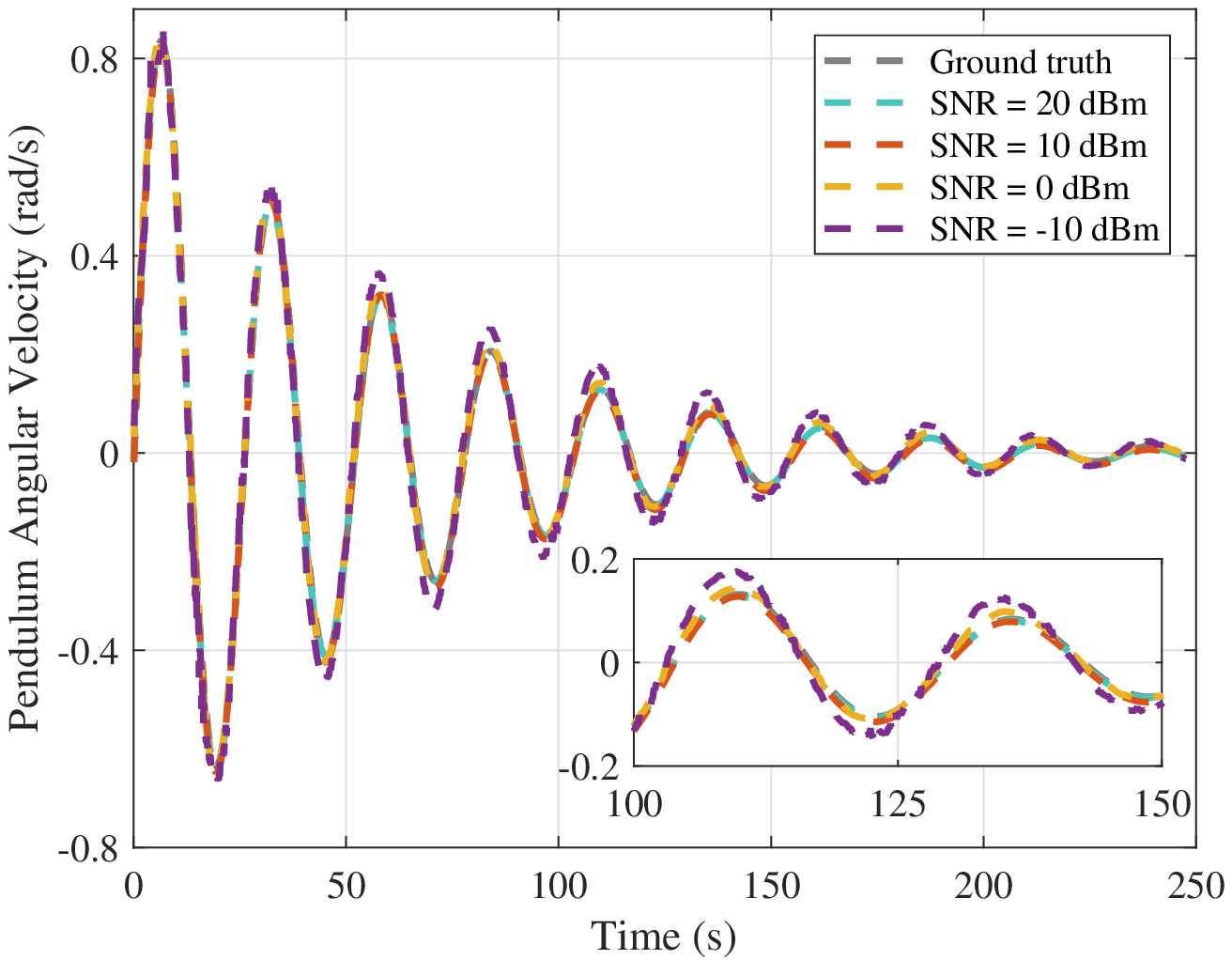}}
   \caption{Time-series of the predicted inverted cart-pole system states utilizing the proposed sensing Koopman autoencoder before observing the ground truth.}
    \label{fig state_prediction}
\end{figure*} 
 
  \begin{figure}[t]
   \centering
   \includegraphics[ width=0.4\textwidth]{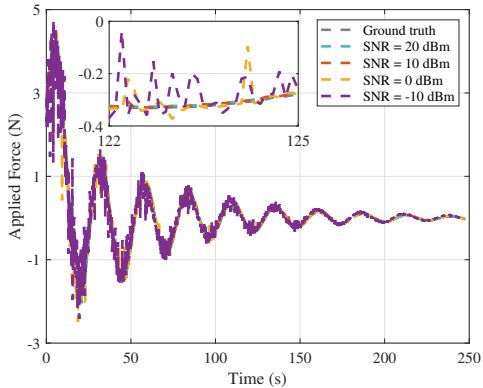} \\
   \caption{Time-series of predicted force applied on the cart utilizing the proposed controlling Koopman autoencoder before observing the ground truth.} \label{fig_force}
  \end{figure}
 
 \subsection{Evaluation Metrics}
 \label{Evaluaion Metric}
  In this subsection, we present the evaluation metrics used to evaluate the proposed sensing Koopman AE and the controlling Koopman AE according to different evaluation metrics in terms of prediction accuracy, wireless communication efficiency, and control stability.

 \begin{table*}[t]
    \centering
   \caption{ State and Action Prediction NRMSE for different latent state representation dimensions and SNR.}
    \label{Table_1}
\begin{tabularx}{1\linewidth}{p{1.5cm} p{2cm} p{2cm} p{4cm} p{4cm}}
    \toprule[1pt]
    & Representation Dimensions $d$   & SNR [dBm]  & State Prediction NRMSE [\%] & Action Prediction NRMSE [\%] \\     
    \cmidrule(r){2-2} \cmidrule(r){3-3} \cmidrule(r){4-4} \cmidrule(r){5-5}
  & \qquad $1$   &  $ \quad -10$   & $14.38$ \hspace{5pt}  \tikz{
        \fill[fill=color1] (0.0,0) rectangle (2.7,0.2);
    } &  $8.97$ \hspace{9pt}  \tikz{
        \fill[fill=color6] (0.0,0) rectangle (1.56,0.2);
        }\\
  &   & $\qquad \; 0$  & $ 14.15$ \hspace{5pt}  \tikz{
        \fill[fill=color2] (0.0,0) rectangle (2.65,0.2);
    } & $3.87$ \hspace{9pt}  \tikz{
        \fill[fill=color7] (0.0,0) rectangle (0.78,0.2);
    } \\
   &   & $\qquad 10$ & $13.89$ \hspace{5pt}  \tikz{
        \fill[fill=color3] (0.0,0) rectangle (2.60,0.2);
     } & $0.99$ \hspace{9pt}  \tikz{
        \fill[fill=color8] (0.0,0) rectangle (0.27,0.2);
     } \\
 &    &  $\qquad  20$  & $\mathbf{9.20}$ \hspace{7pt}  \tikz{
        \fill[fill=color4] (0.0,0) rectangle (2.1,0.2);
    }   & $\mathbf{0.55}$ \hspace{7.5pt}  \tikz{
        \fill[fill=color9] (0.0,0) rectangle (0.21,0.2);
    } \\ \hline
 & \qquad  $2$ &  $ \quad-10$  & $6.69$\; \hspace{7pt}  \tikz{
        \fill[fill=color1] (0.0,0) rectangle (1.16,0.2);
    }  & $9.08$\; \hspace{7pt}  \tikz{
        \fill[fill=color6] (0.0,0) rectangle (1.61,0.2);
    } \\
  &    &  $\qquad \; 0$  & $5.79$\; \hspace{7pt}  \tikz{
        \fill[fill=color2] (0.0,0) rectangle (1.06,0.2);
    } & $3.85$\; \hspace{7pt}  \tikz{
        \fill[fill=color7] (0.0,0) rectangle (0.8,0.2);
    }  \\ 
   &  &  $\qquad 10$  & $5.18$\; \hspace{7pt}  \tikz{
        \fill[fill=color3] (0.0,0) rectangle (0.98,0.2);
    } & $0.99$\; \hspace{7pt}  \tikz{
        \fill[fill=color8] (0.0,0) rectangle (0.27,0.2);
    } \\ 
 &    &  $\qquad 20$  & $\mathbf{4.93}$ \hspace{7pt}  \tikz{
        \fill[fill=color4] (0.0,0) rectangle (0.95,0.2);
    }  & $\mathbf{0.52}$ \hspace{7pt}  \tikz{
        \fill[fill=color9] (0.0,0) rectangle (0.19,0.2);
    }  \\ \hline
   & \qquad $3$ &  $\quad -10$  & $3.84$\; \hspace{7pt}  \tikz{
        \fill[fill=color1] (0.0,0) rectangle (0.86,0.2);
    }   & $8.91$\; \hspace{6.5pt}  \tikz{
        \fill[fill=color6] (0.0,0) rectangle (1.55,0.2);
    } \\
  &    &  $\qquad \; 0$  & $2.95$\; \hspace{7pt}  \tikz{
        \fill[fill=color2] (0.0,0) rectangle (0.74,0.2);
    } & $3.57$\; \hspace{6.5pt}  \tikz{
        \fill[fill=color7] (0.0,0) rectangle (0.74,0.2);
    } \\
   &  &  $\qquad 10$  & $2.18$\; \hspace{7.5pt}  \tikz{
        \fill[fill=color3] (0.0,0) rectangle (0.63,0.2);
    }  & $0.82$\; \hspace{7pt}  \tikz{
        \fill[fill=color8] (0.0,0) rectangle (0.23,0.2);
    }\\
 &     &  $\qquad 20$  & $\mathbf{2.06}$ \hspace{7.5pt}  \tikz{
        \fill[fill=color4] (0.0,0) rectangle (0.60,0.2);
    } & $\mathbf{0.51}$ \hspace{7.5pt}  \tikz{
        \fill[fill=color9] (0.0,0) rectangle (0.19,0.2);
    } \\ \hline
 & \qquad $4$   &  $ \quad -10$ &  $2.82$ \hspace{9pt}  \tikz{
        \fill[fill=color1] (0.0,0) rectangle (0.7,0.2);
    }  & $2.61$\;\hspace{10pt}  \tikz{
        \fill[fill=color6] (0.0,0) rectangle (0.64,0.2);
    }  \\
  &    & $\qquad  0$  & $ 1.67$\hspace{12pt}  \tikz{
        \fill[fill=color2] (0.0,0) rectangle (0.55,0.2);
    }   & $0.87$\;\hspace{10.5pt}  \tikz{
        \fill[fill=color7] (0.0,0) rectangle (0.24,0.2);
    }\\
   &   & $\qquad 10$ & $1.23$\;\hspace{10pt}  \tikz{
        \fill[fill=color3] (0.0,0) rectangle (0.4,0.2);
     } & $0.41$\;\hspace{10.5pt}  \tikz{
        \fill[fill=color8] (0.0,0) rectangle (0.16,0.2);
     } \\
   & &  $\qquad 20$  & $\mathbf{1.03}$ \hspace{8pt}  \tikz{
        \fill[fill=color4] (0.0,0) rectangle (0.3,0.2);
    }  & $\mathbf{0.38}$ \hspace{8pt}  \tikz{
        \fill[fill=color9] (0.0,0) rectangle (0.14,0.2);
    }\\
&    & &     \ \ \ \ \ \ \ \ \hspace{-6pt} \qquad \tikz{
        \draw[black] (2.3,0) -- (5.2,0);
        \draw[black] (2.3,-2pt) -- (2.3,2pt)node[anchor=north] {\tiny$0$};
        \draw[black] (3.75,-2pt) -- (3.75,2pt)node[anchor=north] {\tiny$7.5$};
        \draw[black] (5.2,-2pt) -- (5.2,2pt)node[anchor=north] {\tiny$15$};
    } \vspace{0mm} &  \ \ \ \ \ \ \ \ \hspace{-6pt} \qquad \tikz{
        \draw[black] (2.3,0) -- (5.2,0);
        \draw[black] (2.3,-2pt) -- (2.3,2pt)node[anchor=north] {\tiny$0$};
        \draw[black] (3.75,-2pt) -- (3.75,2pt)node[anchor=north] {\tiny$7.5$};
        \draw[black] (5.2,-2pt) -- (5.2,2pt)node[anchor=north] {\tiny$15$};
    } \vspace{0mm} \\
    \bottomrule[1pt]
\end{tabularx}
\end{table*}

   \begin{table*}[t]
    \centering
    \caption{State and Action Prediction NRMSE for different latent state representation dimensions and training periods.} 
    \label{Table_3}
\begin{tabularx}{1\linewidth}{p{1.5cm} p{2cm} p{2.4cm} p{4cm} p{4cm}}
    \toprule[1pt]
    & Representation Dimension $d$ & Trajectory Length [s] & State Prediction NRMSE [\%] & Action Prediction NRMSE [\%]\\     
    \cmidrule(r){2-2} \cmidrule(r){3-3} \cmidrule(r){4-4}  \cmidrule(r){5-5}
 & $\qquad \quad 1$   &  $\qquad \quad 150$   & $10.71$\hspace{5pt}  \tikz{
        \fill[fill=color1] (0.0,0) rectangle (2.3,0.2);
    } & $1.54$\hspace{8pt}  \tikz{
        \fill[fill=color6] (0.0,0) rectangle (0.59,0.2);
    } \\
 &    & $\qquad \quad 250$  & $9.20$\hspace{9pt}  \tikz{
        \fill[fill=color2] (0.0,0) rectangle (2.01,0.2);
    } & $0.55$\hspace{8pt}  \tikz{
        \fill[fill=color7] (0.0,0) rectangle (0.21,0.2);
    } \\
  &    & $\qquad \quad 350$& $\textbf{7.99}$\hspace{9.75pt}  \tikz{
        \fill[fill=color3] (0.0,0) rectangle (1.77,0.2);
     } & $\textbf{0.29}$ \hspace{6.04pt}  \tikz{
        \fill[fill=color8] (0.0,0) rectangle (0.13,0.2);
    } \\ \hline
   & $\qquad \quad 2$ &  $\qquad \quad 150$  & $5.04$ \hspace{5.75pt}  \tikz{
        \fill[fill=color1] (0.0,0) rectangle (1.04,0.2);
    } & $1.54$ \hspace{5.5pt}  \tikz{
        \fill[fill=color6] (0.0,0) rectangle (0.59,0.2);
    }  \\
    &   &  $\qquad \quad 250$  & $4.93$ \hspace{5.75pt}  \tikz{
        \fill[fill=color2] (0.0,0) rectangle (0.97,0.2);
    } & $0.52$ \hspace{5.75pt}  \tikz{
        \fill[fill=color7] (0.0,0) rectangle (0.19,0.2);
    }\\
    & &  $\qquad \quad 350$  & $\textbf{3.73}$ \hspace{7.0pt}  \tikz{
        \fill[fill=color3] (0.0,0) rectangle (0.87,0.2);
    } & $\textbf{0.35}$ \hspace{6.75pt}  \tikz{
        \fill[fill=color8] (0.0,0) rectangle (0.16,0.2);
    }  \\ \hline
  & $\qquad \quad 3$ &  $\qquad \quad 150$  & $3.99$ \hspace{6.75pt}  \tikz{
        \fill[fill=color1] (0.0,0) rectangle (0.91,0.2);
    } & $1.51$ \hspace{5.75pt}  \tikz{
        \fill[fill=color6] (0.0,0) rectangle (0.58,0.2);
    } \\
   &  &  $\qquad \quad 250$  & $2.06$ \hspace{7pt}  \tikz{
        \fill[fill=color2] (0.0,0) rectangle (0.60,0.2);
    } & $0.51$ \hspace{6pt}  \tikz{
        \fill[fill=color7] (0.0,0) rectangle (0.19,0.2);
    } \\
    &  &  $\qquad \quad 350$  & $\textbf{1.22}$ \hspace{8pt}  \tikz{
        \fill[fill=color3] (0.0,0) rectangle (0.42,0.2);
    } & $\textbf{0.29}$ \hspace{7.5pt}  \tikz{
        \fill[fill=color8] (0.0,0) rectangle (0.13,0.2);
    } \\ \hline
    & $\qquad \quad 4$ &  $\qquad \quad 150$  & $1.28$ \hspace{7pt}  \tikz{
        \fill[fill=color1] (0.0,0) rectangle (0.48,0.2);
    } & $1.35$ \hspace{6.5pt}  \tikz{
        \fill[fill=color6] (0.0,0) rectangle (0.53,0.2);
    } \\
     & &  $\qquad \quad 250$  & $1.03$ \hspace{7pt}  \tikz{
        \fill[fill=color2] (0.0,0) rectangle (0.3,0.2);
    } & $0.38$ \hspace{6.75pt}  \tikz{
        \fill[fill=color7] (0.0,0) rectangle (0.14,0.2);
    } \\
     & &  $\qquad \quad 350$  & $\textbf{0.96}$ \hspace{8pt}  \tikz{
        \fill[fill=color3] (0.0,0) rectangle (0.29,0.2);
    } & $\textbf{0.22}$ \hspace{8pt}  \tikz{
        \fill[fill=color8] (0.0,0) rectangle (0.10,0.2);
    } \\
     
 &    & &\ \ \ \ \ \ \ \ \hspace{-9pt} \qquad \tikz{
        \draw[black] (2.3,0) -- (5.2,0);
        \draw[black] (2.3,-2pt) -- (2.3,2pt)node[anchor=north] {\tiny$0$};
        \draw[black] (3.75,-2pt) -- (3.75,2pt)node[anchor=north] {\tiny$7.5$};
        \draw[black] (5.2,-2pt) -- (5.2,2pt)node[anchor=north] {\tiny$15$};
    } \vspace{0mm} &\ \ \ \ \ \ \ \ \hspace{-9pt} \qquad \tikz{
         \draw[black] (2.3,0) -- (5.2,0);
        \draw[black] (2.3,-2pt) -- (2.3,2pt)node[anchor=north] {\tiny$0$};
        \draw[black] (3.75,-2pt) -- (3.75,2pt)node[anchor=north] {\tiny$7.5$};
        \draw[black] (5.2,-2pt) -- (5.2,2pt)node[anchor=north] {\tiny$15$};
    } \vspace{0mm} \\
    \bottomrule[1pt]
\end{tabularx}
\end{table*}

 \vspace{3pt}\noindent\textbf{Prediction accuracy:}\, Since the main goal of the proposed sensing Koopman AE and the controlling Koopman AE is to predict future system states and control action commands, respectively, we use the normalized root mean square error $\left( \mathrm{NRMSE}\right)$ to evaluate the quality of prediction performance. The $\mathrm{NRMSE}$ is the root mean squared error between the predicted and observed signals in a prediction window of length $M_{p}$ normalized by the norm of the difference between the minimum and maximum vectors of the observed signals, \begin{align}
     \label{eq_RMSE_state}
     \mathrm{NRMSE}_{M_{p}} = \frac{ \sqrt{ \frac{1}{M_{p}} \sum_{ m  = M_{\scriptscriptstyle \text{s}+1} }^{M_{\scriptscriptstyle \text{s}} + M_{p}}  \| \bar{\mathbf{x}}_{m} - \hat{\mathbf{x}}_{m} \|_{2}^2 }} { \| \max (\hat{\mathbf{x}}) -  \min (\hat{\mathbf{x}}) \|_{2}} \times 100,
     \end{align}where $M_{p}$ is the prediction time horizon at test that the remote controller makes in the second phase of the remote controlling. Note that if the second phase becomes long, there will be error propagation that will degenerate the prediction performance. As a result, we have two options to enhance the prediction performance: 1) sending a new latent state representation to initialize the prediction, and 2) shifting to the first phase of remote controlling to fine-tune the trained Koopman operator.

 \vspace{3pt}\noindent\textbf{Wireless communication metric:}\, To validate the performance of the proposed Koopman AE in terms of communication efficiency, we measure the maximum number of consecutive lost packets $M_{\text{lost}}$ until the most recent successful transmission. Hence, the number of consecutive lost packets linearly increases with time if the transmitter-receiver pair experiences adverse channel conditions or the prediction accuracy of the Koopman AE reaches a predefined threshold.     
 
 \vspace{3pt}\noindent\textbf{Control stability:}\, is a control performance metric in terms of the control system response to any initial condition. The control system is said to be asymptotically stable if, for  every $\epsilon > 0$, $m \geq 0 $, and a given control action command, there exists $\| \mathbf{x}_{m} - \mathbf{x}_{d} \|^{2}_{2} < \epsilon$ as $ m \rightarrow \infty$, where $\mathbf{x}_{d} \in \mathbb{R}^{p}$ is the desired system state~\cite{hangos2004analysis}. Then, the control stability along the prediction time horizon at test is the mean squared control error, i.e., the time-averaged of squared system states deviation from its desired state along the prediction time horizon at the test, given as \begin{align}
     \label{eq_MSCE}
     \mathrm{MSCE}_{M_{p}} =  \frac{1}{M_{p}} \sum_{m=1}^{M_{p}}  \| \mathbf{x}_{m} - \mathbf{x}_{d} \|_{2}^{2}.
 \end{align} Note that the control stability in the proposed approach depends on the prediction accuracy that are affected by the communication efficiency, highlighting the importance of jointly designing communication, learning, and control operations.

 \begin{figure*}
    \centering
    \subfigure[]{\includegraphics[width=0.4\textwidth]{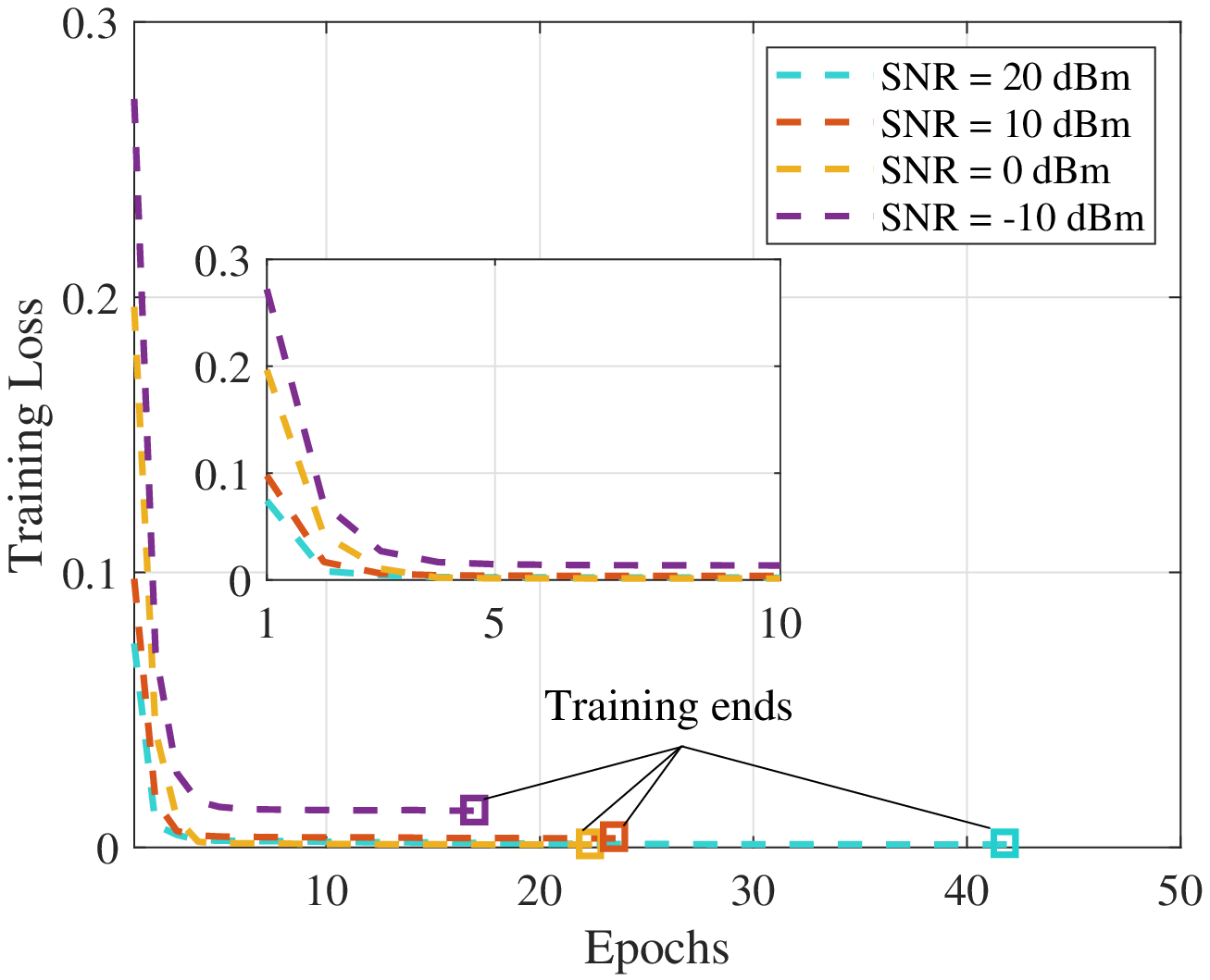}} 
    \subfigure[]{\includegraphics[width=0.4\textwidth]{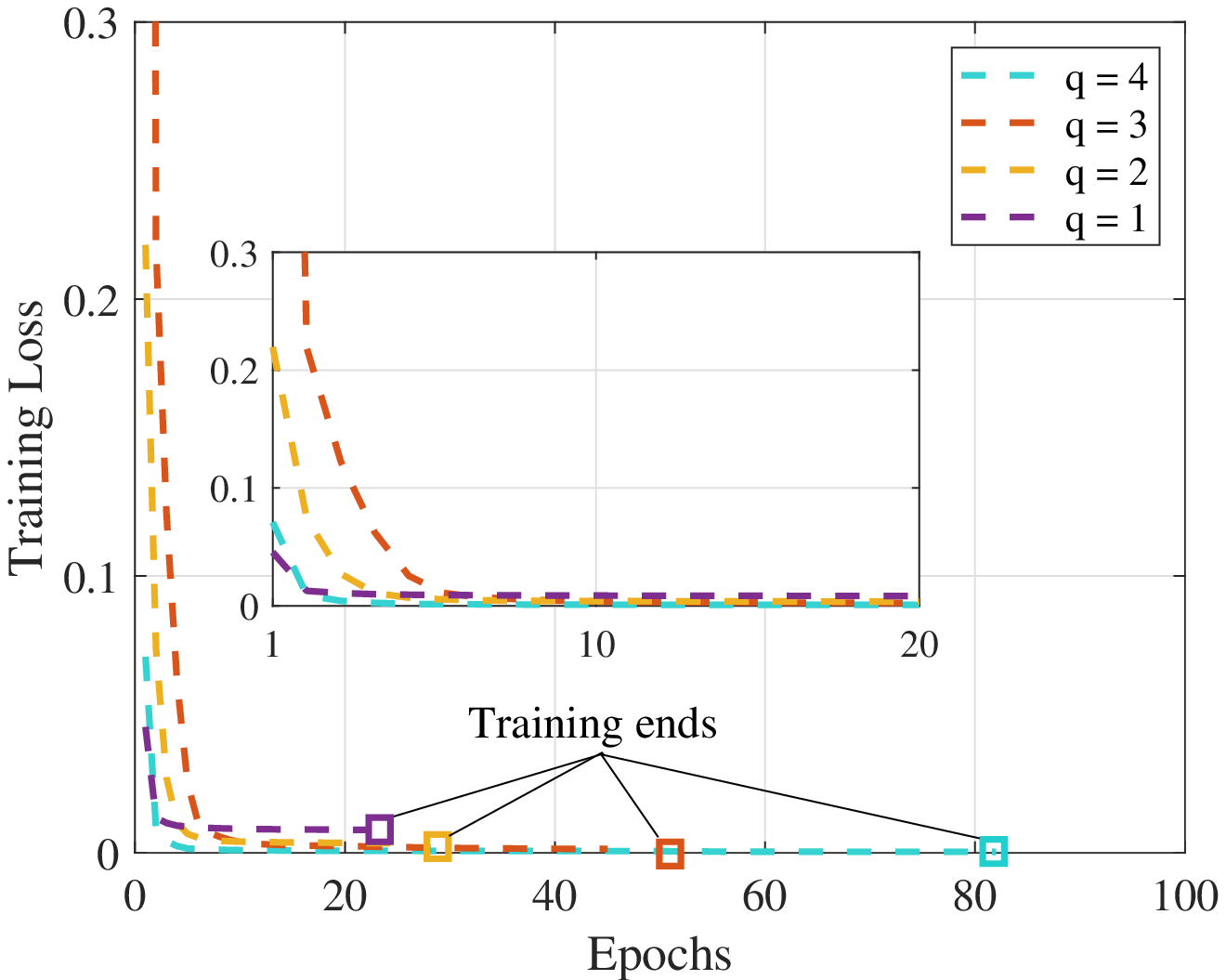}} 
    \caption{Training loss of the sensing Koopman AE with (a) different SNR and (b) different latent state representation dimensions. }
    \label{fig Train_Loss}
  \end{figure*} 

  \subsection{Performance Evaluation of Split Koopman AE}
  \label{Performance_Evaluation}
  In this subsection, we evaluate the performance of the proposed two-way split Koopman AE for different communication, prediction, and control parameters.    

   \vspace{3pt}\noindent\textbf{SNR  Vs. State Prediction Accuracy.}\quad Fig.~\ref{fig state_prediction} demonstrates the prediction performance of the proposed sensing Koopman AE compared to the non-predictive remote monitoring of the non-linear inverted cart-pole system. The remote controller in the proposed sensing Koopman AE predicts the future time-series system states with different SNR and one-step target prediction depth considered in training compared to the remote controller in the non-predictive remote monitoring that receives the non-linear system states at each time slot over an ideal channel. It can be seen that the predicted system states of the proposed sensing Koopman AE with high SNR, i.e., $\text{SNR} \in \{10,20\}$ $\mathrm{dBm}$  match closely the observed system states in the non-predictive remote monitoring compared to the predicted system states with low SNR, i.e., $\text{SNR} \in \{-10,0\}$ $\mathrm{dBm}$.      
   
   The reason behind this result is that the stability of training the sensing Koopman AE relies on the communication reliability of the observed system states at the remote controller, highlighting the importance of co-designing the communication and deep learning operations. As a result, the state prediction accuracy is improved at the cost of increasing communication resources in terms of transmission power and allocated channel bandwidth. Moreover, the state prediction results in Fig.~\ref{fig state_prediction} emphasize the ability of the sensing Koopman AE in discovering the Koopman invariant subspace with four latent state representations and one-step target prediction depth considered in training. Then, the sensing Koopman AE globally linearizes the non-linear system dynamics of the inverted cart-pole and predicts the future system states at the remote controller based on the trained Koopman matrix, improving the forward link communication efficiency.

   \vspace{3pt}\noindent\textbf{SNR  Vs. Action Prediction  Accuracy.}\quad  For the proposed controlling Koopman AE depicted in Fig.~\ref{fig_force}, the actuator predicts the future control action commands with different SNR values, and one-step target prediction depth considered in training compared to the actuator in the non-predictive remote controlling of the non-linear inverted cart-pole system that observes the control action commands calculated over Jacobian linearization method at each time slot over an ideal channel. Here, the transmission power in both the forward and reverse communications are linked together since the observed control action command at the actuator is based on the observed system states affecting the control action prediction performance and the control stability. It is clear that the predicted control action commands in the controlling Koopman AE with high SNR, i.e., $\text{SNR} \in \{10,20\}$ $\mathrm{dBm}$ approximately coincides with the observed control action commands compared to the predicted control action commands with low SNR, $\text{SNR} \in \{-10,0\}$ $\mathrm{dBm}$.

   The rationale behind this result is that the learning stability of the controlling Koopman AE depends on the communication reliability of the observed control action commands at the actuator, highlighting the importance of jointly designing the communication, learning, and control operations. Hence, the action prediction accuracy is improved which yields stabilized control system at the expense of utilizing high communication resources in terms of transmission power and allocated channel bandwidth in the first phase of remote controlling. Moreover, the action prediction results in Fig.~\ref{fig_force} depict the ability of the controlling Koopman AE in capturing the non-linear action dynamics. Hence, the controlling Koopman AE  has the ability to predict the future control action commands, improving both the forward and reverse links communication efficiency.   
   
 \begin{figure*}
    \centering
    \subfigure[State prediction performance.\label{fig_pred_depth_a}]{\includegraphics[width=0.4\textwidth]{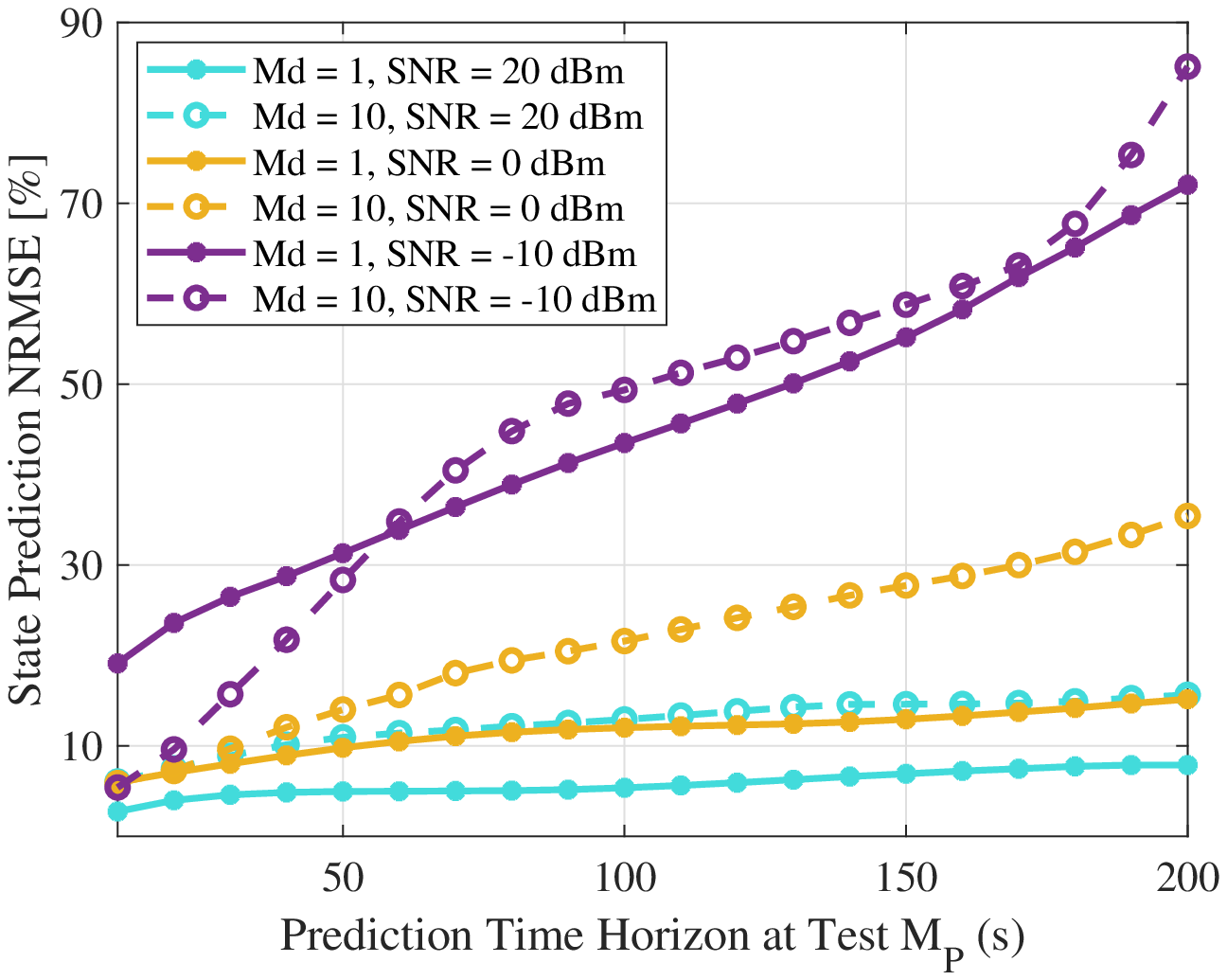}} 
    \subfigure[State training completion time.\label{fig_pred_depth_b}]{\includegraphics[width=0.4\textwidth]{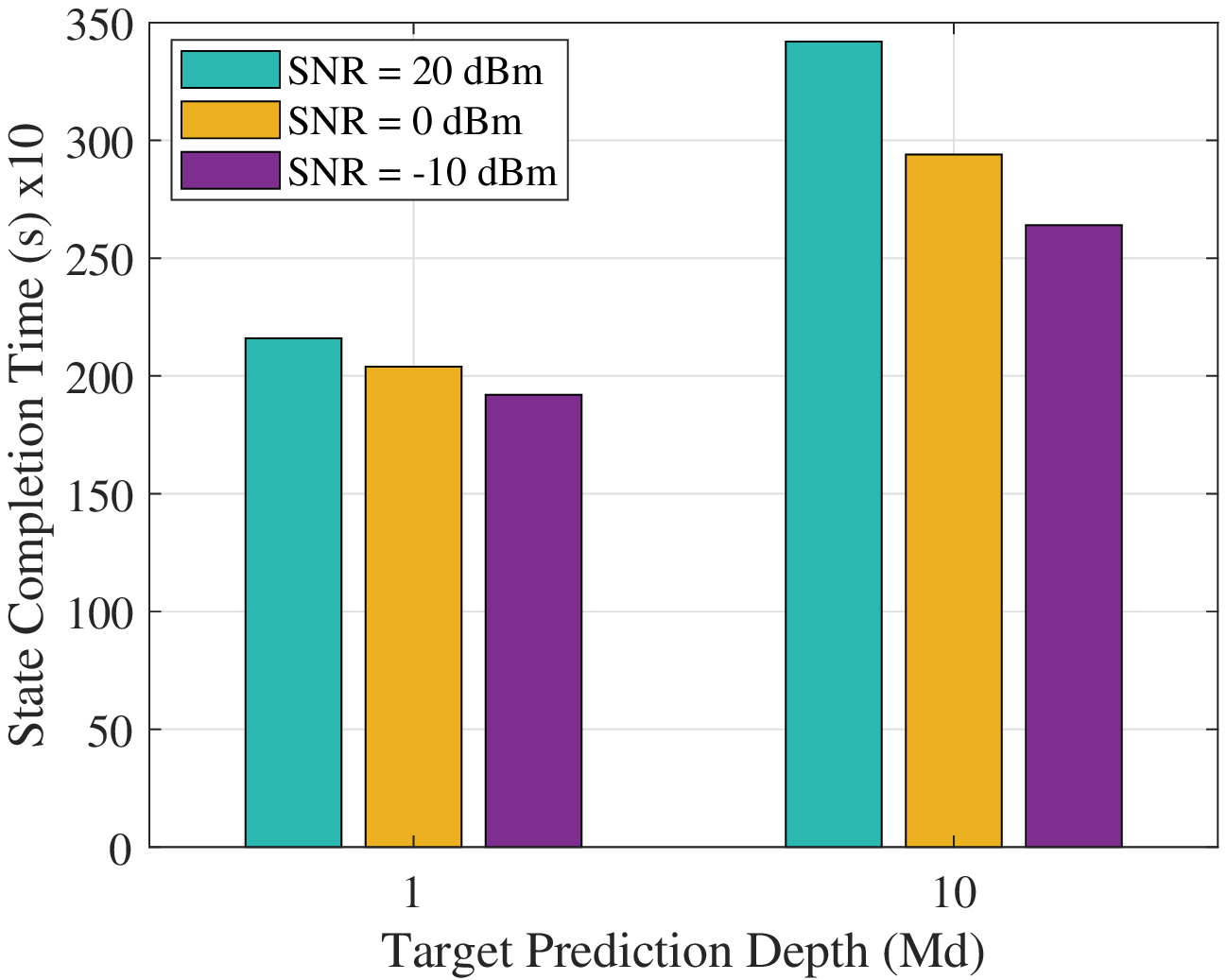}}
    \subfigure[Action prediction performance.\label{fig_pred_depth_c}]{\includegraphics[width=0.4\textwidth]{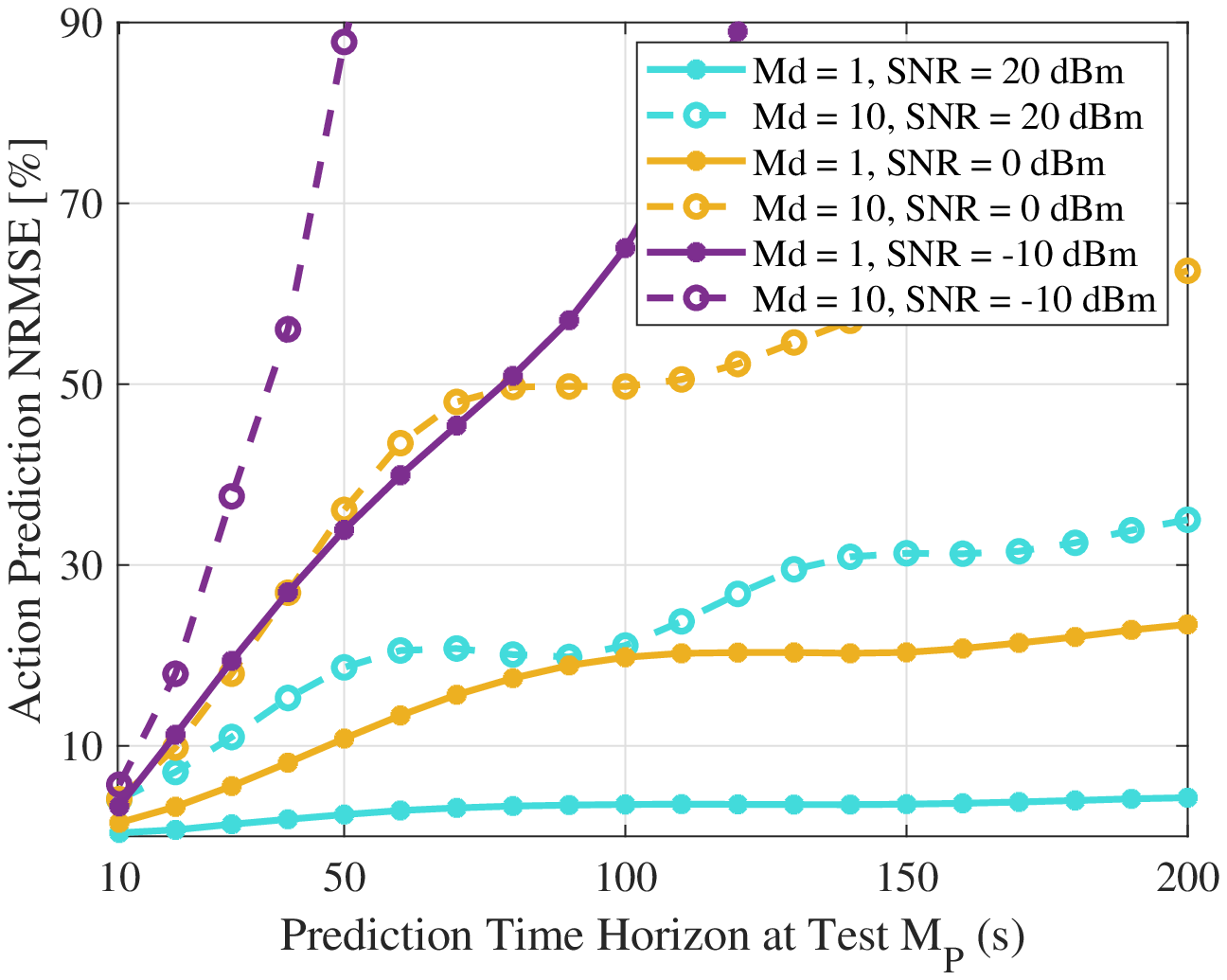}} 
    \subfigure[Action training completion time.\label{fig_pred_depth_d}]{\includegraphics[width=0.4\textwidth]{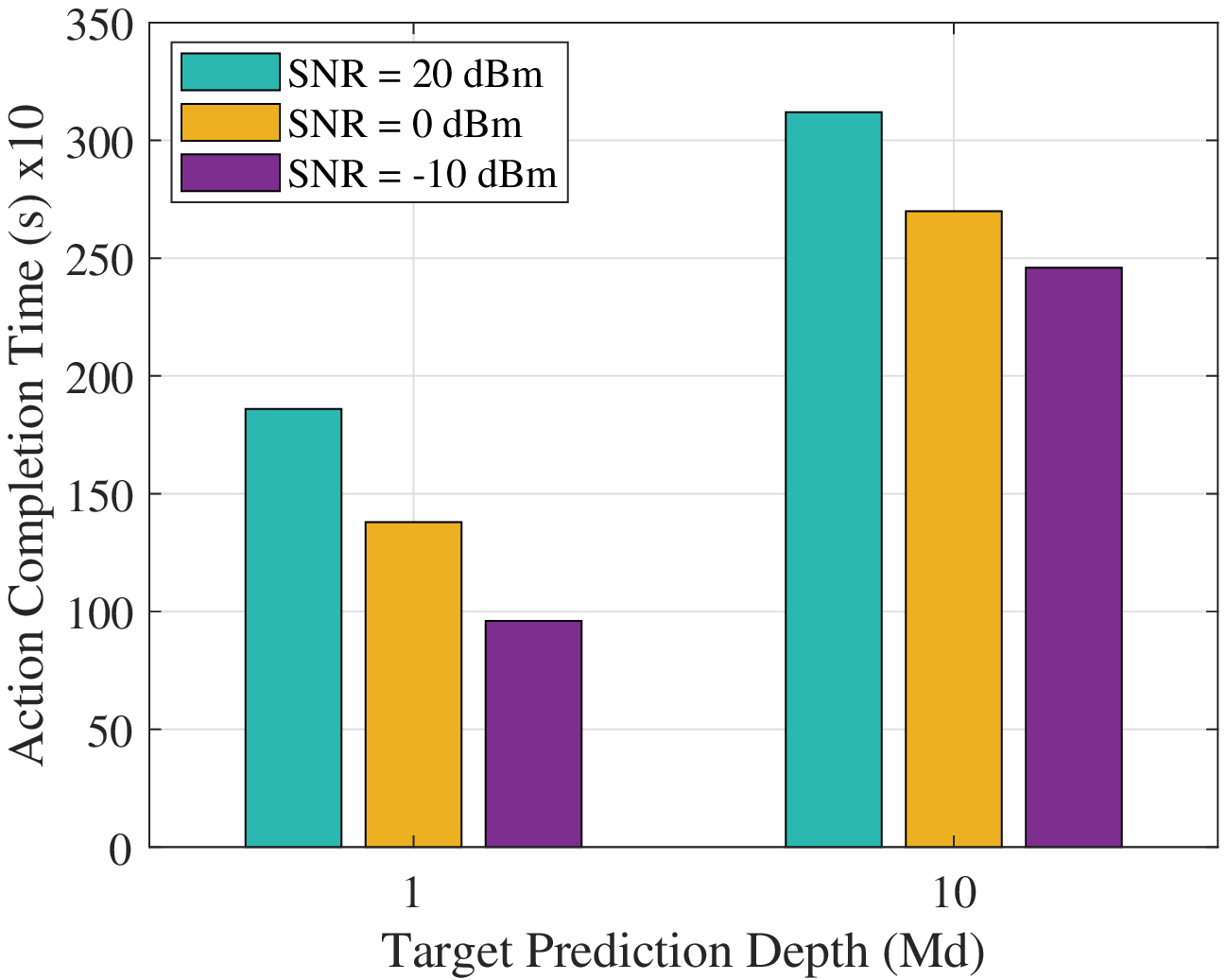}} 
    \caption{Prediction performance and training completion time of both the proposed split sensing Koopman learning and the proposed split controlling closed-loop Koopman learning for different target prediction depths and different SNR values.}
    \label{fig_pred_depth}
  \end{figure*}
 
   \vspace{3pt}\noindent\textbf{SNR  \& Representation Dim Vs. Prediction Accuracy.}\quad As shown in Fig.~\ref{fig state_prediction}, the state prediction performance of the sensing Koopman AE is improved at the cost of increasing the communication resources in terms of four-dimensional latent state representations and a high SNR value. Table~\ref{Table_1} describes the state prediction performance of the sensing Koopman AE for different SNR values and different latent state representation dimensions. From this table, it is clear that the state prediction accuracy is improved at the cost of increasing the SNR for the same latent state representation dimension, the same sensor transmission power, the same distance from the sensor to the controller, one-step target prediction depth considered in training, and the same training period. The state prediction accuracy is almost the same in the lowest latent state representation dimension, i.e., $d = 1$, for low SNR values, i.e., $\text{SNR} \in \{-10,0,10\}$ dBm. In contrast, the state prediction accuracy is improved by increasing the SNR values in the other latent state representation dimensions $d \in \{2,3,4\}$. The rationale behind the state prediction results in Table~\ref{Table_1} is that the lowest latent state representation dimension is not enough to discover the Koopman invariant subspace in the observables space of the non-linear inverted cart-pole system affecting the state prediction performance. Hence, for obtaining high state prediction performance, the linearized Koopman subspace reflecting the non-linear system dynamics is obtained at the expense of increasing the communication resources in terms of selecting a high SNR value and a large latent state representation dimension, leading to a trade-off between the communication cost and the prediction accuracy.

   Table.~\ref{Table_1} demonstrates also the control action prediction performance of the controlling Koopman AE compared to the non-predictive remote controlling of the non-linear inverted cart-pole system. The controlling Koopman AE located at the actuator locally predicts the future control action commands for different SNR values and different latent state representation dimensions. Note that the control action prediction performance is improved at the cost of increasing the SNR value for the same latent state representation dimension, the same controller transmission power, the same distance from the controller to the actuator, one-step target prediction depth considered in training, and the same training period. Additionally, for the same SNR value, we notice that the control action prediction performance is almost the same for the low latent state representation dimensions of $d  \in \{1,2,3 \} $ compared to the four-dimensional latent state representation. This demonstrates the trade-off between the communication payload size and the control action prediction accuracy, in addition to the fact that the four-dimensional latent state representations quickly improve the control action prediction accuracy as a result of reflecting the non-linear action dynamics in the controlling Koopman invariant subspace.
  
  \vspace{3pt}\noindent\textbf{Trajectory Length \& Representation Dim Vs. Prediction Accuracy.}\quad As a result of the difficulty in identifying the non-linear dynamics in an interpretable way, the controller in the non-predictive remote controlling case receives the system states at each time slot and calculates the control action using the non-linear control theory. In contrast, our proposed sensing Koopman AE identifies the non-linear system dynamics in an interpretable linear form that is utilized by the remote controller to calculate the optimal control action using the linear control theory and to predict the future system states even without communication. Table.~\ref{Table_3} demonstrates the state prediction accuracy of the sensing Koopman AE for the same sensor transmission power, the same distance from the sensor to controller, different training periods, different latent state representation dimensions, and one-step target prediction depth considered in training. The state prediction accuracy is improved as the training period increases for the same SNR value and the same latent state representation dimensions, highlighting the trade-off between the state prediction accuracy and the communication payload size. The state prediction accuracy in Table~\ref{Table_3} is improved as a result of increasing the number of observed system states per trajectory for the same SNR value and the same latent state representation dimension. The rationale behind this result is that increasing the number of observed system states per trajectory ensures the training dataset is rich enough to represent the non-linear system dynamics and guarantees the robustness of discovering the Koopman invariant subspace at the expense of increasing the communication resources.       
  
  \begin{figure*}[t!]
    \centering
    \subfigure[State prediction performance. \label{fig_state_repesentations_a}]{\includegraphics[width=0.4\textwidth]{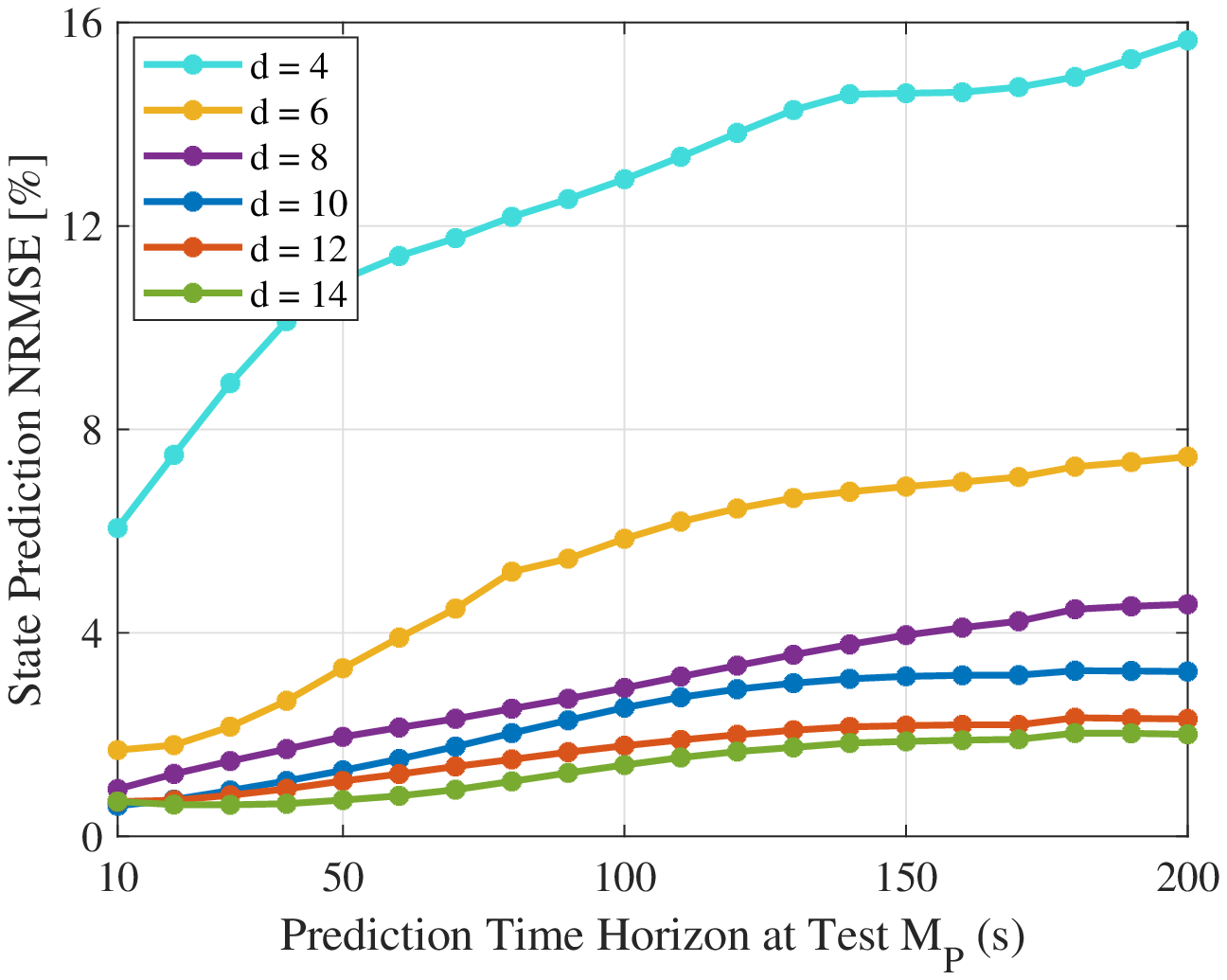}} 
    \subfigure[ State training completion  time.\label{fig_state_repesentations_b}]{\includegraphics[width=0.4\textwidth]{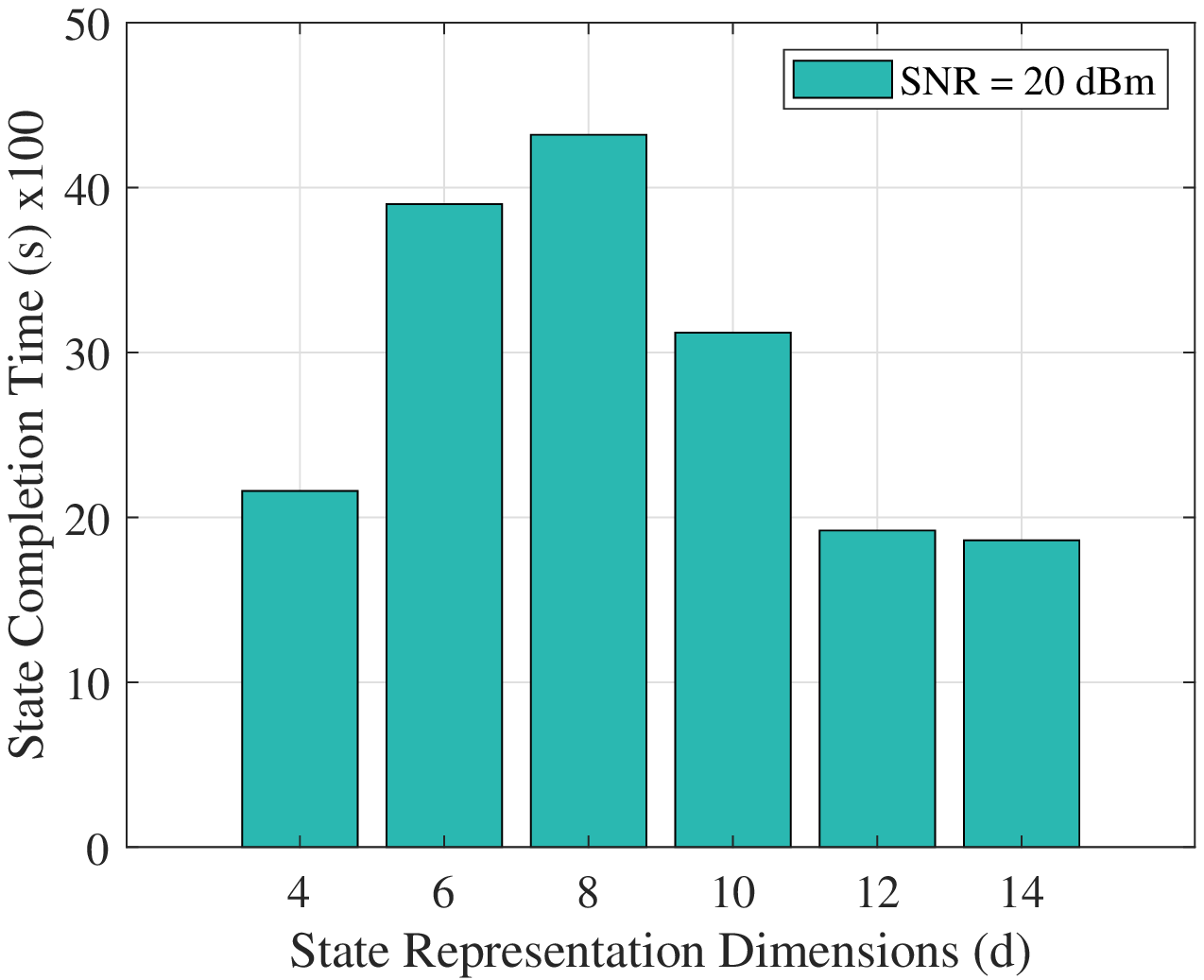}} 
    \caption{Prediction performance and training completion time of the proposed sensing Koopman AE with a ten-step target prediction depth, different latent state representation dimensions $ d \in \{ 4, 6, 8, 10, 12, 14\} $, and  $\mathrm{SNR} = 20 \, \mathrm{dBm}$.}
    \label{fig_state_repesentations}
  \end{figure*}  
  
  Table.~\ref{Table_3} demonstrates also the control action prediction performance of the controlling Koopman AE for the same controller transmission power, the same distance from the sensor to controller, different training periods, different latent state representation dimensions, and one-step target prediction depth considered in training. It is clear that the control action prediction performance is improved at the cost of increasing the number of received control action commands per trajectory in the training dataset for the same SNR value and the same latent state representation dimension, highlighting the trade-off between the action prediction performance and the communication payload size. Moreover, the action prediction performance is almost the same for low latent state representation dimensions $d \in \{1,2,3\} $, the same target SNR, and the same trajectory length compared to the action prediction performance for the four-dimensional latent state representation. The reason behind this result is that the low latent state representation dimensions are insufficient to discover the controlling Koopman invariant subspace of the non-linear action dynamics. Additionally, for the same latent state representation dimension, increasing the trajectory length increases the action prediction performance since the high sampling rate helps in describing the non-linear action dynamics and discovering the controlling Koopman invariant subspace.

  \vspace{3pt}\noindent\textbf{SNR \& Representation Dimensions vs. Training Loss.}\quad Fig.~\ref{fig Train_Loss} demonstrates the overall training loss of the sensing Koopman AE for different SNR values and different latent state representation dimensions. For the same latent state representation dimension, i.e., $d= 3$, it is clear that the training loss with the lowest SNR value, i.e., $\mathrm{SNR} = -10$ $\mathrm{dBm}$ converges faster compared to the other SNR values at the cost of achieving low training accuracy. In contrast, the training loss with the high SNR, i.e., $\mathrm{SNR} = 20$ $\mathrm{dBm}$ converges slowly to the minimum value compared to the other SNR values, i.e., $\text{SNR} \in \{-10, 0, 10\}$ $\mathrm{dBm}$.  This in turn shows the impact of the SNR value on the training accuracy in which the received data samples to the another part of the sensing Koopman AE affects the training accuracy, convergence speed, and communication payload size.    
 
  For the same SNR value, we can see that the training loss of the sensing Koopman AE slowly converges as the representation dimension increases. More specifically, the communication payload size of the forward and backward propagation signals is proportional to the number of weights parameters in the Koopman hidden layer. As a result, increasing the latent state representation dimensions results in a high latency for transmitting the forward and backward propagation signals which leads to high prediction accuracy, highlighting the trade-off between the prediction accuracy, transmission latency, and communication payload size.
  
 \begin{figure*}[t!]
    \centering
    \subfigure[State prediction performance.\label{fig_state_traj_a}]{\includegraphics[width=0.4\textwidth]{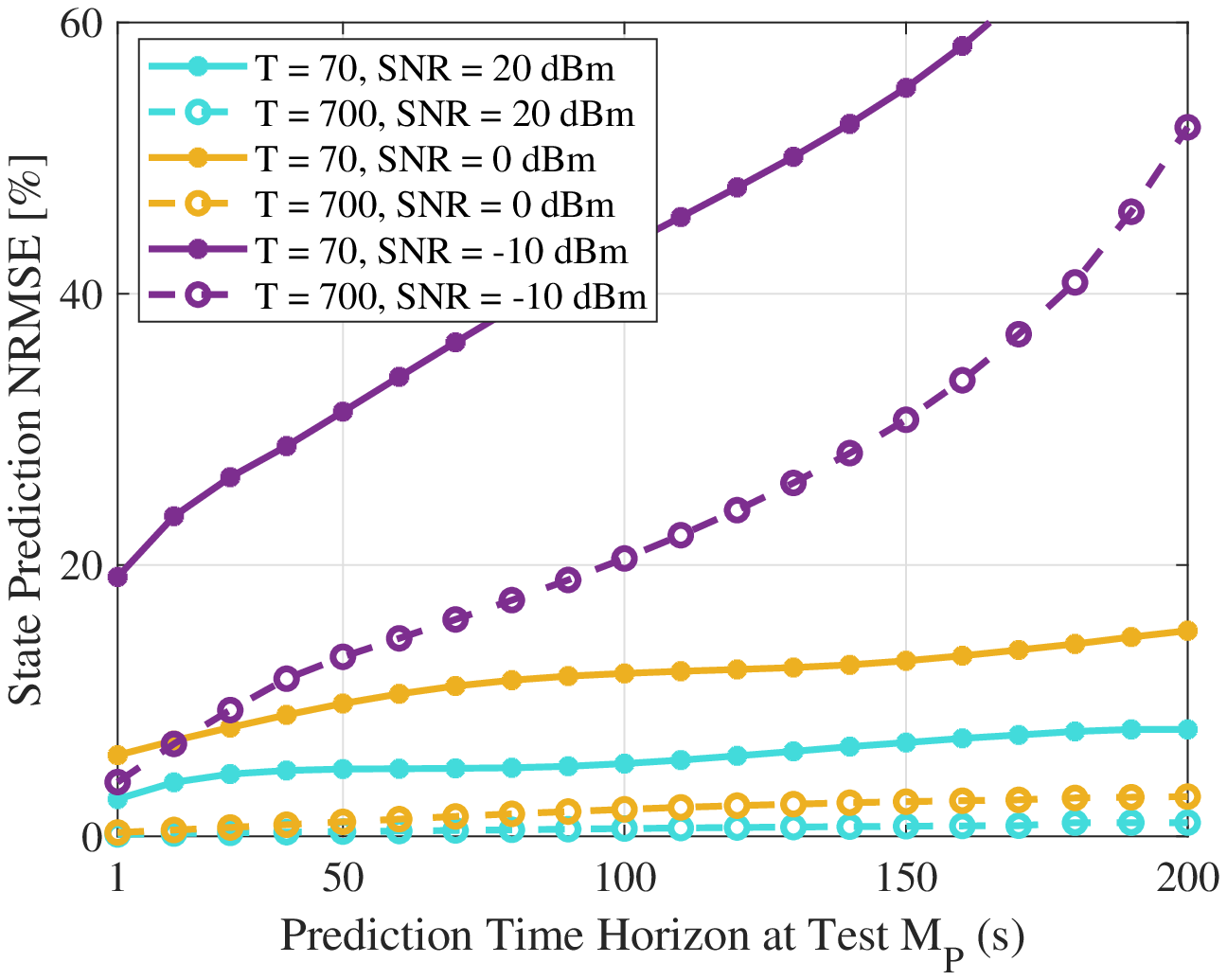}} 
    \subfigure[State training completion time. \label{fig_state_traj_b}]{\includegraphics[width=0.4\textwidth]{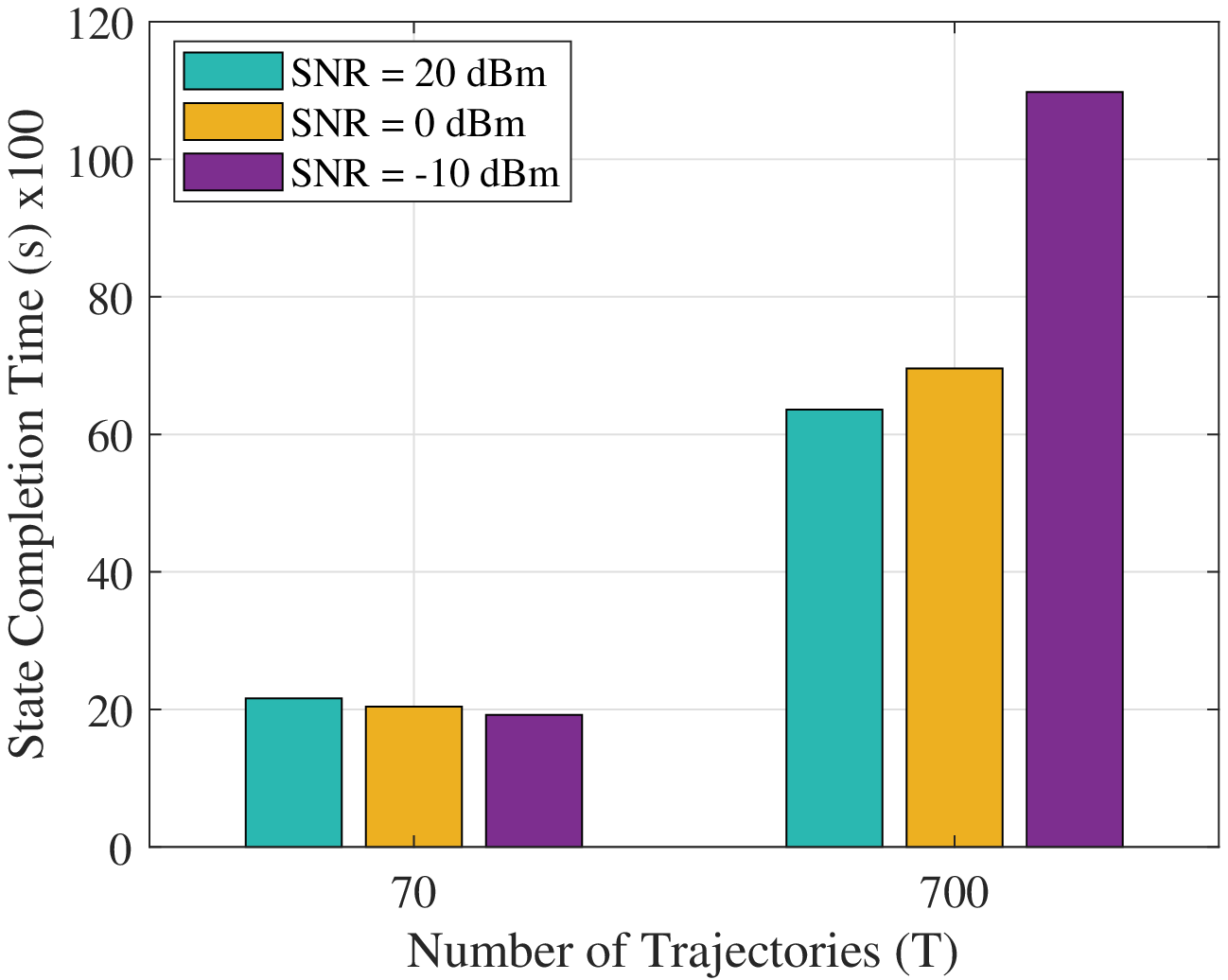}} 
    \subfigure[Action prediction performance. \label{fig_state_traj_c}]{\includegraphics[width=0.4\textwidth]{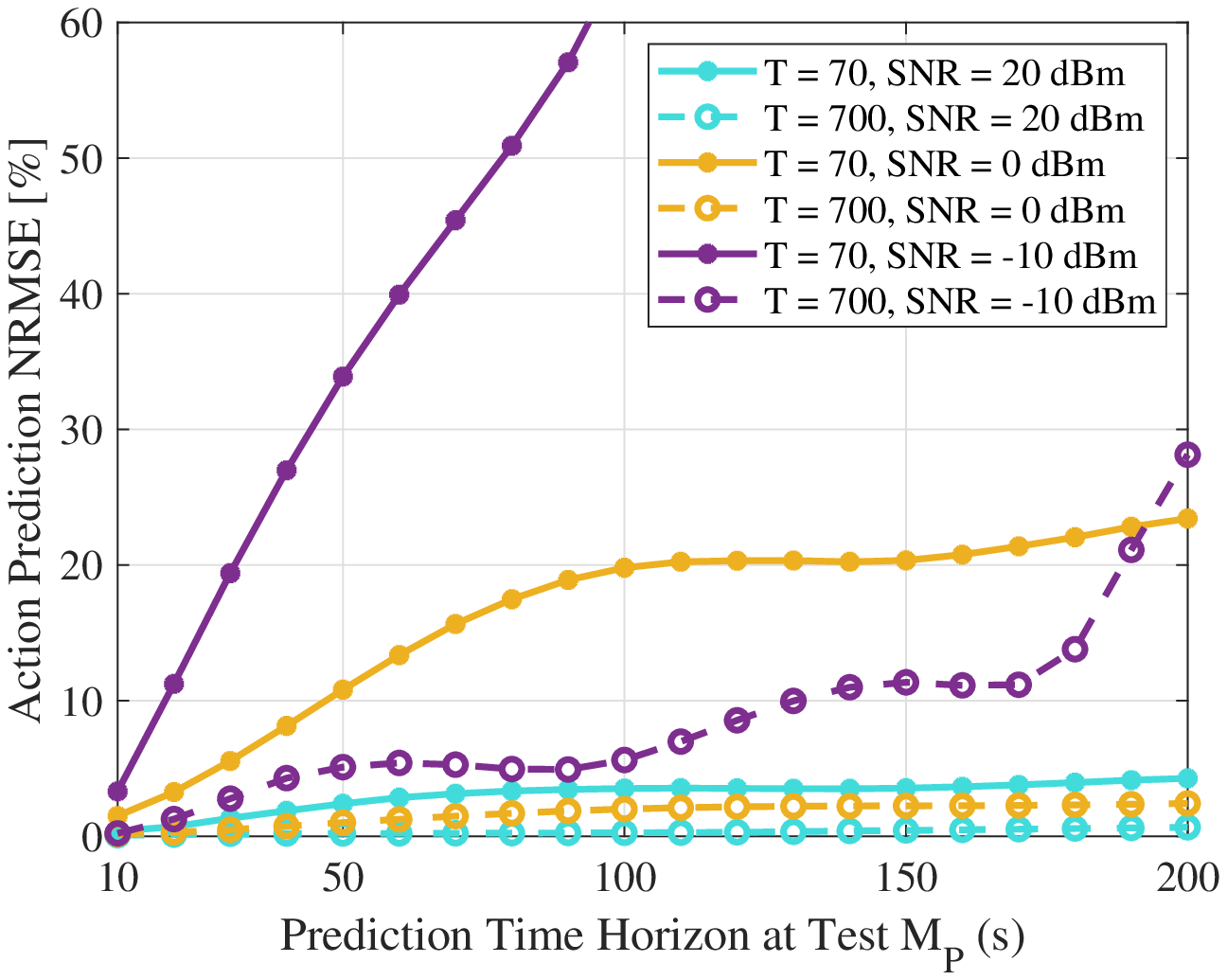}} 
    \subfigure[Action training completion time. \label{fig_state_traj_d}]{\includegraphics[width=0.4\textwidth]{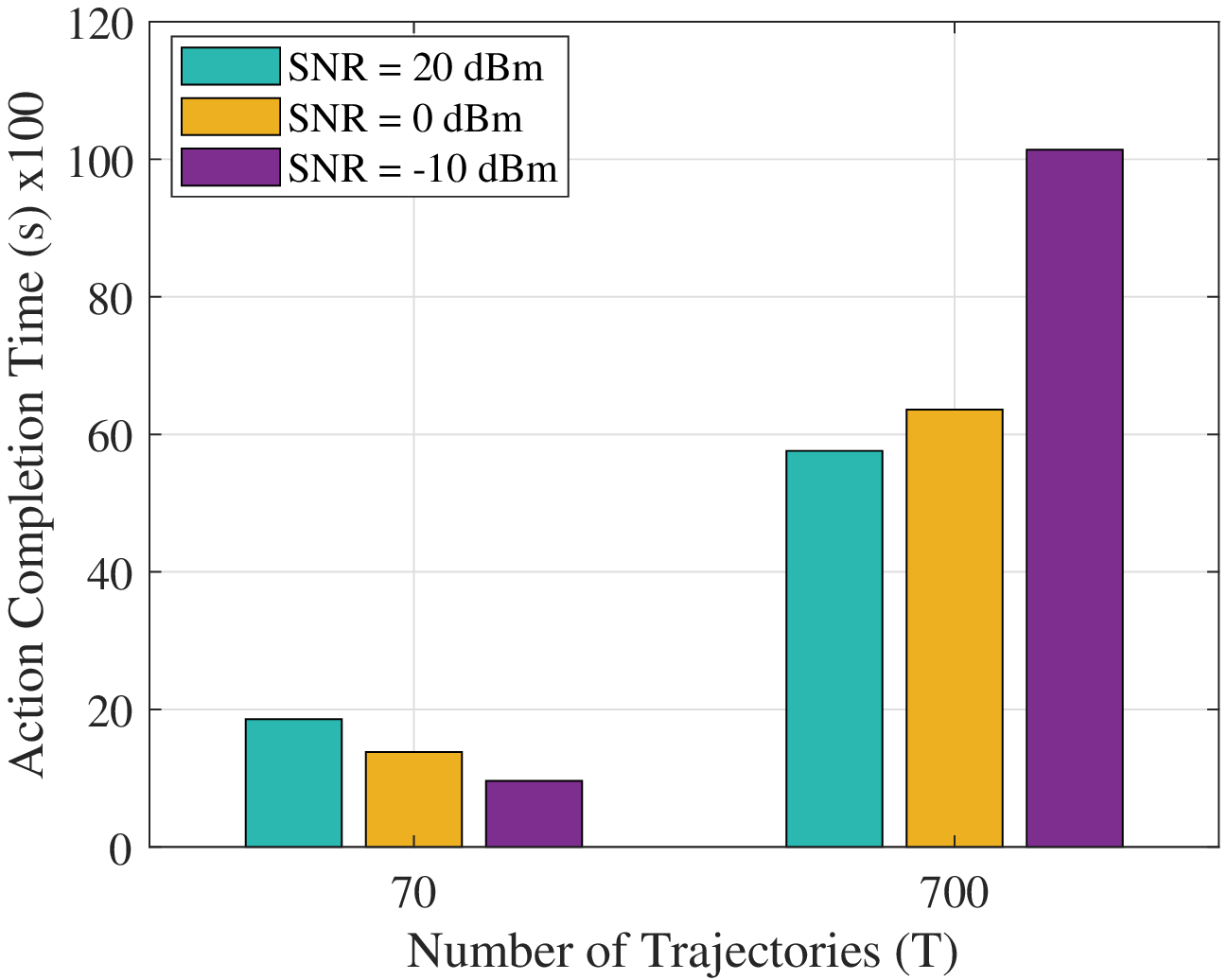}} 
    \caption{Prediction performance and training completion time of the proposed sensing Koopman AE with a one-step target prediction depth, different number of trajectories in the training dataset $T_{t} = \{ 70, 700 \} $, different SNR values $\mathrm{SNR} = \{ 20, 0, -10  \} \, \mathrm{dBm}$. }
    \label{fig_state_traj}
  \end{figure*}

 \vspace{3pt}\noindent\textbf{Target Prediction Depth Vs.  Prediction Performance \& Training Completion Time.}\quad To dive deeper into the benefits of both the sensing Koopman AE and the controlling Koopman AE, we increase the range of random initial conditions to generate the training dataset as $[-2.0,2.0]$ for each state dimension and also increase the trajectory length as $\mathcal{M}_{\mathrm{Train}} = 500 \, \mathrm{s}$. Here, the results are obtained based on averaging several training runs with different seeds. Fig.~\ref{fig_pred_depth_a} and Fig.~\ref{fig_pred_depth_c} show the prediction performance of the sensing Koopman AE and the controlling Koopman AE, respectively for different target prediction depths considered in training, different SNR values, and four-dimensional latent state representations. For the same target prediction depth considered in training, we can see that the state prediction performance decreases as the prediction time horizon at the test increases due to the error propagation as the prediction time increases. As a result, based on a predefined prediction threshold, we send a new latent state representation or shift to the first phase of remote control to enhance the state prediction performance.

 For the same SNR value, it is clear that the prediction performance of the split Koopman AE with a one-step target prediction depth considered in training is better than that with a ten-step target prediction depth over different prediction time horizons at the test time, i.e., increasing the target prediction depth deteriorates the prediction performance over different prediction time horizons at the test. The rationale behind this result is that the small errors in training the Koopman operator with a high target prediction depth during the first phase of remote control lead to large errors when evaluating the prediction performance for a large prediction time horizon at the test.   
 
 Fig.~\ref{fig_pred_depth_b} and Fig.~\ref{fig_pred_depth_d} show the training completion time required to train the sensing Koopman AE and the controlling Koopman AE, respectively, for different SNR values and different target prediction depths. For the same SNR value, it is clear that the training completion time associated with a ten-step target prediction depth is larger than that with a one-step target prediction depth due to the high computational complexity associated with a ten-step target prediction depth considered in training. Hence, increasing the target prediction depth considered in training leads to a high training completion time and a high communication payload size due to the necessity to transmit the system states in the first phase of remote control until the sensing Koopman AE is well-trained.

 By analyzing both the prediction performance and the training completion time of the split Koopman AE, we can see that the split Koopman AE with one-step target prediction depth and a low SNR value has better prediction performance compared to that with ten-step target prediction and a high SNR value. In addition, the split Koopman AE with one-step target prediction depth has a low communication cost in terms of transmission power and communication payload size. For instance, the split Koopman AE with one-step target prediction depth and $\mathrm{SNR} = 0 \, \mathrm{dBm}$ has better prediction performance compared to that with ten-step target prediction depth and $\mathrm{SNR} = 20 \, \mathrm{dBm}$. Overall, increasing the target prediction depth with the current hyperparameter setting deteriorates the prediction performance compared to the low target prediction depth. Hence, we next study the effect of changing the latent state representation dimensions on the prediction performance with a ten-step target prediction depth for the same SNR value.

 \vspace{3pt}\noindent\textbf{Representation Dim Vs. Prediction Performance \& Training Completion Time.}\quad Fig.~\ref{fig_state_repesentations} presents the prediction performance and training completion time of the sensing Koopman AE for different latent state representation dimensions, ten-step target prediction depth considered in training, $\mathrm{SNR} = 20 \, \mathrm{dBm}$, and the same previous hyperparameter setting. First, it is clear from Fig.~\ref{fig_state_repesentations_a} that the prediction performance of the sensing Koopman AE with a ten-step target prediction depth is improved at the cost of increasing the latent state representation dimensions, highlighting the trade-off between the target prediction depth considered in training and the latent state representation dimensions. Furthermore, we can see that the prediction performance with a ten-step target prediction depth starts to saturate after twelve-dimensional latent state representations over different prediction time horizons at the test. The reason behind this result is that increasing the latent state representation dimensions in the high target prediction depth setting ensures discovering the Koopman invariant subspace that guarantees high prediction performance for a long prediction time horizon at the test. Hence, selecting a large dimension of latent state representations with a large target prediction depth is instrumental in casting non-linear system dynamics in a linear form at the expense of increasing the communication payload size, leading to a trade-off between the communication cost and the prediction performance with a high target prediction depth considered in training.   
 
 Fig.~\ref{fig_state_repesentations_b} shows the training completion time required to train the sensing Koopman AE with a ten-step target prediction depth considered in training and different latent state representation dimensions. We can see that the training completion time increases as the latent state representation dimensions increase until eight-dimensional latent state representations, after which it starts to decrease until twelve latent state representation dimensions, then it is almost the same. The reason behind this result is that increasing the latent state representation dimensions in the large target prediction depth considered in training hastens the convergence speed of the sensing Koopman AE training to discover the Koopman invariant subspace. As a result, reducing the communication payload size required to observe the system states in the first phase of remote control until the sensing Koopman AE is well trained.

 \vspace{3pt}\noindent\textbf{Trajectories Number Vs.  Prediction Performance \& Training Completion Time.}\quad Fig.~\ref{fig_state_traj_a} and Fig.~\ref{fig_state_traj_c} demonstrate the prediction performance of the sensing Koopman AE and the controlling Koopman AE, respectively, with a one-step target prediction depth considered in training, four-dimensional latent state representations, the same previous hyperparameter setting,  different SNR values, and a different number of trajectories in the training dataset. For the same SNR value, it is clear that the prediction performance of the split Koopman AE is improved over different prediction time horizons at the test by increasing the number of trajectories in the training dataset as a result of well representing the non-linear system dynamics by increasing the number of trajectories in the training dataset. For the same number of trajectories, we can see the impact of the SNR on the prediction performance, in which the prediction performance with a high SNR value is higher than that with a low SNR value. For example, the controlling Koopman AE with $T_{t} = 700$ and $\mathrm{SNR} = 20 \, \mathrm{dBm}$ has higher prediction performance compared to that with the other SNR values, i.e., $\mathrm{SNR} \in \{0, -10 \} \, \mathrm{dBm}$. In addition, the controlling Koopman AE with $T_{t} = 70$ and $\mathrm{SNR} = 20 \mathrm{dBm}$ has almost the same prediction performance compared to that with $T_{t} = 700$ and $\mathrm{SNR} = 0 \, \mathrm{dBm}$ while it has better prediction performance compared to the others with different number of trajectories and low SNR values. This presents the relationship between the SNR value and the number of trajectories in the training dataset, in which a large number of trajectories in the training dataset with a high SNR value has a much larger prediction performance compared to those that have the same number of trajectories with low SNR values. Otherwise, choosing a low number of trajectories in the training dataset with a high SNR value has better prediction performance compared to those with a large number of trajectories in the training dataset with a low SNR value, highlighting the trade-off between the communication payload size and the SNR value in improving the prediction performance.

 Fig.\ref{fig_state_traj_b} and Fig.~\ref{fig_state_traj_d} show the training completion time required to train the sensing Koopman AE and the controlling Koopman AE, respectively for a different number of trajectories in the training dataset, different SNR values, and the same hyperparameter setting. For the same SNR value, it is clear that the training completion time increases as the number of trajectories in the training dataset increases as a result of increasing the payload size of the forward and backward propagation signals to train the split Koopman AE. For the same number of trajectories in the training dataset, it is clear that the training completion time of the split Koopman AE with a large number of trajectories in the training dataset decreases as the SNR value increases. The reason behind this result is that observations received at another part of the split Koopman AE are sufficient and reliable to discover the Koopman invariant subspace. In contrast, decreasing the SNR value for the same large number of trajectories in the training dataset increases the training completion time, leading to a trade-off between the training completion time and the communication cost in terms of the SNR value and the number of trajectories in the training dataset.
  
 \begin{figure}
 \centering
 \includegraphics[ width=0.4\textwidth]{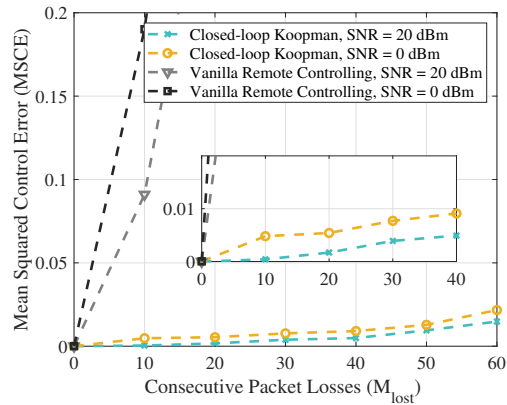} \\
 \caption{Control stability for the split Koopman AE and the non-predictive remote controlling over different values of consecutive packet losses with different SNR values.} \label{fig_MSCE}
 \end{figure}
 
 \begin{figure*}[t!]
    \centering
    \subfigure[Cart position.]{\includegraphics[width=0.4\textwidth]{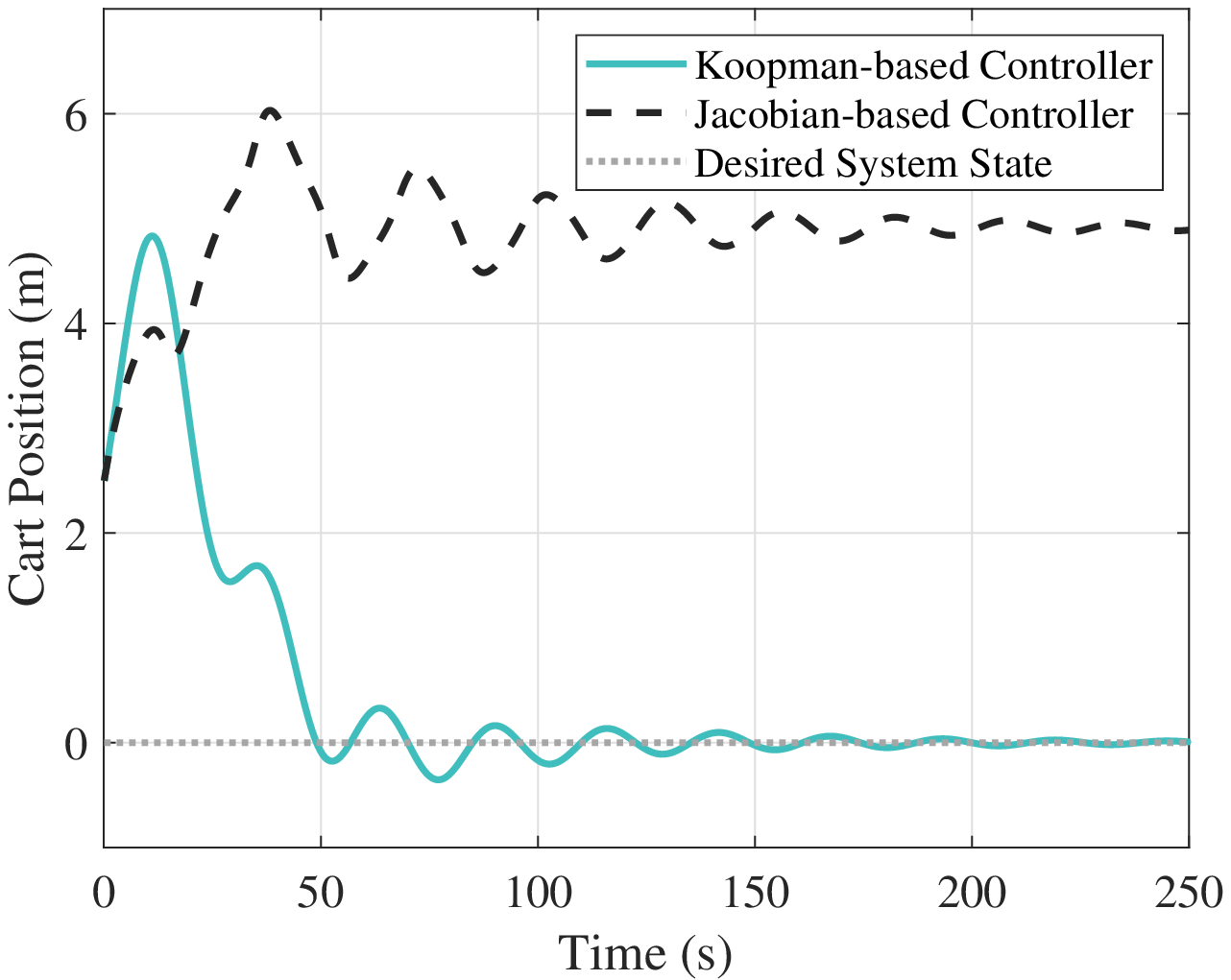}} 
    \subfigure[Cart velocity]{\includegraphics[width=0.4\textwidth]{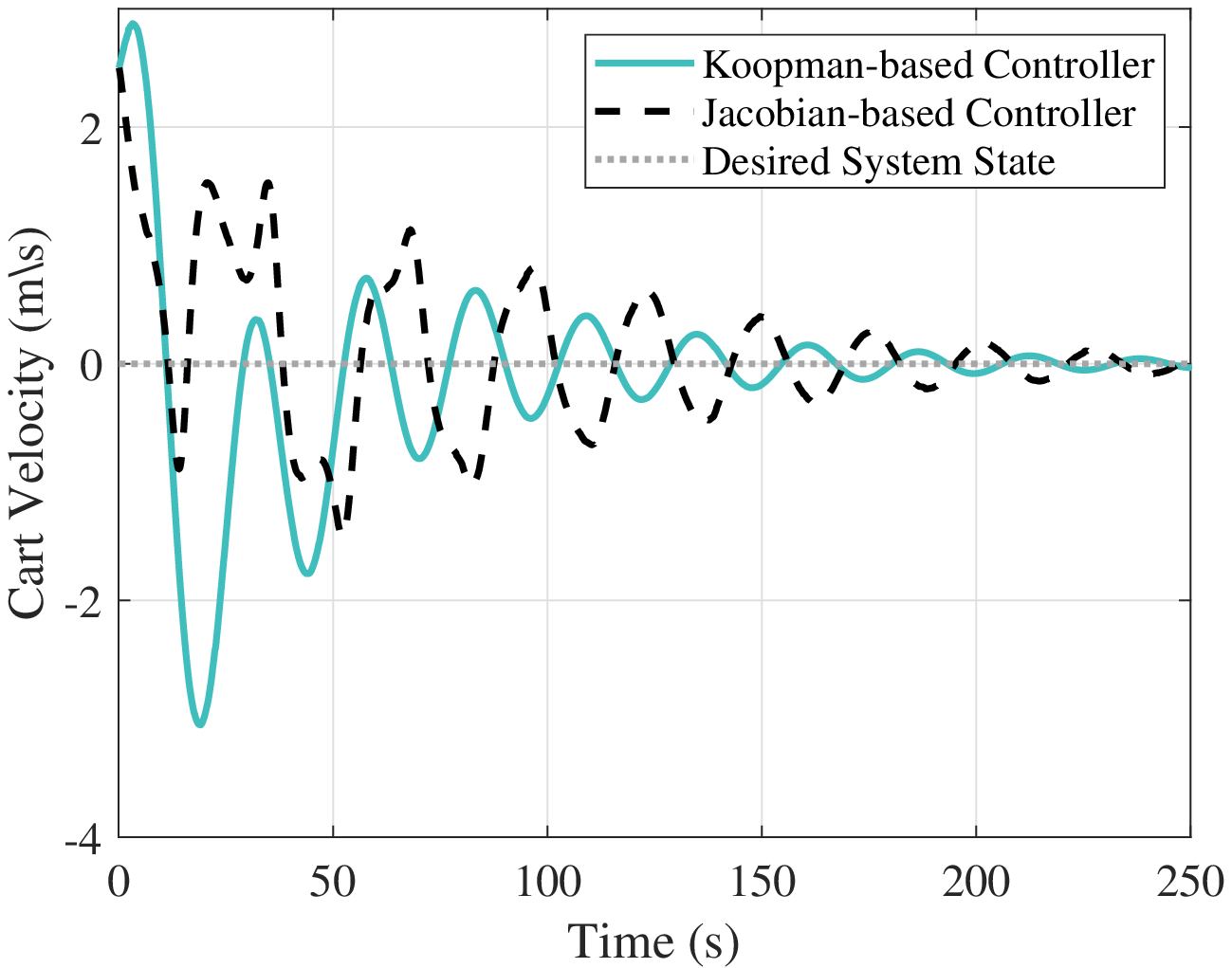}} 
    \subfigure[Pendulum angular position.]{\includegraphics[width=0.4\textwidth]{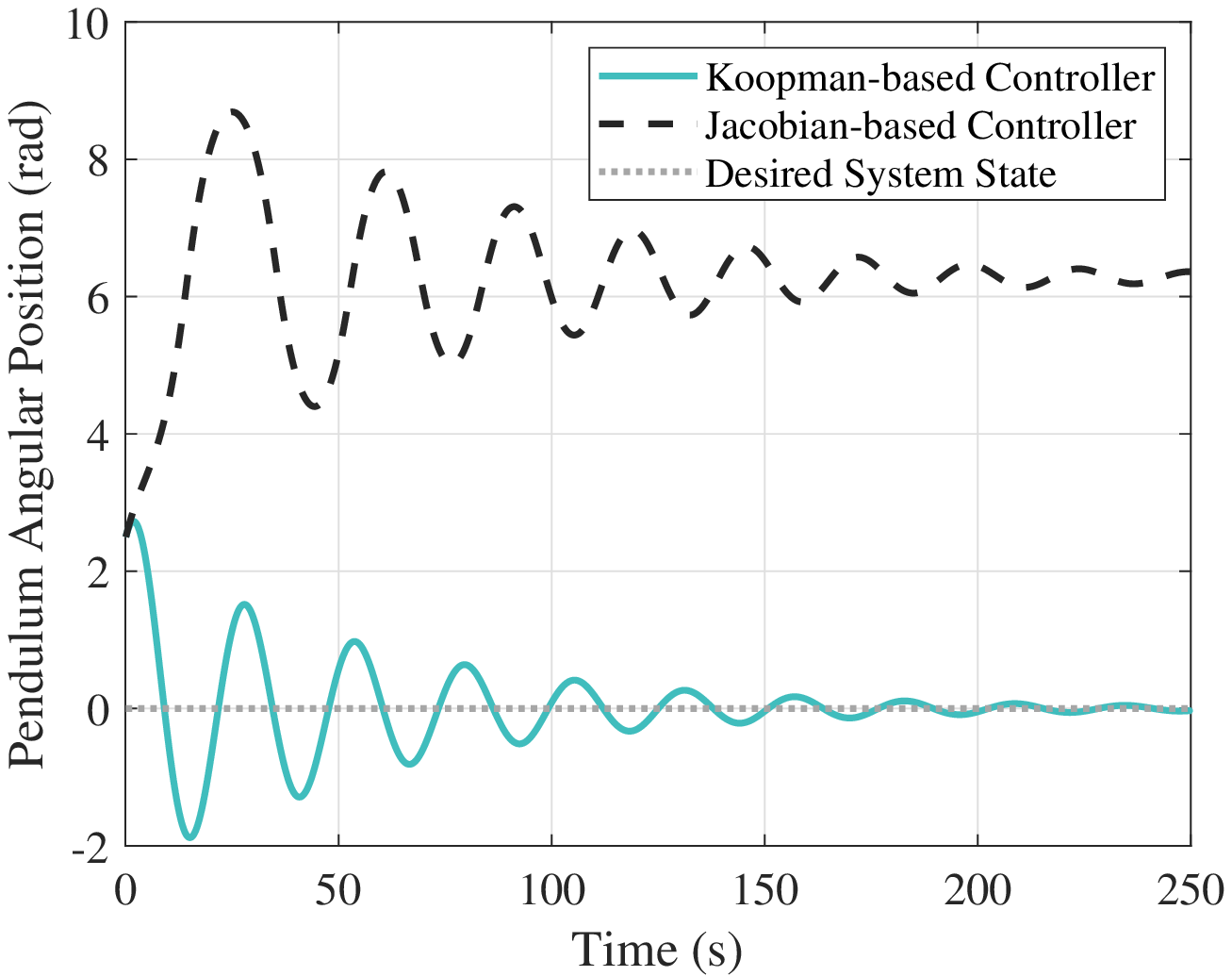}} 
    \subfigure[Pendulum angular velocity.]{\includegraphics[width=0.4\textwidth]{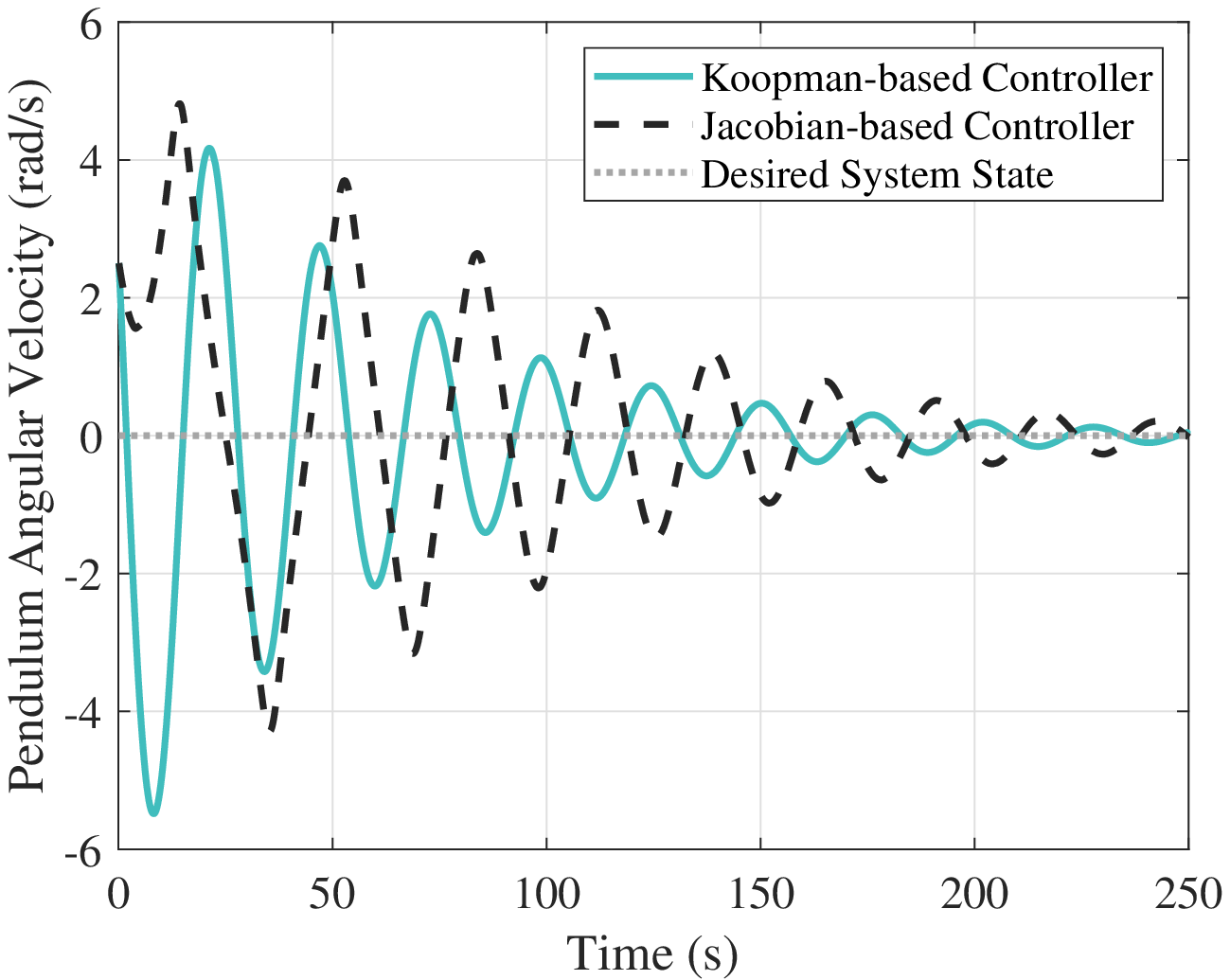}} 
    \caption{Comparison of Koopman-based Controller versus Jacobian-based Controller (a) Cart Position (b) Cart Velocity (c) Pendulum angular position and (d) Pendulum angular velocity. }
    \label{fig_Controller}
\end{figure*}   
   
 \vspace{3pt}\noindent\textbf{Control Stability Vs. Consecutive Packet Losses.}\quad Fig.~\ref{fig_MSCE} illustrates the control stability, i.e., the mean squared control error, over different values of consecutive packet losses with different SNR values. It is clear that the mean squared control error for the split Koopman AE is almost the same as the number of consecutive packet losses increases compared to the non-predictive remote control. The rationale behind this result is that the sensing Koopman AE at the remote controller compensates for the missing system states by locally predicting them after observing a sufficient number of system states along with their latent state representations over favorable channel conditions with different SNR values. Moreover, the actuator locally predicts the missing control action commands utilizing the trained controlling Koopman AE after observing a sufficient number of control action commands, thereby improving the communication efficiency and the control performance. The control performance of the proposed split Koopman AE with $\mathrm{SNR} = 0 \, \mathrm{dBm}$ is less than that of one with $\mathrm{SNR} = 20 \, \mathrm{dBm}$ as a result of the communication reliability impact on the prediction accuracy.

 \vspace{3pt}\noindent \textbf{Koopman-based Vs. Jacobian-based Linear Controller} In Fig.~\ref{fig_Controller}, we show the inverted cart-pole system states for the Koopman-based controller and Jacobian-based controller. Here, the inverted cart-pole system is assumed to be operated from an initial condition far from the equilibrium point, i.e., $\mathbf{x}_{0} = [2.5 \; 2.5 \; 2.5 \; 2.5 ]^{\mathsf{T}}$. Note that the Koopman-based linear controller drives the inverted cart-pole system to the desired states compared to the Jacobian-based controller. The rationale behind failing the Jacobian-based controller in stabilizing the non-linear system is that the initial condition is far from the equilibrium point, $\mathbf{x}_{d} = [0 \; 0 \; 0 \; 0 ]^{\mathsf{T}}$ compared to the Koopman-based linear controller that is robust against the initial condition as a result of obtaining a linearized Koopman invariant subspace reflecting the non-linear dynamics with high accuracy.

%% file: main.bbl
\begin{thebibliography}{10}
\providecommand{\url}[1]{#1}
\csname url@samestyle\endcsname
\providecommand{\newblock}{\relax}
\providecommand{\bibinfo}[2]{#2}
\providecommand{\BIBentrySTDinterwordspacing}{\spaceskip=0pt\relax}
\providecommand{\BIBentryALTinterwordstretchfactor}{4}
\providecommand{\BIBentryALTinterwordspacing}{\spaceskip=\fontdimen2\font plus
\BIBentryALTinterwordstretchfactor\fontdimen3\font minus
  \fontdimen4\font\relax}
\providecommand{\BIBforeignlanguage}[2]{{%
\expandafter\ifx\csname l@#1\endcsname\relax
\typeout{** WARNING: IEEEtran.bst: No hyphenation pattern has been}%
\typeout{** loaded for the language `#1'. Using the pattern for}%
\typeout{** the default language instead.}%
\else
\language=\csname l@#1\endcsname
\fi
#2}}
\providecommand{\BIBdecl}{\relax}
\BIBdecl

\bibitem{girgis2021split}
A.~M. Girgis, H.~Seo, J.~Park, M.~Bennis, and J.~Choi, ``Split learning meets
  {K}oopman theory for wireless remote monitoring and prediction,'' in
  \emph{2021 IEEE 32nd Annual International Symposium on Personal, Indoor and
  Mobile Radio Communications (PIMRC)}.\hskip 1em plus 0.5em minus 0.4em\relax
  IEEE, 2021, pp. 1191--1196.

\bibitem{yang20196g}
P.~Yang, Y.~Xiao, M.~Xiao, and S.~Li, ``6{G} wireless communications: Vision
  and potential techniques,'' \emph{IEEE Network}, vol.~33, no.~4, pp. 70--75,
  2019.

\bibitem{saad2019vision}
W.~Saad, M.~Bennis, and M.~Chen, ``A vision of 6{G} wireless systems:
  Applications, trends, technologies, and open research problems,'' \emph{IEEE
  network}, vol.~34, no.~3, pp. 134--142, 2019.

\bibitem{latva2020key}
M.~Latva-aho, K.~Lepp{\"a}nen, F.~Clazzer, and A.~Munari, ``Key drivers and
  research challenges for 6{G} ubiquitous wireless intelligence,'' 2020.

\bibitem{park2017wireless}
P.~Park, S.~Coleri~Ergen, C.~Fischione, C.~Lu, and K.~H. Johansson, ``Wireless
  network design for control systems: A survey,'' \emph{IEEE Communications
  Surveys Tutorials}, vol.~20, no.~2, pp. 978--1013, Dec. 2017.

\bibitem{meng2004remote}
C.~Meng, T.~Wang, W.~Chou, S.~Luan, Y.~Zhang, and Z.~Tian, ``Remote surgery
  case: robot-assisted teleneurosurgery,'' in \emph{IEEE International
  Conference on Robotics and Automation, 2004. Proceedings. ICRA'04. 2004},
  vol.~1.\hskip 1em plus 0.5em minus 0.4em\relax IEEE, 2004, pp. 819--823.

\bibitem{liu2019taming}
C.-F. Liu and M.~Bennis, ``Taming the tail of maximal information age in
  wireless industrial networks,'' \emph{IEEE Communications Letters}, vol.~23,
  no.~12, pp. 2442--2446, 2019.

\bibitem{zeng2019joint}
T.~Zeng, O.~Semiari, W.~Saad, and M.~Bennis, ``Joint communication and control
  for wireless autonomous vehicular platoon systems,'' \emph{IEEE Transactions
  on Communications}, vol.~67, no.~11, pp. 7907--7922, 2019.

\bibitem{razzaghi2021real}
P.~Razzaghi, E.~Al~Khatib, S.~Bakhtiari, and Y.~Hurmuzlu, ``Real time control
  of tethered satellite systems to de-orbit space debris,'' \emph{Aerospace
  Science and Technology}, vol. 109, p. 106379, 2021.

\bibitem{Docomo:18}
M.~Iwabuch, A.~Benjebbour, Y.~Kishiyama, and Y.~Okumura, ``Field experiments on
  {5G} ultra-reliable low-latency communication {(URLLC)},'' \emph{NTT Docomo
  Technical Journal}, vol.~20, no.~1, pp. 14--23, Jul. 2018.

\bibitem{angjelichinoski2019statistical}
M.~Angjelichinoski, K.~F. Trillingsgaard, and P.~Popovski, ``A statistical
  learning approach to ultra-reliable low latency communication,'' \emph{IEEE
  Transactions on Communications}, vol.~67, no.~7, pp. 5153--5166, 2019.

\bibitem{PetarURLLC:17}
P.~Popovski, J.~J. Nielsen, C.~Stefanovic, E.~de~Carvalho, E.~G. Str\"{o}m,
  K.~F. Trillingsgaard, A.~Bana, D.~Kim, R.~Kotaba, J.~Park, and R.~B.
  S\o{}rensen, ``Wireless access for ultra-reliable low-latency communication
  ({URLLC}): Principles and building blocks,'' \emph{IEEE Network}, vol.~32,
  no.~2, pp. 16--23, Mar. 2018.

\bibitem{bennis2018ultrareliable}
M.~Bennis, M.~Debbah, and H.~V. Poor, ``Ultra-reliable and low-latency wireless
  communication: Tail, risk, and scale,'' \emph{Proceedings of the IEEE}, vol.
  106, no.~10, pp. 1834--1853, 2018.

\bibitem{park2017coverage}
J.~Park and P.~Popovski, ``Coverage and rate of downlink sequence transmissions
  with reliability guarantees,'' \emph{IEEE Wireless Communications Letters},
  vol.~6, no.~6, pp. 722--725, 2017.

\bibitem{bahdanau2014neural}
D.~Bahdanau, K.~Cho, and Y.~Bengio, ``Neural machine translation by jointly
  learning to align and translate,'' \emph{arXiv preprint arXiv:1409.0473},
  2014.

\bibitem{cho2014learning}
K.~Cho, B.~Van~Merri{\"e}nboer, C.~Gulcehre, D.~Bahdanau, F.~Bougares,
  H.~Schwenk, and Y.~Bengio, ``Learning phrase representations using rnn
  encoder-decoder for statistical machine translation,'' \emph{arXiv preprint
  arXiv:1406.1078}, 2014.

\bibitem{Virtual_Reality}
M.~S. Elbamby, C.~Perfecto, M.~Bennis, and K.~Doppler, ``Toward low-latency and
  ultra-reliable virtual reality,'' \emph{IEEE Network}, vol.~32, no.~2, pp.
  78--84, 2018.

\bibitem{koopman1931hamiltonian}
B.~O. Koopman, ``Hamiltonian systems and transformation in {H}ilbert space,''
  \emph{Proceedings of the national academy of sciences of the united states of
  america}, vol.~17, no.~5, p. 315, 1931.

\bibitem{birkhoff1932recent}
G.~D. Birkhoff and B.~O. Koopman, ``Recent contributions to the ergodic
  theory,'' \emph{Proceedings of the National Academy of Sciences of the United
  States of America}, vol.~18, no.~3, p. 279, 1932.

\bibitem{Vepakomma:2018:Splita}
P.~Vepakomma, O.~Gupta, T.~Swedish, and R.~Raskar, ``Split learning for health:
  Distributed deep learning without sharing raw patient data,'' in \emph{ICLR
  Wksp. AI for Social Good, New Orleans, Louisiana, USA}, May 2019.

\bibitem{park2019wireless}
J.~Park, S.~Samarakoon, M.~Bennis, and M.~Debbah, ``Wireless network
  intelligence at the edge,'' \emph{Proceedings of the IEEE}, vol. 107, no.~11,
  pp. 2204--2239, 2019.

\bibitem{theis2017lossy}
L.~Theis, W.~Shi, A.~Cunningham, and F.~Husz{\'a}r, ``Lossy image compression
  with compressive autoencoders,'' \emph{arXiv preprint arXiv:1703.00395},
  2017.

\bibitem{lusch2018deep}
B.~Lusch, J.~N. Kutz, and S.~L. Brunton, ``Deep learning for universal linear
  embeddings of nonlinear dynamics,'' \emph{Nature communications}, vol.~9,
  no.~1, pp. 1--10, 2018.

\bibitem{CoCoCo}
G.~Zhao, M.~A. Imran, Z.~Pang, Z.~Chen, and L.~Li, ``Toward real-time control
  in future wireless networks: Communication-control co-design,'' \emph{IEEE
  Communications Magazine}, vol.~57, no.~2, pp. 138--144, 2019.

\bibitem{Gigis21_twoGPR}
A.~M. Girgis, J.~Park, M.~Bennis, and M.~Debbah, ``Predictive control and
  communication co-design via two-way gaussian process regression and aoi-aware
  scheduling,'' \emph{IEEE Transactions on Communications}, pp. 1--1, 2021.

\bibitem{yu2019event}
M.~Yu, S.~Cai, and V.~K. Lau, ``Event-driven sensor scheduling for
  mission-critical control applications,'' \emph{IEEE Transactions on Signal
  Processing}, vol.~67, no.~6, pp. 1537--1549, 2019.

\bibitem{WSN}
J.~Li, P.~Zeng, X.~Zong, M.~Zheng, and X.~Zhang, ``Communication and control
  co-design for wireless sensor networked control systems,'' in
  \emph{Proceeding of the 11th World Congress on Intelligent Control and
  Automation}, 2014, pp. 156--161.

\bibitem{gatsis2015opportunistic}
K.~Gatsis, M.~Pajic, A.~Ribeiro, and G.~J. Pappas, ``Opportunistic control over
  shared wireless channels,'' \emph{IEEE Transactions on Automatic Control},
  vol.~60, no.~12, pp. 3140--3155, 2015.

\bibitem{BoChang}
B.~Chang, L.~Zhang, L.~Li, G.~Zhao, and Z.~Chen, ``Optimizing resource
  allocation in urllc for real-time wireless control systems,'' \emph{IEEE
  Transactions on Vehicular Technology}, vol.~68, no.~9, pp. 8916--8927, 2019.

\bibitem{eisen2019control}
M.~Eisen, M.~M. Rashid, K.~Gatsis, D.~Cavalcanti, N.~Himayat, and A.~Ribeiro,
  ``Control aware radio resource allocation in low latency wireless control
  systems,'' \emph{IEEE Internet of Things Journal}, vol.~6, no.~5, pp.
  7878--7890, 2019.

\bibitem{eisen2019control2}
------, ``Control aware communication design for time sensitive wireless
  systems,'' in \emph{ICASSP 2019-2019 IEEE International Conference on
  Acoustics, Speech and Signal Processing (ICASSP)}.\hskip 1em plus 0.5em minus
  0.4em\relax IEEE, 2019, pp. 4584--4588.

\bibitem{wang2016stabilization}
W.~Wang, R.~Postoyan, D.~Ne{\v{s}}i{\'c}, and W.~M.~H. Heemels, ``Stabilization
  of nonlinear systems using state-feedback periodic event-triggered
  controllers,'' in \emph{2016 IEEE 55th Conference on Decision and Control
  (CDC)}.\hskip 1em plus 0.5em minus 0.4em\relax IEEE, 2016, pp. 6808--6813.

\bibitem{aranda2017design}
E.~Aranda-Escol{\'a}stico, M.~Abdelrahim, M.~Guinaldo, S.~Dormido, and
  W.~Heemels, ``Design of periodic event-triggered control for polynomial
  systems: A delay system approach,'' \emph{IFAC-PapersOnLine}, vol.~50, no.~1,
  pp. 7887--7892, 2017.

\bibitem{6761063}
R.~Postoyan, A.~Anta, W.~Heemels, P.~Tabuada, and D.~Nešić, ``Periodic
  event-triggered control for nonlinear systems,'' in \emph{52nd IEEE
  Conference on Decision and Control}, 2013, pp. 7397--7402.

\bibitem{li2013network}
H.~Li and Y.~Shi, ``Network-based predictive control for constrained nonlinear
  systems with two-channel packet dropouts,'' \emph{IEEE Transactions on
  Industrial Electronics}, vol.~61, no.~3, pp. 1574--1582, 2013.

\bibitem{de2008lyapunov}
D.~M. de~la Pe{\~n}a and P.~D. Christofides, ``Lyapunov-based model predictive
  control of nonlinear systems subject to data losses,'' \emph{IEEE
  Transactions on Automatic Control}, vol.~53, no.~9, pp. 2076--2089, 2008.

\bibitem{fadali2013digital}
M.~S. Fadali and A.~Visioli, \emph{Digital control engineering: analysis and
  design}.\hskip 1em plus 0.5em minus 0.4em\relax Academic Press, 2013.

\bibitem{tailor2011linearization}
M.~R. Tailor and P.~Bhathawala, ``Linearization of nonlinear differential
  equation by {T}aylor's series expansion and use of {J}acobian linearization
  process,'' \emph{International Journal of Theoretical and Applied Science},
  vol.~4, no.~1, pp. 36--38, 2011.

\bibitem{mezic2017koopman}
I.~Mezic, ``Koopman operator spectrum and data analysis,'' \emph{arXiv preprint
  arXiv:1702.07597}, 2017.

\bibitem{schmid2010dynamic}
P.~J. Schmid, ``Dynamic mode decomposition of numerical and experimental
  data,'' \emph{Journal of fluid mechanics}, vol. 656, pp. 5--28, 2010.

\bibitem{proctor2016dynamic}
J.~L. Proctor, S.~L. Brunton, and J.~N. Kutz, ``Dynamic mode decomposition with
  control,'' \emph{SIAM Journal on Applied Dynamical Systems}, vol.~15, no.~1,
  pp. 142--161, 2016.

\bibitem{7799268}
A.~Surana, ``Koopman operator based observer synthesis for control-affine
  nonlinear systems,'' in \emph{2016 IEEE 55th Conference on Decision and
  Control (CDC)}, 2016, pp. 6492--6499.

\bibitem{korda2018linear}
M.~Korda and I.~Mezi{\'c}, ``Linear predictors for nonlinear dynamical systems:
  Koopman operator meets model predictive control,'' \emph{Automatica},
  vol.~93, pp. 149--160, 2018.

\bibitem{proctor2018generalizing}
J.~L. Proctor, S.~L. Brunton, and J.~N. Kutz, ``Generalizing koopman theory to
  allow for inputs and control,'' \emph{SIAM Journal on Applied Dynamical
  Systems}, vol.~17, no.~1, pp. 909--930, 2018.

\bibitem{brunton2021modern}
S.~L. Brunton, M.~Budi{\v{s}}i{\'c}, E.~Kaiser, and J.~N. Kutz, ``Modern
  {K}oopman theory for dynamical systems,'' \emph{arXiv preprint
  arXiv:2102.12086}, 2021.

\bibitem{kaiser2017data}
E.~Kaiser, J.~N. Kutz, and S.~L. Brunton, ``Data-driven discovery of {K}oopman
  eigenfunctions for control,'' \emph{arXiv preprint arXiv:1707.01146}, 2017.

\bibitem{mamakoukas2019local}
G.~Mamakoukas, M.~Castano, X.~Tan, and T.~Murphey, ``Local {K}oopman operators
  for data-driven control of robotic systems,'' in \emph{Robotics: science and
  systems}, 2019.

\bibitem{bonnert2020estimating}
M.~Bonnert and U.~Konigorski, ``Estimating {K}oopman invariant subspaces of
  excited systems using artificial neural networks,'' \emph{IFAC-PapersOnLine},
  vol.~53, no.~2, pp. 1156--1162, 2020.

\bibitem{guastello2009introduction}
S.~J. Guastello and L.~S. Liebovitch, ``Introduction to nonlinear dynamics and
  complexity.'' 2009.

\bibitem{koda2020communication}
Y.~Koda, J.~Park, M.~Bennis, K.~Yamamoto, T.~Nishio, M.~Morikura, and
  K.~Nakashima, ``Communication-efficient multimodal split learning for mmwave
  received power prediction,'' \emph{IEEE Communications Letters}, vol.~24,
  no.~6, pp. 1284--1288, 2020.

\bibitem{brunton2016koopman}
S.~L. Brunton, B.~W. Brunton, J.~L. Proctor, and J.~N. Kutz, ``Koopman
  invariant subspaces and finite linear representations of nonlinear dynamical
  systems for control,'' \emph{PloS one}, vol.~11, no.~2, p. e0150171, 2016.

\bibitem{bemporad2002explicit}
A.~Bemporad, M.~Morari, V.~Dua, and E.~N. Pistikopoulos, ``The explicit linear
  quadratic regulator for constrained systems,'' \emph{Automatica}, vol.~38,
  no.~1, pp. 3--20, 2002.

\bibitem{xiao2020deep}
Y.~Xiao, X.~Zhang, X.~Xu, X.~Liu, and J.~Liu, ``A deep learning framework based
  on koopman operator for data-driven modeling of vehicle dynamics,''
  \emph{arXiv preprint arXiv:2007.02219}, 2020.

\bibitem{sutskever2013importance}
I.~Sutskever, J.~Martens, G.~Dahl, and G.~Hinton, ``On the importance of
  initialization and momentum in deep learning,'' in \emph{International
  conference on machine learning}.\hskip 1em plus 0.5em minus 0.4em\relax PMLR,
  2013, pp. 1139--1147.

\bibitem{hangos2004analysis}
K.~M. Hangos, J.~Bokor, and G.~Szederk{\'e}nyi, \emph{Analysis and control of
  nonlinear process systems}.\hskip 1em plus 0.5em minus 0.4em\relax Springer,
  2004, vol.~13.

\end{thebibliography}
